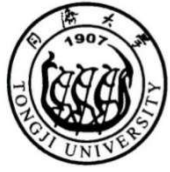

# 同濟大學
## TONGJI UNIVERSITY

## 硕士学位论文

# 基于探地雷达的半刚性基层
# 脱空识别及注浆效果评价

姓　　名：　郑家麒

学　　号：　1433944

所在院系：　交通运输工程学院

学科门类：　工学

学科专业：　交通运输工程

指导教师：　钱劲松　副教授

二〇一七年五月

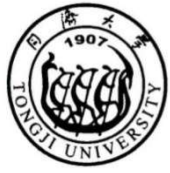

TONGJI UNIVERSITY

A dissertation submitted to

Tongji University in conformity with the requirements for

the degree of Master of Philosophy

# GPR-based Detection of Voids and Evaluation of Grouting Under Semi-rigid Basement

Candidate: Jiaqi ZHENG

Student Number: 1433944

School: School of Transportation Engineering

Discipline: Engineering

Major: Transportation Engineering

Supervisor: Associate Prof. Jinsong QIAN

May, 2017

# 学位论文版权使用授权书

本人完全了解同济大学关于收集、保存、使用学位论文的规定，同意如下各项内容：按照学校要求提交学位论文的印刷本和电子版本；学校有权保存学位论文的印刷本和电子版，并采用影印、缩印、扫描、数字化或其它手段保存论文；学校有权提供目录检索以及提供本学位论文全文或者部分的阅览服务；学校有权按有关规定向国家有关部门或者机构送交论文的复印件和电子版；在不以赢利为目的的前提下，学校可以适当复制论文的部分或全部内容用于学术活动。

# 同济大学学位论文原创性声明

本人郑重声明：所呈交的学位论文，是本人在导师指导下，进行研究工作所取得的成果。除文中已经注明引用的内容外，本学位论文的研究成果不包含任何他人创作的、已公开发表或者没有公开发表的作品的内容。对本论文所涉及的研究工作做出贡献的其他个人和集体，均已在文中以明确方式标明。本学位论文原创性声明的法律责任由本人承担。

学位论文作者签名：

年　　　　月　　　　日

# 摘要

道路的半刚性基层脱空是一种常见的隐性缺陷，其检测和维修存在较大难度，注浆修复的效果也不易评价。研究半刚性基层脱空的判别及注浆评价的方法，对于道路病害检测及养护具有重要意义。本论文采用理论分析、数值模拟、实测数据分析等手段，研究了不同类型和尺寸的基底脱空在雷达图像上的特征，提出了基底脱空的判别及尺寸估算方法，并研究了信号处理的流程和效果。本文的主要研究内容如下：

（1）理论分析了不同类型（充气、充水、注浆修复）和尺寸（高度、水平尺寸）的基底脱空在探地雷达 A 扫描及 B 扫描图像上的特征，并使用基于 FDTD 方法的 gprMax 软件数值模拟发射波在道路中的传播，验证了了基底脱空在探地雷达图像上的特征。此外，分析了探地雷达天线频率对脱空探测的影响。

（2）研究了 0.01m-0.3m 高度范围内的基底脱空的判别及高度估算方法。对于充气脱空，分别使用最小二乘系统辨识法和吉洪诺夫正则化解卷积法实现了脱空判别及尺寸估算，并讨论了二者各自的适用条件。对于充水及注浆脱空，使用基于反射波幅的介电常数法实现了脱空判别。

（3）研究了 0.04m-0.52m 水平尺寸范围内基底脱空的水平尺寸估算方法。依据充气脱空的仿真结果，选取了基底脱空的水平尺寸的计算指标，并回归得到了基底脱空水平尺寸计算式。

（4）讨论了信号处理的流程，并依据实测数据研究了各种信号处理技术在去除干扰及补偿衰减等方面的实际效果，并讨论了这些预处理手段对基底脱空在图像上的特征的影响。

**关键词：** 路面、半刚性基层、脱空、探地雷达、脱空判别、脱空尺寸估算、信号处理

# Abstract

The void underneath semi-rigid base is a common defect in roads. There are some difficulties in the detection and repair for this kind of hidden damage, as well as in the evaluation of the effects of grouting treatment. For the detection and maintenance of roads, it is essential to study the detection and judging for voids underneath base and the evaluation of the spread of grout. Through theoretical analysis, numerical simulation and analysis of real data, this research generated the characteristics of under-base voids of different types and dimensions on GPR images, proposed the detecting and dimension-measuring methods for under-base voids, and studied the process and effects of data analysis techniques.

(1) The characteristics of under-base voids of different types (air-filled, water-filled or grout-treated) and dimensions (height and horizontal dimensions), on A-scan and B-scan GPR image respectively, were analyzed theoretically. The gprMax software which is based on the FDTD method was employed to simulate the transmission of GPR wave within the road structure, which certified the conclusion of theoretical analysis of the image characteristics of voids. In addition, the influence of antenna frequency on the detection for voids are also analyzed.

(2) Approaches for detecting voids and for estimating its height were studied, focusing on voids with a height ranging from 0.01m to 0.3m. The Least Squares Method of System Identification and the Tikhonov Regularized Deconvolution were both successfully applied to the detection and dimension estimation of air-filled voids, and their application conditions were discussed. As for water-filled and grout-treated voids, the reflection-amplitude-based dielectric constant method was used for void detection.

(3) The approach for estimating the horizontal dimension of voids was studied, focusing on voids with a length ranging from 0.04m to 0.52m. According to the simulating results of air-filled voids, the estimation index was selected, and the linear calculation formula for length of voids was generated by regression analysis.

(4) The data processing process was discussed. Also, the effects of different data processing techniques were studied in terms of noise filtering and attenuation compensation, and their influence on the image characteristics was also discussed.

# 目录

# 第一章 引言

## 1.1 研究背景

在道路养护运营阶段，需要及时掌握全面、准确的道路设施状况信息，从而在病害扩展前尽早采取养护措施，降低养护费用，并避免由于道路病害造成的交通事故。

半刚性基层是我国沥青路面中广泛采用的基层形式，具有抗压强度高、造价低等优点。但半刚性基层在土基不均匀沉降等的作用下会产生基层脱空等病害，这些损坏具有隐蔽性，不易被发现和及时处理，最终可能导致面层的病害。基底脱空上方的半刚性基层局部受力状态类似于简支梁，其底部受拉，容易产生疲劳裂缝，进而可能导致路面局部出现病害。

由于半刚性基层底部脱空是一种隐性病害，其检测和修复存在一定的难度。一方面，基底脱空隐藏在面层和基层下方，无损检测脱空必然受到面层和基层的干扰；另一方面，基底脱空成因较为复杂，脱空形态的差异性较大，脱空内充水或充气情况也会对测试结果产生影响。此外，为更精确判断病害程度，需要检测脱空的面积和高度，对检测技术提出更高的要求。

目前，水泥混凝土路面的板下脱空的无损检测方面相对有一些研究。大多数研究集中在脱空情况下电磁波的正演分析，实际工程中用探地雷达检测板下脱空的研究成果较少。很大一部分原因是混凝土板下脱空高度只有毫米级，超出了现有探地雷达的测量精度。James C. Ni[1]等人研究了注浆前后脱空位置的雷达图像的区别，发现注浆前的"倒V"形注浆后消失，但该结论只是定性描述，缺乏定量分析。

类似地，沥青路面半刚性基层与土基之间也可能出现脱空。同济大学的陈南[2,3]等人研究了基于 FWD 弯沉盆的半刚性基层脱空判识方法，提出采用弯沉比和面积指标作为基层脱空判定指标和相应的判定标准，并且建立了估算脱空面积的经验回归模型。而探地雷达作为另一种无损检测手段，在半刚性基层底部脱空的检测中有应用的潜力，故本文以此作为研究内容。

因此，研究探地雷达在半刚性基层底脱空及注浆效果评价中的应用，能够与现有弯沉盆判别方法互为补充，为半刚性基层底脱空提供快速、精确的检测手段。

## 1.2 国内外研究现状

针对本课题的研究问题，下面分别从结构层脱空识别技术、结构层脱空注浆效果评价两个方面进行文献综述。

### 1.2.1 结构层脱空识别技术

半刚性基层底的脱空是一种隐性病害，用传统手段很难达到及时有效的检测。往往路面出现反射裂缝、沉陷等病害，通过钻芯取样才会发现基层脱空。从探地雷达检测识别的角度来说，沥青路面半刚性基层的脱空与水泥混凝土路面面层和基层之间的脱空具有相似性，因此以下也总结刚性路面板底脱空的识别方法，作为本研究的参考。

（1）沥青半刚性基层底脱空识别方法

由于基层底脱空的隐蔽性，其检测相对困难。传统检测方法钻芯取样法为有损检测，速度慢、不能连续检测、对路面有损坏。

根据 FWD 检测得到的弯沉盆可以用来检测基层脱空。同济大学的陈南[2,3]研究了基于弯沉盆的半刚性基层脱空识别方法。基于 FWD 弯沉盆的方法在水泥混凝土面板下方脱空检测中也有相关研究[4-8]。

探地雷达技术（Ground Penetrating Radar，GPR）是道路隐性病害无损检测的常用手段之一，国内外对于探地雷达在道路工程中的应用主要集中在厚度估算[9-13]、沥青层松散剥落识别[14-17]、沥青层压实检测[18-20]、沥青混合料离析检测[21,22]、沥青路面裂缝识别[23-26]、桥面板钢筋检测[27-30]等方面。对于结构层脱空有一些研究，但尚不成熟[31]。

根据电磁波在层状介质中的传播原理，当雷达波从高介电常数的材料传播至低介电常数的材料时，会在界面上产生一个负峰值的反射波，而当从低介电常数的材料传播至高介电常数的材料时，会在界面上产生一个正峰值的反射波。根据这个原理，当雷达波进入介电常数较小的气充脱空（介电常数接近 1）时，会在波形图上显示为一个负波峰，而当雷达波进入介电常数较大的水充脱空（介电常数接近 80）时，会在波形图上显示为一个正波峰。根据该理论，可以实现脱空的探地雷达识别。

同济大学的陈南[3]在实际道路上给出了脱空路段的雷达图像，展示了脱空位置的雷达图像的特点，但没能提出探地雷达识别脱空的方法和标准。

（2）刚性路面板底脱空识别方法

刚性路面板底脱空识别研究相对较多，主要有外观评定法、贝克曼梁弯沉判断、FWD 检测法、探地雷达检测法[32, 33]。由于刚性路面板底脱空和半刚性基层

底脱空有一定相似性，本课题可将刚性路面板底脱空的研究方法和成果作为参考。

① 外观评定法

外观评定法是指根据经验判别路面板底是否有脱空，主要有以下 6 种[34]：唧泥现象判别；重车通过时，车内颠簸，水泥板垂直移动或板块松动；相邻板错台，路基下沉一侧脱空；横纵缝填缝料出现空缝状态；板角断裂；重锤敲打有回响。

② 贝克曼梁弯沉判断

贝克曼梁弯沉仪构造操作简单，被广泛用于道路的弯沉检测。贝克曼梁检测法的工作原理[34]为杠杆原理，用支撑在表架上的百分比读取测试梁前臂端点的位移，据此推算出前臂端点所测出的车轮缝隙的回弹弯沉，测试结果为测点静态弯沉的最大峰值。

贝克曼梁弯沉法判别脱空分为单一弯沉值法和相邻板角弯沉差法。我国《公路水泥混凝土路面养护规范》[35]规定，采用标准荷载作用下 0.2mm 弯沉值作为脱空判别标准。美国沥青混凝土加铺层设计手册采用 40KN 轴载作用下相邻板角静弯沉差 0.05mm 作为脱空判别标准。

贝克曼梁检测方法的优点是结构简单，操作方便，但也存在一些缺点。首先，贝克曼梁测试法使用的是静压，而车辆荷载为动载，该差异会影响正确判定结构承载力。其次，只测得单点最大弯沉，不能得到弯沉盆，而梁的支点可能落在弯沉盆里，影响结果精度。最后，弯沉值受温度、湿度、人工操作影响，且测量速度慢。

③ FWD 检测法

落锤式弯沉仪（FWD）是一种脉冲动力弯沉仪。该设备通过使预设重量的落锤自由落地，来模拟车辆对路面结构的瞬时动荷载。通过传感器采集路面的弯沉盆特征，包括落锤点最大弯沉、弯沉盆曲线和弯沉盆时程曲线。操作规程和技术工作条件见《公路路基路面现场测试规程》[36]。

根据动态弯沉盆判断脱空的方法有截距法和弯沉盆变异法[37, 38]。截距法是指分别测得不同冲击荷载水平下板角弯沉大小，回归得到荷载-弯沉关系式并绘制直线，将该线延长，若与坐标轴的截距大于 0.05mm，则认为脱空。弯沉盆变异法的原理是路面板底存在脱空时弯沉盆曲线会发生变异，拟合出无脱空路面板的弯沉盆曲线，进而将实测路面弯沉盆与之对比判断是否脱空。

④ 探地雷达检测法

相对于沥青路面半刚性基层底脱空，国内外对于探地雷达在刚性路面板底脱空的探测中有更多的研究，目前国外的研究者得到了一些研究成果[39]。

长安大学的万捷[40]、柴福斌[41]以及重庆交通大学的魏国杰[34]研究了基于 GPR 的水泥混凝土路面底部脱空检测，对于是否脱空的判断与钻芯的结果一致

性较好，但均通过人工判别图像，没有固定的判别指标及标准，无法进行重复试验验证；且没有给出脱空位置实际扫描雷达图像，无法作为后续研究中雷达图像分析的参考。郑州大学的冯晋利[42]、李婧琳[34]用考虑了介电常数的虚部的电磁波正演方法研究了水泥混凝土层底脱空的图像特征，发现 1cm 以上的气充脱空、2cm 以上的水充脱空在图像上较明显，更小尺寸的脱空在图像上不明显；李婧琳建立了基于模式识别和灵敏度分析的反演方法，但该结论只用于模型试验，没有在实际道路检测中验证，仍需进一步研究。

University of Kentucky 的 M. E. Kalinski[43]等人在实际水泥混凝土道路上用探地雷达检测 CTB 基层顶部的脱空情况，取得了良好效果。具体检测过程为：使用 500MHz 的探地雷达，采用网格状测线，探测水泥混凝土面层下方 CTB 基层中的空洞并绘出空洞厚度等高线图，与多点钻芯得到的空洞厚度等高线图比较。研究发现，从单个波的波形图（图 1.1）[43]来看，脱空位置有一个明显的负波峰，且脱空高度与峰值正相关；从沿测线的扫描图像来看（图 1.2）[43]，脱空位置的颜色明显异于周围颜色，不同颜色代表不同波幅。据此得到脱空厚度分布等高线图，与钻芯得到的脱空厚度分布等高线图相近，可识别的脱空最小高度为 0.25 英寸（6mm）。该研究证明了探地雷达在脱空检测中的可行性，但仅用脱空位置的负波峰定性表示脱空程度，没有定量建立脱空高度和波形图指标的关系。

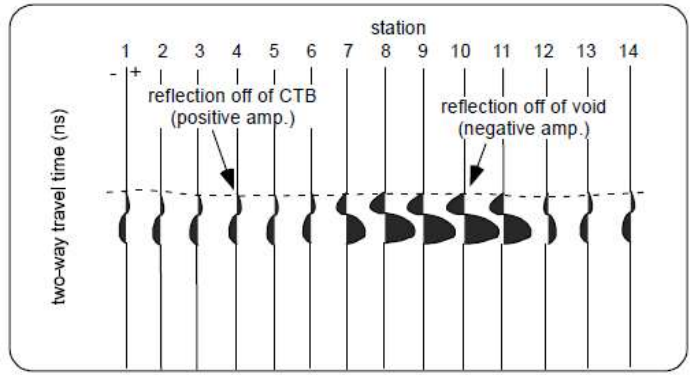

图 1.1 脱空区域雷达反射波瀑布图 [43]

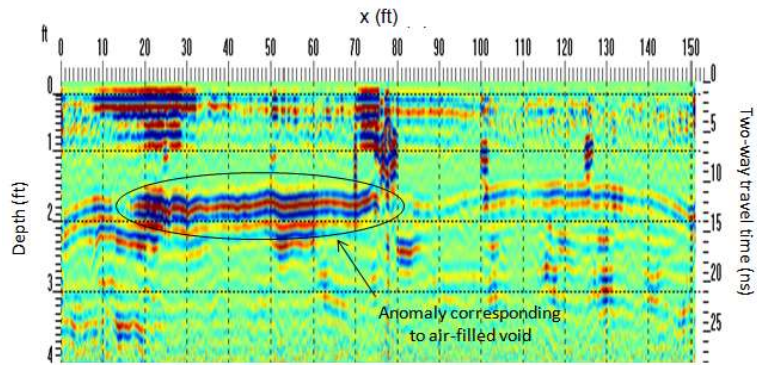

图 1.2 脱空区域沿测线雷达扫描图像[43]

当路基中有输水管道泄露时，土基中的颗粒可能逐渐随水流被冲走，造成较大的土基空洞。这种空洞通常比脱空的尺寸大很多，也更容易被探地雷达识别。

Texas Department of Transportation 的 Dar Hao Chen[44]通过三个实际案例研究了路面下方空洞的探测，使用 400MHz 的雷达成功探测了 50mm 至 2m 的空洞，并通过钻芯、开挖验证。

综上，相对于外观评定法、贝克曼梁法、FWD 法，探地雷达在道路结构层的检测中具有快速、连续、直观、更容易与其他病害区分等优点，值得进一步研究。目前在脱空病害的雷达波正演、实测图像判断等方面有相关研究，但对于实测图像的脱空病害判断目前仍无指标和标准，也缺乏对脱空尺寸和实测雷达图像特征之间关系的研究。

### 1.2.2 结构层脱空注浆效果评价

注浆工程属于隐蔽工程，我国对道面注浆工程效果评价方法尚不完善，实际工程中，一般采用注浆前后 FWD 测试弯沉的对比作为评价。用探地雷达评价注浆效果有一些研究，但尚不完善。

（1）FWD 评价注浆效果

西南交通大学的阳恩慧等人[45]研究了用弯沉评价路基压力注浆效果的方法。为加固路基，阳恩慧等人将水泥粉煤灰浆液充填路基孔隙中，以解决不均匀沉降的问题。通过测定 9 个位置上的路表瞬时变形，得到动载作用下的动态弯沉和弯沉盆，从而反算得到各材料层的动态弹性模量，通过比较土基注浆前后的模量变化来对注浆效果作出评价。结果表明，路基和底基层模量明显提高，注浆效果明显；弯沉模量反算方法能对注浆效果进行结构性评价，但弯沉测试受环境、温度及认为因素的影响。

柴震林[46]等人研究了机场复合道面的注浆效果评价。不同于一般水泥混凝土道面，沥青混凝土材料属于温度敏感性材料，材料模量会随着温度改变而明显变化，因此采用有限元数值模拟确定了温度修正系数用于修正实测弯沉数据；通过比较实测折减系数与理论折减系数，判断基层顶面反应模量是否达到注浆工程设计要求。工程验证表明，该方法可较准确评价复杂条件下的道面注浆效果。

长沙理工大学的曾胜[47]等人以 Winkler 地基下威斯特卡德板角弯沉理论解为基础，利用板中弯沉盆参数求解相关参数，并将脱空定性评价、脱空区充填率评价和不均匀支撑定性评价三大指标作为板底板底注浆效果的检验标准，在试验中得到了验证。

（2）GPR 评价注浆效果

探地雷达识别脱空本质上是利用空气和路面材料的介电常数差异造成的电

磁波反射。类似地，由于注浆材料[48]与路面材料的介电常数不同，因此可使用探地雷达探测注浆材料的扩散情况。

University of Kentucky 的 M. E. Kalinski[43]使用 500MHz 的探地雷达，测得注浆前脱空厚度分布等高线图；注浆后再次检测，原脱空位置的负波峰变为正波峰，说明空洞被注浆材料填充，但注浆后检测是在降雨后进行的，雨水对检测结果的影响无法确定，仍需进一步研究。

National Taipei University of Technology 的 James C. Ni[49, 50]等人对比注浆前后的 GPR 图像，评价注浆效果。该研究发现注浆前脱空位置的雷达波形为一个"倒 V 形"，注浆后该异常波形消失，表明注浆材料填充了脱空。

盾构隧道壁后注浆效果的 GPR 检测与路面结构脱空注浆的 GPR 评价有相似之处，可作为参考。同济大学的黄宏伟[51]等人使用 200MHz 的探地雷达探测壁后注浆效果，发现雷达图像上可以清晰地反映管片及其背后的注浆和土体的情况，但是，由于钢筋混凝土管片的内外表面对电磁波的多次反射，对剖面图的解释造成一定的影响，在路面结构中也可能存在类似问题，需要在研究中注意。Xiongyao Xie[52]等人使用 GPR 和 3D 激光对隧道状况进行了检测，Hai Liu[53]等人研究了隧道壁后注浆效果的 GPR 评价，使用 CMP 方法测算注浆层厚度，采用 FK 滤波处理雷达信号，发现注浆厚度为 16-17cm，略小于设计注浆厚度 20cm。

综上，与目前常用的 FWD 弯沉盆法相比，探地雷达在注浆检测中有受温度影响小、受路面材料影响小、直观等优点，值得进一步研究。

## 1.3  主要研究内容与技术路线

### 1.3.1  主要研究内容

（1）基层脱空病害的电磁波正演

文献综述中已提到，目前对于水泥混凝土道面板底脱空的电磁波正演模拟已经有了相关研究，但缺乏半刚性基层沥青路面的相关研究。本研究基于时域有限差分理论进行探地雷达回波仿真，按照典型的半刚性基层沥青路面的几何尺寸、边界条件和材料参数，预设不同高度、不同面积的基层底部脱空，进行电磁波正演，得到一系列探地雷达 A 扫描和 B 扫描图像；考虑到脱空位置可能有水积存，以及注浆后脱空被注浆材料填充，因此分别正演充气脱空、充水脱空、填充注浆材料的脱空的回波图像，为后面图像分析、指标提取和标准制定奠定基础。

（2）脱空的识别方法及尺寸估算

目前的脱空识别方法主要是依靠技术人员经验，定性判断是否存在脱空。本

研究提出基底脱空的判别及高度估算方法。对于充气脱空，将最小二乘系统辨识的方法和吉洪诺夫正则化解卷积的方法应用于脱空判别及高度估算。对于充水脱空和注浆修复脱空，研究基于反射波幅的介电常数法在脱空判别中的效果。此外，提出基底脱空的水平尺寸的估算方法和回归计算公式。

（3）探地雷达信号处理流程

针对实际扫描信号中的干扰及衰减，讨论信号处理的流程，并依据实测数据研究了各种信号处理技术在去除干扰及补偿衰减等方面的实际效果，包括信道调整、去直流漂移、去直达波、时域滤波、频域滤波、空域滤波、信号增益等处理技术。

### 1.3.2 研究方法和技术路线

论文主要采用以下研究方法：

（1）资料收集和理论框架建立。主要通过对国内外文献资料收集和整理，对基层脱空位置的雷达回波特点进行理论分析，并建立论文的研究框架。

（2）数值仿真。对不同脱空情况的半刚性基层沥青路面结构进行数值仿真。

（3）统计分析方法。根据仿真得到的雷达回波 A 扫描、B 扫描，用相关性分析、数值回归的方法提取指标并得到脱空识别方法、脱空尺寸估算方法。

主要技术路线如图 1.3：

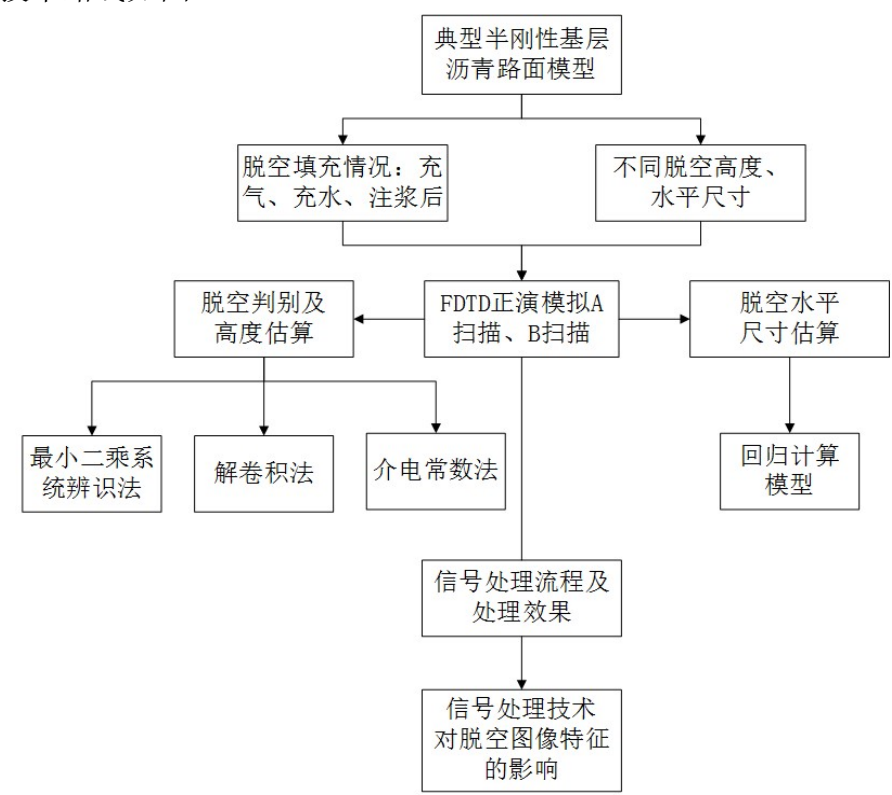

图 1.3 技术路线图

# 第二章 探地雷达理论

探地雷达检测技术的原理为，雷达天线向地面发射电磁波，电磁波在空气及道路材料等介质中传播，在不同介电常数的介质的界面上发生部分发射。通过采集探地雷达回波并分析其特征，可以推断被测结构的几何及材料特性。因此，分析探地雷达设备类型及原理、电磁传播理论以及复合材料的介电常数理论，对于探地雷达在道路检测中的应用研究具有重要意义。此外，道路结构层厚度检测是探地雷达一个相对成熟的应用领域，同时，路面分层以及厚度检测也是道路病害检测中确定病害层位及深度的前提，因此，本章也对 GPR 在道路结构层厚度检测中的应用予以讨论。

## 2.1 探地雷达设备

各种探地雷达都是利用超高频电磁波探测地下介质的设备，但是不同种类的设备发射的电磁波形式各有不同，设备的部件组成也有所差异。

### 2.1.1 探地雷达的分类

从不同的方面考虑，探地雷达有不同的分类方法。根据调制技术的不同，探地雷达可分为三类：调频 GPR，合成脉冲 GPR（也称步进式 GPR），调幅 GPR（也称脉冲式 GPR）。

（1）调频 GPR

通过压控振荡器控制，连续调频雷达的发射频率随时间有规律地变化，因此可以将频率高低作为时间的标记。某一时刻发射频率和接收频率的差值称为拍频。根据发射频率的变化规律，可以将拍频转化为接收到的信号距离发射时的时间差即时延。由该时延和电磁波的传播速度，可以计算探地雷达天线与探测目标之间的距离。

该种类型的探地雷达较适用于浅层单一目标的探测。而对于道路结构等多层体系，多个层位的反射波会导致接收到的信号为叠加信号，数据分析困难。

（2）合成脉冲 GPR（步进式 GPR）

合成脉冲 GPR 也被称为步进式 GPR。与调频 GPR 相同，步进式 GPR 也是对频率调制，区别在于步进式 GPR 的发射波频率是离散变化的,而非像调频 GPR 的发射波的频率是连续变化的。步进式 GPR 根据固定的频率段分别接收雷达回波的相位和幅值，相当于得到反射信号的傅里叶变换。之后，通过反向快速傅里

叶变换，可将信号重新转化为时间域的信号，用以计算时延等。目前市场上的步进式探地雷达主要有美国 3D-Radar 等公司的设备。

（3）调幅 GPR（脉冲式 GPR）

脉冲式雷达是目前最为常用的探地雷达类型。脉冲式探地雷达的发射波为一个时长短（纳秒级）、频谱宽（从数百 MHz 到数 GHz）的脉冲波。由于频率范围很宽，调幅 GPR 也被称为超宽带（UWB，ultra-wide bandwidth）探地雷达。脉冲式 GPR 的工作原理是：探地雷达天线发射电磁脉冲波，该电磁波从空气中向道路结构内部传播，遇到介电常数变化的位置发生反射，若道路结构内部存在多个分界面，则天线接收到的信号包含各个界面上的反射波。接收信号中各个反射脉冲之间的时延，为分界面之间的信号双向传播时长，由此可算得路面结构层的尺寸。

由于脉冲式 GPR 的信号处理相对容易，适合道路结构多各反射面的特点，因此是市面上最为常见的类型。常见的主要有美国 GSSI 公司的 SIR 系列雷达、Penetradar 公司的 EP 系列雷达、加拿大 EKKO 雷达、瑞典的 RAMAC/GPR 探地雷达等。本研究主要针对脉冲式探地雷达，部分涉及步进式雷达。

除了根据调制技术类型的分类，探地雷达也可根据雷达天线是否与地表接触分为空气耦合式雷达和地面耦合式雷达。空耦式雷达在测量中将雷达天线悬于地表约十五厘米到半米的高度，不与地面接触，可以悬挂于车上，在不影响交通的情况下正常行驶测量；通常发射波的中心频率较高，可以用于较浅层位的测量。然而，空耦式雷达的穿透深度有限，无法测量较深位置的情况。地耦式雷达在测量过程中天线与地面接触，因此需要拖行，速度较慢，影响交通；但另一方面，地耦式雷达可以达到较深的探测深度，且由于其与地面接触，阻挡了来自空中的其他来源的电磁波，接收信号的噪声水平较低。由于日常的道路状况调查关注浅层路面结构的状况，因此近些年空耦式雷达由于检测操作方便快捷，逐渐被广泛采用；地耦式雷达主要用于对于整个道路结构的检测，如对于路表以下三米左右范围内路基路面结构的检测。

此外，根据发射天线和接收天线是否为一体以及天线对数，探地雷达还可分为单站式雷达、双站式雷达、多站式雷达及阵列天线雷达。目前较为常见的是单站式雷达，但单站式雷达无法使用测量介电常数的共中点法。阵列天线雷达成本较高，但可以用于生成三维测量图像。

根据以上分析，不同的探地雷达类型具有不同的工作原理和技术特点，应当依据实际的工程需求选用。

### 2.1.2 探地雷达设备的组成

探地雷达设备主要由雷达主机、发射天线和接收天线、电源、存储和显示设备等部分组成。以脉冲式探地雷达为例，图 2.1 展示了典型的路用探地雷达组成结构。

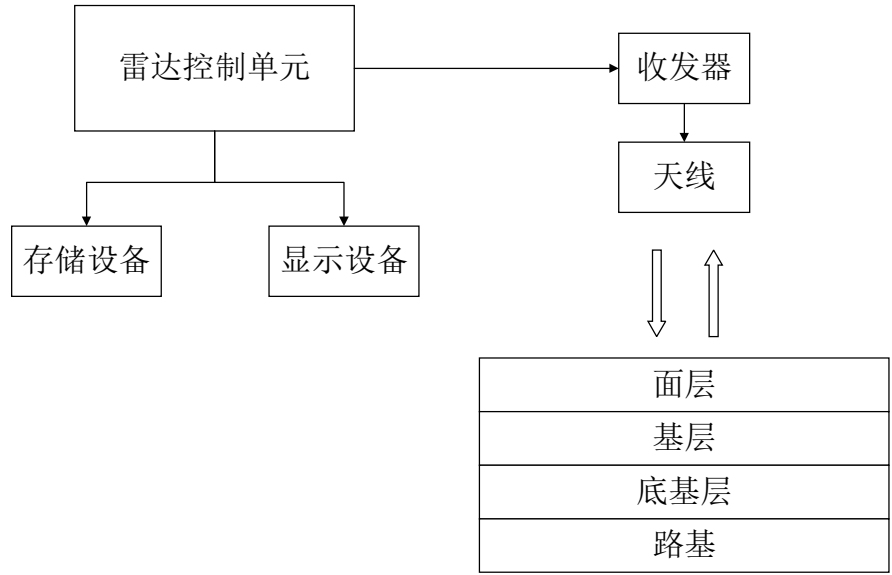

图 2.1 路用雷达设备组成示意图

雷达控制单元产生超带宽脉冲电磁波，经由收发器调制、放大，通过天线发射至路面。该电磁波传播至路表及道路内部各个层位时产生反射波，向上传播至接收天线（图 2.1 中以单站式雷达天线为例）。雷达控制单元将天线采集到的模拟信号转化为数字信号，并存储、显示。

探地雷达的发射频率可以按照时间等间隔，也可按照距离等间隔。如果采用距离等间隔发射，还需配备一个测距传感器（DMI），安装在汽车或推车的轮子上，控制雷达设备每移动一定距离进行一次发射接收。

## 2.2 探地雷达设备的评价指标

探地雷达采集信号的分析处理建立在天线发射信号稳定、信噪比高等特点的基础上，因此在使用探地雷达设备前应当对其性能进行评价。国内目前尚无探地雷达性能标准的规范，大多数研究引用美国德克萨斯州交通研究院提出的指标体系[54-55]。理想的雷达设备应当使单个雷达回波信号上各个反射脉冲波之间不出现噪声或杂波，即采集信号上的任何幅值均代表真实的目标物。该标准体系的制定思路是衡量被测试的雷达设备采集到的真实信号与理想的雷达设备采集到的信号之间的接近程度，具体指标简要阐述如下：

（1） 信噪比（Noise to Signal Ratio）

信噪比 $N/S$ 用于衡量噪声相对于信号的显著水平，计算公式为噪声大小与信号大小的比值（2.1）。测试过程为：探地雷达设备按照厂商建议预热后，正对地面上放置的一块较大金属板（大于雷达波抵达地面位置时的扩散范围）的中心，发射雷达测试波并采集信号。由于铁板反射近似为全反射，所以雷达波不会投射入金属板下方结构，因此接收天线只会接收到一个反射脉冲波，即铁板表面的反射波，该反射波之后接收到的其他信号皆可看作噪声。

$$N/S = \frac{A_n}{A_{mp}}$$

（2.1）

$A_n$ 代表噪声水平，取铁板表面反射之后一半的时窗（如 1GHz 天线一般取 20ns 时窗，则考虑铁板反射后的 10ns 的范围）之内的信号最大幅值。

$A_{mp}$ 代表信号水平，取铁板反射波的最大幅值。

计算 100 个接收信号的比值，然后取平均值。该值不应超过 5%。在雷达图像分析中过高的信噪比可能导致较弱的分界面反射波被噪声淹没，或噪声被误认为是分层。

（2） 信号稳定性（Signal Stability）

信号稳定性反映信号中幅值抖动的程度。幅值抖动（amplitude jitter）指当雷达设备对同一测点静止测量时，短时间内采集到的不同 A 扫描轨迹上的信号强度的变化。测试方法为：仍将探地雷达设备正对地面上放置的一块较大金属板的中心静止测量，以设备最大采集频率连续采集 100 条雷达测迹。

$$J_{Amp} = \frac{A_{max} - A_{min}}{A_{ave}}$$

（2.2）

$A_{max}$ 代表采集到的 100 条雷达测迹线上金属板反射波的最大幅值，$A_{min}$ 代表其中的金属板反射最小幅值，$A_{ave}$ 代表其中的金属板反射幅值的平均值。

计算得到的信号稳定性指标 $J_{Amp}$ 应当小于 1%。由于雷达图像的解析要用到道路相邻测迹之间、铁板测迹与道路测迹之间的对比，若幅值抖动偏高会导致计算出的道路材料介电常数误差偏大以及雷达图像上出现干扰。

时间抖动（time jitter）指采集 100 条测迹所花费时间长短的变化。计算公式为采集 100 条测迹花费时间的最大差值与总时间的比值。

$$J_{Time} = \frac{A_{max} - A_{20}}{A_{20}}$$

（2.3）

$\Delta t_{max}$ 代表采集 100 条测迹花费的最大时长，$\Delta t_{min}$ 代表采集 100 条测迹花费的最短时长。计算得到的 $J_{Time}$ 也应当小于 1%。

（3） 长时间信号稳定性（Long-term Signal Stability）

实际工程检测中可能会使用探地雷达连续扫描达一两个小时。因此，较长时间内发射波的幅值稳定性对于准确计算路面材料的介电常数很重要。计算两个指标：长时间波幅变异性（long term amplitude variation，LAV）和长时间时窗变异性（long term time-window shifting，LTS）。长时间波幅变异性测试方法为：将探地雷达设备正对地面上放置的一块较大金属板的中心静止测量，按照每两分钟采集一条雷达迹，采集 2 个小时。

$$LAV = \frac{A_{\max} - A_{\min}}{A_{20}}$$

（2.4）

$A_{20}$ 代表测试 20 分钟时的金属板反射波峰值，$A_{\max}$ 代表测试 20 分钟开始到 2 小时测试时间结束前金属板反射波峰值的最大值，$A_{\min}$ 代表测试 20 分钟开始到 2 小时测试时间结束前金属板反射波峰值的最小值。计算得到的 $LAV$ 应当小于 3%。

长时间时窗变异性的测试方法为：连续两小时测量，每分钟采集一条测迹，共 120 条。计算每条测迹从开端到波幅最大值之间的时间差。

$$LTS = \frac{(t_{\max} - t_{\min})}{\text{Time-range}}$$

（2.5）

$t_{\max}$ 表示最大波幅左侧的时长的最大值，$t_{\min}$ 为其最小值，Time-range 为总时长，这里为两小时。

（4） 穿透性能

较高频率的雷达天线有更高的精度，但同时穿透深度有所降低。为研究探地雷达的穿透性能，需要对设备进行如下测试。

在一个水池底部放置一块金属板，水池注固定深度的水。探地雷达悬于上方扫描。计算指标 WPI（water penetration index）。

$$WPI = \frac{A_{water}}{A_{air}}$$

（2.6）

$A_{water}$ 为注水情况下的金属板反射波幅值，$A_{air}$ 为无水情况下的金属板反射波幅值。

此外还有衡量耦合波大小的耦合波信号比（end reflection ratio，ERR）等指标。这些指标仅作为比较不同雷达设备的参考，可灵活使用，如 WPI 指标的测试过程可将水换为一块均质混凝土板。

## 2.3 相关电磁学理论

探地雷达在道路中的应用的研究，建立在电磁学理论的基础上。以下分为两部分阐述：电磁波在同种均匀介质中的传播理论，以及电磁波在不同介质交界面上的反射理论。

### 2.3.1 电磁场及电磁传播理论

麦克斯韦方程组和本构关系式为电磁学中最为基础的两组关系式，它们是定量分析探地雷达波在各向同性均匀介质中的传播规律的基础。其中，麦克斯韦方程组描述了电磁场的物理性质，本构关系式描述了材料的电磁性质。

麦克斯韦方程组为：

$$\nabla \times H = J + \frac{\partial D}{\partial t} \tag{2.7}$$

$$\nabla \times E = -\frac{\partial B}{\partial t} \tag{2.8}$$

$$\nabla \bullet B = 0 \tag{2.9}$$

$$\nabla \bullet D = \rho \tag{2.10}$$

式中 $H$ 代表磁场强度（A/m），$D$ 代表电位移矢量（C/m$^2$），$t$ 代表时间（s），$E$ 代表电场强度（V/m），$B$ 代表磁感应强度（Wb/m$^2$），$J$ 代表外加电源的电流密度（A/m$^2$），$\rho$ 代表电荷密度（C/m$^3$）。

本构方程组描述了电磁场作用下材料的电磁响应，对于探地雷达检测来说，材料的电磁性质有重要意义：

$$D = \varepsilon E \tag{2.11}$$

$$B = \mu H \tag{2.12}$$

$$J = \sigma_e E + J' \tag{2.13}$$

式中 $\varepsilon$ 代表介质的介电常数（F/m），$\mu$ 代表介质的磁导率（H/m），$\sigma_e$ 表示介质的电导率（S/m），$J'$ 表示产生时变电磁场的电流源或非电的外源。

此外，电荷与电流密度遵从电荷守恒方程：

$$\nabla \times J = -\frac{\partial \rho}{\partial t} \tag{2.14}$$

以上关系式将辐射场中每一点的电场和磁场场量，与场源和介质的特性联系起来。探地雷达探测的是反射电磁波的电压值，因此下面推导电磁场中的电场表达式。

在一个角频率为 $\omega$ 的时谐电磁场中（时谐因子为 $e^{i\omega t}$），矢量 **r** 定义的空间点

位置上的复数电场 $\mathbf{E(r)}$ 满足公式：

$$\nabla^2 \mathbf{E(r)} + \omega^2 \varepsilon \mu (1 - j\frac{\sigma}{\omega\varepsilon}) \mathbf{E(r)} = 0 \tag{2.15}$$

如果假设电场平行于 X 轴且仅由 z 坐标决定，则（2.15）式可简化为

$$\frac{d^2}{dz^2} E_x(z) + \omega^2 \varepsilon \mu (1 - j\frac{\sigma}{\omega\varepsilon}) E_x(z) = 0 \tag{2.16}$$

（2.16）式的一个解为

$$E_x(z) = E_0^+ e^{-jkz} + E_0^- e^{jkz} \tag{2.17}$$

其中

$$k = \omega \sqrt{\varepsilon\mu(1 - j\frac{\sigma}{\omega\varepsilon})} = \beta - j\alpha, \quad \alpha, \beta \geq 0 \tag{2.18}$$

式中 $k$ 为电磁波在介质中传播的波数，$\alpha$ 为衰减系数（Np/m），$\beta$ 为相位常数（rad/m）。

由（2.18）式可得

$$\alpha = \omega \sqrt{\frac{\mu\varepsilon}{2}} \sqrt{\sqrt{\left(\left(\frac{\sigma}{\omega\varepsilon}\right)^2 + 1\right)} - 1} \tag{2.19}$$

$$\beta = \omega \sqrt{\frac{\mu\varepsilon}{2}} \sqrt{\sqrt{\left(\left(\frac{\sigma}{\omega\varepsilon}\right)^2 + 1\right)} + 1} \tag{2.20}$$

波传播的相速度是 $v = \omega / \beta$，因此

$$v = \frac{\omega}{\beta} = \frac{1}{\sqrt{\frac{\mu\varepsilon}{2}} \sqrt{\sqrt{\left(\left(\frac{\sigma}{\omega\varepsilon}\right)^2 + 1\right)} + 1}} \tag{2.21}$$

由于道路材料的电导率很低，因此 $\left[(\sigma/\omega\varepsilon)^2 + 1\right]$ 可近似看作 1。则（2.21）式简化为

$$v = \frac{c}{\sqrt{\varepsilon_r}} \tag{2.22}$$

其中 $\varepsilon_r = \dfrac{\varepsilon}{\varepsilon_0}$

式中 $c$ 为真空中的光速，约为 $3 \times 10^8$ m/s，$\varepsilon_r$ 为介质的相对介电常数，$\varepsilon_0$ 为真空中的介电常数，$\varepsilon$ 为介质的介电常数。

### 2.3.2 电磁波在两种介质交界面上的反射与折射

以上电磁场理论讨论的是电磁波在均匀介质中的传播。在路面检测中，关注点主要在于不同结构层的分层，以及病害等异常区域的识别。因此，这里讨论电磁波在不同介质交界面上的反射和折射。为简化问题，这里先讨论最简单的模型：电磁波为平面波，两种接触的介质的分界面是无限大的平面。

电磁波传播至两种介电常数不同的介质的分界面时，会发生折射和反射。入射波、折射波、反射波三者之间的方向关系，遵循电磁波的反射和折射定律。因此，对于任何介质，入射角 $\theta_i$ 等于反射角 $\theta_r$。反射定律：

$$\theta_i = \theta_r \tag{2.23}$$

折射波的方向与相接触的两种介质的性质有关。折射定律：

$$\sin\theta_i / \sin\theta_t = v_1 / v_2 = n \tag{2.24}$$

式中 $v_1$、$v_2$ 分别表示电磁波在介质 1 和介质 2 中传播的波速，$\sin\theta_t$ 为折射角，$n$ 称为折射率。当 $n > 1$ 时，$\theta_i > \theta_t$，$v_1 > v_2$；$n < 1$ 时，$\theta_i < \theta_t$，$v_1 < v_2$。

电磁波在不同介质交界面上会发生能量的重新分配。根据能量守恒定律，电磁波在界面处折射、反射前和折射、反射后总能量不变。因此，入射波能量等于反射波能量和折射波能量之和。以下讨论反映三者电场强度幅值之间关系的反射系数和折射系数。

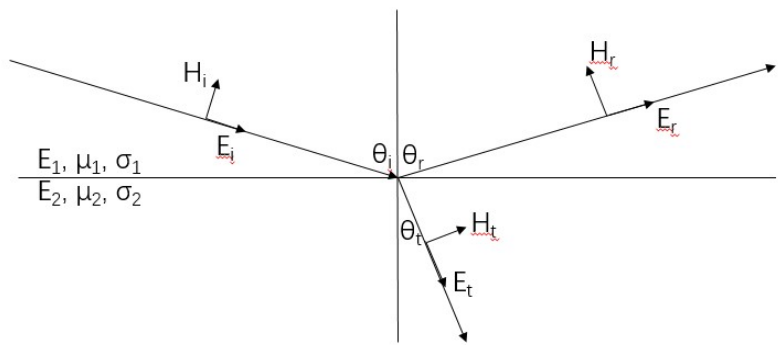

图 2.2 电磁波的界面反射和折射

如图 2.2，$E_i$、$E_r$、$E_t$ 分别表示入射波、反射波和折射波的电场强度幅值。$H_i = E_i / \eta_1$，$H_r = E_r / \eta_1$，$H_t = E_t / \eta_2$ 分别为入射波、反射波和折射波的磁场强度。其中，$\eta_1 = \sqrt{\mu_1 / \varepsilon_1}$，$\eta_2 = \sqrt{\mu_2 / \varepsilon_2}$ 分别为上下两种介质的波阻抗。

根据能量守恒定律及电磁波切向分量相等理论，有

$$E_i + E_r = E_t \tag{2.25}$$

$$H_i \cos\theta_i - H_r \cos\theta_r = H_t \cos\theta_t \qquad (2.26)$$

定义反射系数 $R_{12} = E_r / E_i$，折射系数 $T_{12} = E_t / E_i$。代入式（2.25）和式（2.26）可求得

$$R_{12} = \frac{-\eta_1 \cos\theta_i + \eta_2 \cos\theta_t}{\eta_1 \cos\theta_i + \eta_2 \cos\theta_t} \qquad (2.27)$$

$$T_{12} = \frac{2\eta_2 \cos\theta_i}{\eta_1 \cos\theta_i + \eta_2 \cos\theta_t} \qquad (2.28)$$

将折射率 $n = \eta_1 / \eta_2$，$n\cos\theta_t = \sqrt{n^2 - \sin^2\theta_i}$ 代入式（2.27）和式（2.28），得

$$R_{12} = \frac{\cos\theta_i - \sqrt{n^2 - \sin^2\theta_i}}{\cos\theta_i + \sqrt{n^2 - \sin^2\theta_i}} \qquad (2.29)$$

$$T_{12} = \frac{2\cos\theta_i}{\cos\theta_i + \sqrt{n^2 - \sin^2\theta_i}} \qquad (2.30)$$

在大多数的探地雷达操作流程中，要求将天线固定使其底部与地面平行。因此，电磁波为垂直入射，即 $\theta_i = 0$。反射系数和折射系数公式得到进一步简化：

$$R_{12} = (1-n)/(1+n) \qquad (2.31)$$

$$T_{12} = 2/(1+n) \qquad (2.32)$$

由两式可知，当 $n < 1$，$\varepsilon_2 < \varepsilon_1$，即电磁波从介电常数大的介质进入介电常数小的介质时，$R_{12}$ 大于零，$T_{12}$ 大于 1，$E_r$ 与 $E_i$ 同向，$H_r$ 与 $H_i$ 反向，$E_r > E_i$，$H_t < H_i$。当 $n > 1$，$\varepsilon_2 > \varepsilon_1$，即电磁波从介电常数小的介质进入介电常数大的介质时，$R_{12}$ 小于零，$T_{12}$ 小于 1，$E_r$ 与 $E_i$ 反向，$H_r$ 与 $H_i$ 同向，$E_t < E_i$，$H_t > H_i$。

对于道路结构的雷达图像来说，地表反射是电磁波从介电常数较小介质（空气）进入介电常数较大介质（路面），因此地表反射波的方向与雷达发射波的电场方向相反；如果道路结构内部存在干燥脱空，则脱空区域上方界面的反射波的电场方向与发射波相同，也即与地表反射波的电场方向相反；水的介电常数大于道路材料的介电常数，因此如果道路结构内部存在滞水脱空，则脱空区域上方界面的反射波与发射波相反，也即与地表反射波的电场方向相同。

### 2.3.3 电磁波在多层介质中的传播

道路结构是一个多种介质的层状体系。路面、基层和路基由于材料不同，具有不同的介电常数；沥青面层内部上、中、下面层材料基本相同，但由于空隙比等设计指标的差别，也会造成介电常数的差异。因此这里从理论上讨论电磁波在

多层介质中的传播。这里假设道路结构各层内部材料电磁性质均匀，各层水平方向上几何尺寸相对于探地雷达的电磁场范围足够大。

电磁波在道路层状结构中的传播如图 3 所示。雷达信号在每一个介电常数变化的界面上通常会同时发生部分入射和部分反射（金属介质例外，可认为金属界面上电磁波发生全反射）。当探地雷达发射的电磁波入射到地表时，就会在该电磁界面发生反射和折射，其反射波返回至雷达接收天线，折射波进入道路最上层继续传播。当该折射波传递至道路最上方两层材料的交界面时，该电磁波相当于第二个电磁界面的入射波，再次发生反射和折射。第二个界面上的折射波继续向下传播，之后在第三个电磁界面再次发生折射和反射；而第二个界面上的反射波向上传播至地表，发生折射和反射，折射部分透过路表传递至雷达接收天线，反射部分在该道路结构层内部传播，不断发生折射、反射。

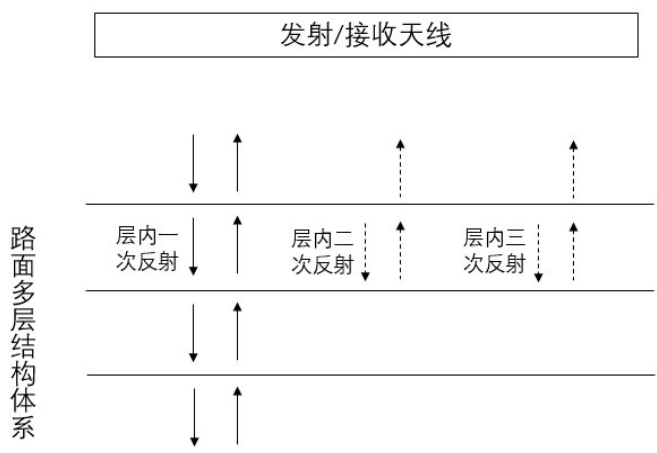

图 2.3 雷达波在道路层状体系中的传播

因此，电磁波实际上会在结构层上下两界面之间发生多次反射，并最终使接收天线接收到同一界面的多个回波。但由于道路内部界面不是强反射界面，第一次反射之后的反射很弱，所以相关研究一般不作考虑。

由于探地雷达发出的电磁波在道路结构中的传播距离较短，脉冲波的形状基本不发生变化，只是信号强度不断减小。因此，接收天线采集到的信号在时域上可以看作是一系列经过振幅缩小、添加时延后的发射信号的叠加，表达式[56]如下：

$$y_r(t) = \sum_{i=0}^{N-1} A_i x(t-\tau_i) + n(t) \qquad (2.33)$$

式中，$x(t)$ 为入射脉冲波，$N$ 为道路结构的总层数，$A_i$ 为第 $i$ 个分界面上的反射波的相对振幅，$n(t)$ 为噪声，$\tau_i$ 为脉冲波从发射天线传播至第 $i$ 个分界面再反射回接收天线所经过的双程时长。该式没有包含道路结构层内部的多次反射，

原因在上文中已说明。

## 2.4 介电常数理论

在一般性的工程研究中，道路材料常被当作电介质，即电的绝缘体。而实际的道路材料，尤其是含水率较高时，内部有自由电荷，如电子和离子等，会一定程度上削弱电磁波。因此，道路材料被描述为"有损电介质"更为精确。当道路材料含水率很高时，材料中的自由电荷密度很高，电磁波削减则更为明显。

### 2.4.1 介电常数 $\varepsilon$

介电常数也称电容率（permittivity），反映一种材料以电荷的形式储存电磁能量的能力。相对介电常数（relative permittivity），$\varepsilon_r$，是某种材料的介电常数与真空的介电常数的比值。真空的相对介电常数为 1，其他材料的相对介电常数均大于 1。

不同材料的相对介电常数相差很大；一些材料在不同频率的电磁波中相对介电常数也不同，被称为色散性材料。道路材料的相对介电常数一般为复数，实数部分表示电容率，虚数部分表示损耗。由于大多数道路材料对电磁波的损耗较小，所以相关文献往往忽略相对介电常数的虚部，只考虑其实部。但对于电磁损耗较大的道路材料，如水泥混凝土、高含水率黏土等，在对探地雷达图像的精确分析中不应忽略相对介电常数的虚部。

以下简单阐述雷达波传递到道路材料（均匀介质）中某一位置时，介质内部发生的变化。在没有外加电磁场时，介质内部的正负电荷不存在极化，介质作为整体电荷数为零。当电磁波传递至介质时，介质内部的正负电荷发生移动（极化）：一方面，原子内部的电子云中心与原子核的位置发生偏移，另一方面，从介质整体来说，如果没有周围环境中的电荷的中和作用，介质内的正负电荷分别集聚到介质的表面。当前半段上升趋势的雷达波进入介质时，介质内的正负电荷距离增大，能量被储存；当后半段下降趋势的雷达波进入介质时，外加电场强度下降，能量逐渐释放。这个过程中介质内部产生偶极距，偶极距密度与施加的电场强度成正比，该比例就被称为电容率或介电常数。如果分离开的电荷可以自由移动和物理接触，则该极化过程中部分电磁波能量转化为热能，相应的介质材料被称为有耗介质。因此，介电常数还有虚数部分，用于代表能量的耗散。介电常数的实数部分和虚数部分都受雷达波的频率影响。雷达波把一部分能量用于移动电荷，而电荷的移动又会产生电流从而发射电磁波。该电磁波的相位略微滞后于入射的雷达波，与入射雷达波叠加相当于使原有的雷达波减速。这就解释了电磁波的传

播速度与相对介电常数负相关。

## 2.4.2 混合材料的介电常数

实际道路材料并不是单一种类的物质，而是多种物质的混合物，如沥青路面是沥青、石料和空气等物质的混合物，土基是土颗粒、水、空气等物质的混合物。因此，作为探地雷达应用于道路工程的基础，有必要讨论混合材料的介电常数模型。混合材料的介电常数模型有两类：经验回归模型和半理论半经验模型。

在众多针对土体的经验回归模型中，Annan[57]提出的一个模型被广泛采用（式）。

$$\varepsilon' = 3.03 + 9.3\theta_V + 146(\theta_V)^2 - 76.6(\theta_V)^3 \qquad (2.34)$$

式中 $\theta_V$ 为体积含水率。该经验模型要求材料为低耗介质且干燥状态的介电常数大致在 3 和 4 之间。该式可以有效用于计算土体的天然含水率。但式在黏土及有机质土中不适用[58]。而且该模型没有考虑介电常数的虚部，仅适用于低电磁损耗的介质。

另一类模型可以被称为半经验半理论模型。这类模型通过混合物的组成成分的特性，来推算混合物整体的性质。Tsui 和 Mattews[59]提出的 CRIM 模型（complex refractive index model）较适用于道路结构，而且可同时用于介电常数实部和虚部的计算，见式（2.35）。

$$\varepsilon_{mix}^e = \left( \sum_{i=1}^{N} f_i \sqrt{\varepsilon_i} \right)^2 \qquad (2.35)$$

式中 $\varepsilon_{mix}^e$ 为混合物的整体介电常数（复数形式），$f_i$ 为 第 $i$ 种成分的体积比，$\varepsilon_i$ 为第 $i$ 种成分的介电常数（复数形式）。

很多情况下，混合物可以看做三相物质的组合，如土可以看做土颗粒、水和空气的组合。若组成混合物的三相物质的介电常数分别已知，可以将 CRIM 公式写作：

$$\varepsilon_{mix}^e = \left[ \left( \phi S_W \sqrt{\varepsilon_W} \right) + \left( (1-\phi)\sqrt{\varepsilon_m} \right) + \left( \phi(1-S_W)\sqrt{\varepsilon_g} \right) \right]^2 \qquad (2.36)$$

式中 $\phi$ 为空隙率，$S_W$ 为饱和度（空隙中水的填充体积比），$\varepsilon_W$、$\varepsilon_m$ 和 $\varepsilon_g$ 为水、固态物质和空气的介电常数，$\varepsilon_{mix}^e$ 为混合物的介电常数。

## 2.5 探地雷达图像的种类

探地雷达设备采集到的信号具有不同的展示方式，主要包括 A 扫描、B 扫

描、C扫描和"瀑布图"。其中，最基本的GPR图像为A扫描。单个A扫描图像表示探地雷达单次发射、接收得到的信号，是接收天线接收到的电场强度在一定长度的时窗范围内在时间轴上的波动。本论文第四章中对于脱空的识别及高度计算主要基于A扫描。A扫描只能表示某一点位的地下状况，为了展示一条连续测线下道路截面的状况，尤其是为了识别地下探测目标物的形状，可以用颜色或灰度表示A扫描上各时刻的波幅，然后将一条GPR测线上各点的扫描依次紧密堆叠起来，称为B扫描。本文第四章中脱空水平尺寸的计算以及第五章中现场实测数据的噪声去除主要基于B扫描。在一些研究中，需要对比同一深度处不同水平位置的信号差异。将某一深度范围内的信号波幅的均值表示在代表水平面的二维坐标系上，被称为C扫描。此外，还有一些研究者把一些相邻位置的A扫描按照固定间隔堆叠起来，这种图像被称为"瀑布图"。

## 2.6 GPR 路面厚度检测方法

GPR 在道路中最广泛的应用为路面结构层厚度的检测。探地雷达在路面厚度方面的应用包含以下几种情况：首先，GPR可以用于施工质量的检测评价，如检测新建道路的路面厚度是否达到设计厚度，或者检测既有道路加铺工程的加铺层厚度是否达到要求；其次，用于既有道路的路况调查，由于旧路的厚度存在不均匀现象，且原始资料往往不完整，需要使用探地雷达对旧路的厚度情况做调查，为后续养护、修护、重建工程提供参考资料；此外，落锤式弯沉仪（FWD）计算道路结构层模量的输入参数之一为各结构层厚度，因此有时 GPR 和 FWD 配合使用，用于道路结构层模量检测。

### 2.6.1 GPR 测量路面厚度计算公式

以脉冲式探地雷达为例，介绍路面结构层厚度探测计算方法。一个状况完好的道路结构层可视作一个均匀介质层，脉冲波传播至上下两个界面，分别会产生一个探地雷达回波，两个回波的时间间隔即为脉冲波在介质层内传播的双向往返时长（式（2.37））。再根据电磁波传播速度计算公式（式（2.38）），可以求得介质层，即路面结构层的厚度（式（2.39））。

$$\Delta d = \frac{v \Delta t}{2} \tag{2.37}$$

$$v = \frac{c}{\sqrt{\varepsilon_r}} \tag{2.38}$$

$$\Delta d = \frac{c\Delta t}{2\sqrt{\varepsilon_r}}$$ （2.39）

式中 $\Delta d$ 为探测的结构层的厚度，$\Delta t$ 为脉冲波在该层上下两个界面发生反射之间的时间差，$v$ 为雷达波在该层介质中的传播速度，$c$ 为电磁波在真空中的传播速度，为常数约等于 $3 \times 10^8 m/s$，$\varepsilon_r$ 为该结构层材料的介电常数。

由式（2.39）可知，计算道路结构层厚度，需要先计算两个未知参数，一是该层上下两界面的反射脉冲（即采集到的回波信号上两相邻反射脉冲波）之间的时间差 $\Delta t$，二是该层的介电常数。

### 2.6.2 层内传播时间差的获取

首先讨论时间差 $\Delta t$ 的获取。获取 $\Delta t$ 最直接的方法是在雷达扫描迹上找到该结构层顶部和底部两界面处的反射波，读取两反射脉冲的时间间隔（选取对应特征位置，如两波峰之差或两波各自的第一个波谷之差）。要确定两相邻反射脉冲波之间的时间差，在不采用特殊的信号处理方法的情况下，需要保证这两个相邻回波不重叠，否则两相邻信号重叠导致波形变化，波峰等特征位置移动，无法直接计算间隔时长。由此，探地雷达检测层厚的理论计算精度为

$$\Delta d = \frac{cT}{2\sqrt{\varepsilon_r}}$$ （2.40）

式中 $T$ 表示发射脉冲波周期时长。

以 1GHz 探地雷达、道路材料介电常数为 9 为例计算，则探地雷达测层厚的理论精度为 5cm。即该种道路材料用 1GHz 探地雷达探测，只能测量计算厚度大于 5cm 的结构层。如果道路材料的介电常数小于 9，或使用的雷达天线频率小于 1GHz，探测精度还会更低。实际道路中，沥青上、中、下面层有可能小于该尺寸，这种情况下需要额外的信号处理技术和算法。

### 2.6.3 结构层介电常数的计算

其次讨论结构层介电常数 $\varepsilon_r$ 的计算。介电常数是探地雷达道路检测中最重要的一个指标。一方面，电磁波在介电常数不同的两种介质的界面上发生反射，因此电磁波可用来识别结构层界面和病害；另一方面，介电常数决定电磁波在介质中的传播速度，因此对于计算结构层厚度很重要。获取道路各结构层的介电常数的方法主要有两种：钻芯取样法和反射波幅法。

（1）钻芯取样法

钻芯取样法是指，在道路上某点先用探地雷达扫描，在回波信号上读取各层

间界面反射波之间的距离，然后在同一点钻芯取样，测量芯样的各层厚度。根据式（2.41），可以求出该位置的各层材料介电常数。认为该取样点附近一定距离内的道路各层材料介电常数一致，在附近路段上使用探地雷达扫描，用取样点计算出的介电常数计算厚度。

$$\varepsilon_r = \frac{c^2}{v^2} = \frac{c^2 \Delta^2 t}{4 \Delta^2 d}$$ （2.41）

这种方法优点在于原理简单，除钻芯外不需额外操作。在道路各层材料均匀的路段，这种方法具有较高精度。但是对于施工质量不佳的道路或旧路，由于不同位置处同层材料性质不一，因此取样点的介电常数不能代表其他位置的介电常数，不应使用此方法。此外，当道路层厚度较小，导致相邻的材料分界面反射波发生叠加，则波峰位置变化，不能再用本方法。

（2）反射波幅法

雷达波在道路各层分界面上发生反射和折射，导致能量再分配。一部分能量以反射波的形式向上传播，另一部分能量以折射波的形式继续向下传播。折射波和反射波的能量分配比例取决于分界面两侧介质的介电常数的相对关系。根据电磁波传播的这一特性，可以通过反射波幅计算道路各层介电常数。这种方法的操作过程为：在地表放置一块足够大的金属板（不小于 1m*1m），雷达天线正对金属板中心发射、收集电磁波，记录金属板反射最大幅值的均值 $A_p$；然后，到需要测量道路介电常数的各位置，发射采集电磁波，读取地表反射反射脉冲最大峰值，以及由上到下 n 个道路材料分界面的反射脉冲峰值。根据美国弗吉尼亚理工学院 Al-Qadi 等人[60]的研究，各结构层的介电常数计算公式为：

$$\varepsilon_{R,1} = \frac{\left(A_p + A_0\right)^2}{\left(A_p - A_0\right)^2}$$ （2.42）

式中 $A_p$ 代表铁板反射最大波幅，$A_0$ 代表地表反射最大波幅，$\varepsilon_{R,1}$ 代表最上层道路结构的介电常数。

$$\varepsilon_{R,n} = \varepsilon_{R,n-1} \left( \frac{1 - \left(\frac{A_0}{A_p}\right)^2 + \sum_{i=1}^{n-2} \gamma_i \frac{A_i}{A_p} + \frac{A_{n-1}}{A_p}}{1 - \left(\frac{A_0}{A_p}\right)^2 - \sum_{i=1}^{n-2} \gamma_i \frac{A_i}{A_p} - \frac{A_{n-1}}{A_p}} \right)^2$$ （2.43）

$$\gamma_i = \frac{\sqrt{\varepsilon_{R,i}} - \sqrt{\varepsilon_{R,i+1}}}{\sqrt{\varepsilon_{R,i}} + \sqrt{\varepsilon_{R,i+1}}}$$ （2.44）

式中 $\gamma$ 代表反射系数，i 为反射界面序号，地表相应的 i = 0，n 代表介电常数待测层，$A_i$ 代表第 i 个层界面上的反射波幅值，$\varepsilon_{R,n}$ 代表第 n 层的介电常数，$\varepsilon_{R,n-1}$

代表第$n-1$的介电常数。

反射波幅法是应用最广泛的道路介电常数测量方法。但是这种方法也有一定缺陷：其计算公式的推导过程假设电磁波在层内的传播不发生能量损耗，即道路材料电导率为0，但实际的道路材料不符合这一条件；该方法只依据各层表面的反射决定相应结构层的整体介电常数，因此要求材料均匀，该条件在旧路中很难满足。

至此，读取了各结构层上下两界面的反射脉冲之间的时间差$\Delta t$，计算得到了各结构层的介电常数，便可以计算各结构层厚度。

除了以上两种分别读取时间差、计算介电常数从而得到厚度的方法外，还有一种共中点法可以用于计算最上层材料的介电常数和厚度。

共中点法要求发射天线和接收天线可以各自移动，因此适用于双站式探地雷达。如图4所示，要测量P点下方第一层道路结构的介电常数，先分别将发射天线和接收天线置于与P点距离相等、水平高度相同的$T_1$和$T_2$位置，进行扫描操作，计算从$T_1$发出电磁波，经第一层道路结构下界面反射，再传播至$R_1$的时长$t_1$。然后，移动收、发天线的位置至$T_2$、$R_2$，仍与P点距离相等、水平高度相同，再次扫描并计算电磁波层内走时$t_2$。

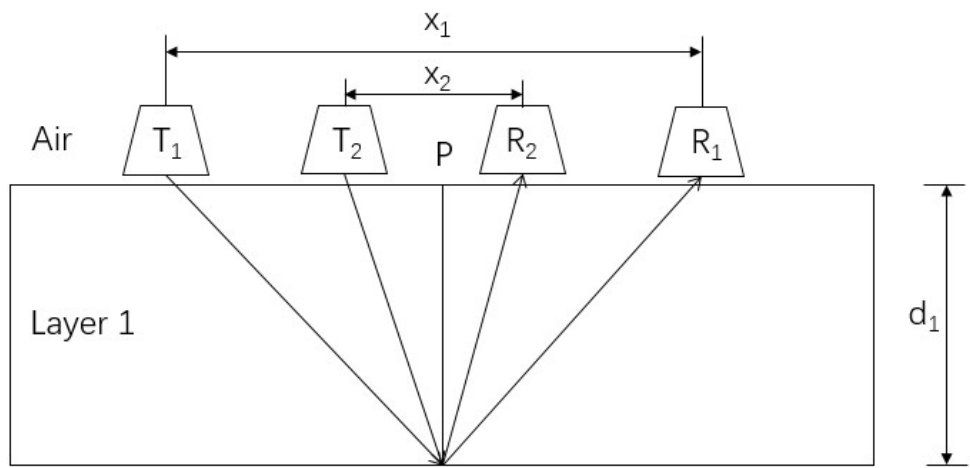

图 2.4 共中点法测量介电常数

根据波的传播速度公式可列（2.45）、（2.46）两个关系式：

$$\frac{c}{\sqrt{\varepsilon_1}} t_1 = 2\sqrt{\left(\frac{x_1}{2}\right)^2 + d_1^2} \tag{2.45}$$

$$\frac{c}{\sqrt{\varepsilon_1}} t_2 = 2\sqrt{\left(\frac{x_2}{2}\right)^2 + d_1^2} \tag{2.46}$$

由（2.45）、（2.46）两式可以求得最上道路结构层的介电常数$\varepsilon_1$和层厚$d_1$：

$$\varepsilon_1 = \frac{c^2 \left(t_1^2 - t_2^2\right)}{x_1^2 - x_2^2} \qquad (2.47)$$

$$d_1 = \sqrt{\frac{x_2^2 t_1^2 - x_1^2 t_2^2}{4\left(t_2^2 - t_1^2\right)}} \qquad (2.48)$$

共中点法优点在于计算的是整层的介电常数,适用于沥青路面压实度检测等方面。但此种方法不能直接计算多层结构的各层介电常数,且要求必须使用双站式天线。

以上为几种常见的道路结构层厚度检测计算方法。计算得到道路结构的各层尺寸后,便可寻找雷达图像中的异常点,结合异常点所在层位,分析道路内部包括脱空在内的病害类型及范围。

## 2.7 本章小结

(1)按照调制技术的不同,探地雷达可分为调频 GPR、步进式 GPR 以及脉冲式 GPR;按照天线与地面的相对位置,可分为空耦式 GPR 和地耦式 GPR。不同类型的探地雷达具有不同的特点,应按照工程探测需求合理选用。探地雷达设备由雷达主机、收发天线、电源、存储和显示设备等部分组成。探地雷达设备质量可靠,是采集准确有效的探测数据和图像的前提条件。衡量探地雷达质量的指标包括信噪比、信号稳定性、长时间信号稳定性、穿透性能等。

(2)探地雷达工程检测建立在电磁场理论和电磁波反射、折射理论的基础上。麦克斯韦方程组和介质的本构方程组描述了电磁场在同种介质中传播过程中任意位置的电磁场量。电磁波的反射和折射理论描述了电磁波在不同介质分界面上的反射、折射与相邻两种介质电磁特性的关系。道路是一个具有多个水平反射面的层状体系,因此接收天线采集到的信号在时域上可以看作是一系列经过振幅缩小、添加时延后的发射信号的叠加。

(3)材料的介电常数是道路检测中最重要的指标,决定道路结构层电磁反射特性以及探地雷达波在道路结构层内的传播速度。道路材料是多种物质的混合物,其介电常数可以根据各组成成分的介电常数及比例,代入混合材料的介电常数模型计算。

(4)在 GPR 道路病害检测前,需要先进行道路分层及各层厚度计算,以便于确定病害的层位和深度。一方面,需要从图形上读取反射界面之间的时间差,另一方面,需要用反射波幅法或共中点法计算层内电磁波传播速度,从而求得道路结构层厚度。

# 第三章 探地雷达数值模拟

由于道路基层脱空的形状、尺寸以及脱空内的填充物质差别很大，雷达图像上的特点也各不相同。因此，很难通过现场检测得到的探地雷达图像归纳脱空病害的图像判别特征。目前道路检测部门使用探地雷达对于道路病害的检测，一般先采用经验法判断病害位置，再派人员前往现场判断病害类型，很难直接根据图像断定病害类型。

若通过探地雷达图像特点直接判别基层脱空，将能显著提高道路检测的工作效率。为了实现这一技术，首先需要从理论上分析不同尺寸、填充情况的脱空病害在雷达图像上的特点。由于现场道路情况、环境因素、探地雷达设备状况的复杂性，这里采用电磁场正演模型研究脱空病害的图像特点。

## 3.1 基层脱空的图像特征

在用 FDTD 方法建模仿真含脱空道路的探地雷达图像前，先从理论上分析基层脱空可能具有的图像特征。常用的探地雷达图像有 A 扫描和 B 扫描，以下分别阐述。

A 扫描（如图 3.1）是指单次发射雷达波收到的单条回波，代表在道路上边移动边扫描过程中单个测点位置的反射波。A 扫描的横轴为时间轴，纵轴为电压值。当道路结构中存在充气脱空时，由于空气的介电常数小于道路材料，因此当雷达波会在进入脱空的界面处发生反射，反射波的电场方向与入射波相同，即产生一个与地表反射电压正负相反的反射波；雷达波在离开脱空再次进入道路材料的界面上会再次发生反射，该反射波与入射波电压方向相反，即电压正负号与地表反射一致。雷达信号分析中地表反射为一个正的波峰，因此，充气脱空在 A 扫描上的特征为一个负波峰后跟着一个正波峰，当脱空高度较小时两波峰叠加。相反地，由于水和注浆材料的介电常数大于道路材料，充水脱空和注浆修复后的原脱空位置的 A 扫描波形特征为一个正波峰后跟着一个负波峰，同样可能叠加；由于道路材料和水的介电常数差异大于和注浆材料的差异，因此充水脱空的波峰更明显。

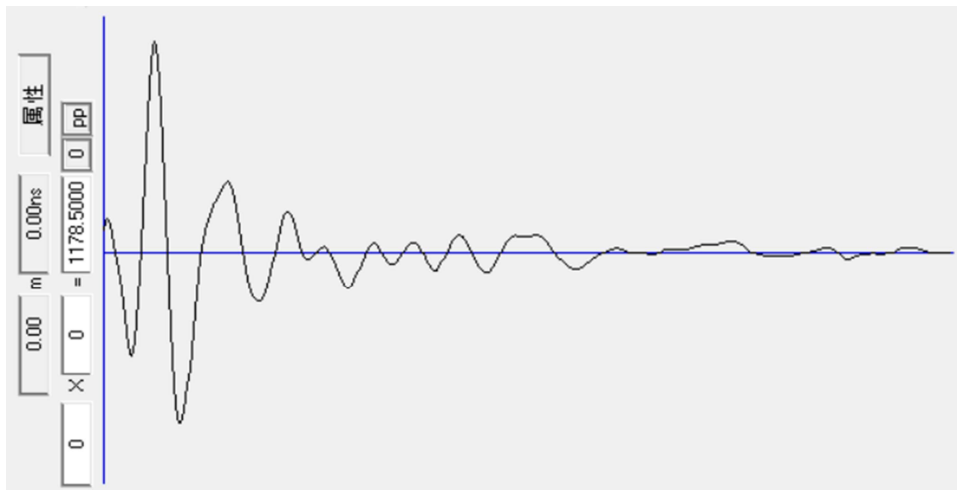

图 3.1 探地雷达图像 A 扫描

B 扫描（如图 3.2）是一条测线上各测点的回波信号的叠加，其横坐标为雷达设备相对起始点移动的距离，纵坐标为时间轴，每条测线上每个时刻的信号电压幅值大小用颜色或灰度表示。

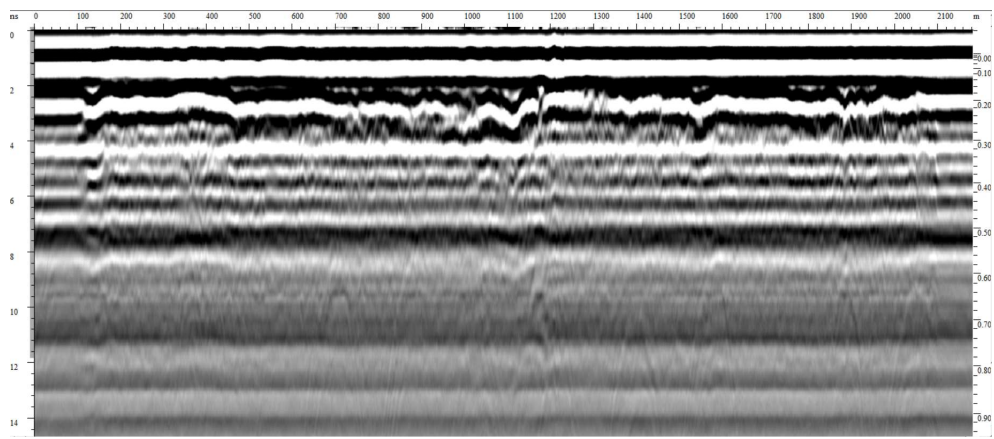

图 3.2 探地雷达图像 B 扫描

较小范围的脱空病害和较大范围的脱空在 B 扫描上特征有区别。首先讨论较小范围的脱空，将其视作一个点状目标。由于探地雷达的辐射形式为一个向外扩展的圆锥而非一条细线，因此当探地雷达不在目标物的正上方，但是该圆锥范围可以覆盖该目标物时，其表面的反射波也会反映在探地雷达回波信号上。因此，如图 3.3 所示，探地雷达从左向右移动测量过程中，在一定移动范围内都会探测到该物体。由于雷达天线和目标物的距离不断变化，目标物的 B 扫描上出现的位置也会变化。假设雷达天线移动轨迹为一条与道路表面平行的直线，记点状目标物与该直线的距离为 $a$ ；自点状目标物引一条与天线轨迹相交的垂线，天线在某一时刻与该垂交点的距离记为 $x$ ，则天线与目标物的距离 $y = \sqrt{a^2 + x^2}$ 。 B 扫描图像上 x 对应横坐标，y 对应纵坐标，a 为常数，满足关系式

$$y^2 = x^2 + a^2 \tag{3.1}$$

变形得

$$\frac{y^2}{a^2} - \frac{x^2}{a^2} = 1 \tag{3.2}$$

可以推得该点状目标物在 B 扫描图像上的轨迹为等轴双曲线的一支，双曲线关于 y 轴对称，点状目标物的位置为双曲线焦点（0，c），如图 3.4 所示。这里为简化计算，忽略折射导致的电磁波传播方向的变化。这就是一些文献[50]根据经验总结出的点状目标物"倒 V 形"或"椭圆形"特征的实际原因。由于介电常数的差异，充气脱空的反射"倒 V 形"颜色或灰度，与冲水或注浆脱空的反射特征相反。

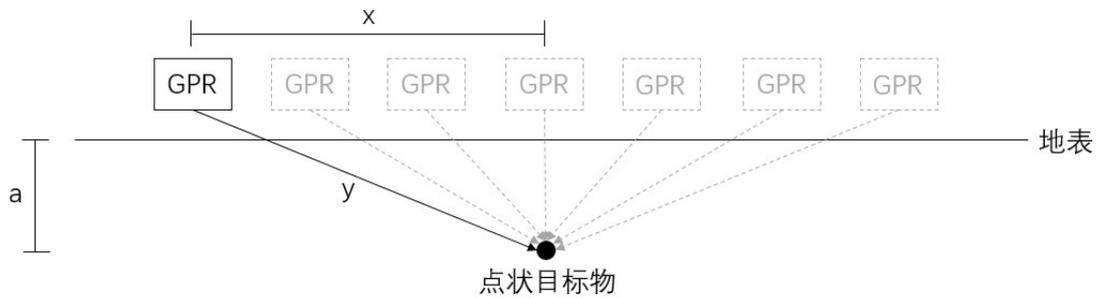

图 3.3 探地雷达与目标物之间距离的变化

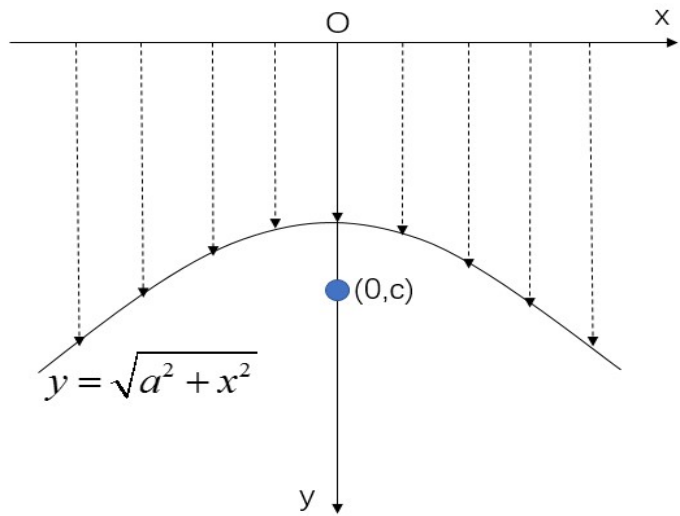

图 3.4 点状目标物呈双曲线形示意图

当脱空范围较大时，脱空位置不再被视作一个点目标，而被当成一定范围内的材料层。在 B 扫描图像上，显示为某一段图像上在相近深度处出现一条反射带，近似于道路结构层界面的反射。当脱空内部为空气时，脱空带上缘反射的反射条带颜色或灰度与地表反射相反；当脱空内部为水或注浆材料时，该反射条带

颜色或灰度与地表反射相同，内部为水时反射更明显。

## 3.2 FDTD 方法及 gprMax 简介

### 3.2.1 有限差分时间域（FDTD）方法

有限差分时域法是计算电磁场场量的一种方法。其计算原理为，把需要计算的空间划分为一定精度的网格单元，认为一个网格单元范围内场量相等。使用麦克斯韦方程的差分近似形式，计算每个网格单元的磁场和电场分量，从而实现对整个仿真范围的电磁场求解。

### 3.2.2 gprMax 数值仿真软件简介

gprMax 是英国爱丁堡大学的 Craig Warren 和 Antonis Giannopoulos 等人开发的一款用于模拟电磁波传播的数值仿真软件[61]。该软件使用有限差分时间域（Finite-Difference Time-Domain，FDTD）的方法，求解三维空间上的麦克斯韦方程组。该软件可用于探地雷达的成像模拟，也可用于其他领域的电磁波传播的仿真。GprMax 可以实现三维的仿真；当需要二维仿真时，只需把不需要的维度上的计算单元数设置为 1 即可。

该软件目前没有 GUI 界面，无法进行可视化操作。用户对于模型空间范围、不同介质的几何和电磁参数、天线类型及移动方式、发射波的波形和中心频率等参数的设置，全部以代码的形式写入.in 输入文件。以 "#" 开头的代码行被作为命令读取，其余代码行被视作注释文本。输入文件中也允许直接写入 Python 代码，在模型仿真过程中执行。该软件的输出文件为.out 格式，可以直接生成模拟的雷达图像，也可调入 MATLAB 绘制。该软件还可把输入的几何参数转化成.geo 格式的几何尺寸文件，可以直接或通过 MATLAB 生成直观的三维图像。输出文件符合 HDF5 格式，所以也可以使用开源的 Visualization Toolkit（VTK）查看和处理。 gprMax 允许设置的参数大致可以分为四类：空间参数、目标物参数、天线参数、发射波参数。空间参数包括：空间范围尺寸坐标，空间尺寸精度，仿真时间窗长短，边界条件。目标物参数包括：各种材料相对介电常数、电导率、相对磁导率、磁损耗，平面形状（边、圆盘、三角形），立体形状（立方体、球体、圆柱、部分圆柱、部分立方体），增加目标物体表面粗糙度、表面水膜层或草丛层等。天线参数包括：天线类型，收、发天线位置，收、发天线移动速度。发射波参数包括：波形，幅值，频率。

## 3.3 模型的建立过程

在实际道路脱空检测中，道路结构材料、厚度，脱空病害的面积、高度、填充情况，检测使用的天线频率、发射波波形等，都会对检测结果造成影响。其中，脱空高度对反射波形的叠加有重要影响，脱空面积对 B 扫描上的脱空位置特征有影响，填充情况直接影响局部介电常数从而影响反射波形，天线频率决定探测精度和深度，因此在建模过程中将这四个指标作为主要变量。其他可能影响脱空检测的指标作为次要因素考虑。由于实际工程中一般采用单天线探地雷达，即雷达图像反映沿扫描迹线的二维道路剖面上的情况。因此，本文与大多数文献一样，采用 2D 仿真，在建模中一个方向上设置单层立方单元。

### 3.3.1 扫描对象参数设置

（1）空间范围。根据经验，探地雷达的电磁波发射到地面上覆盖范围在 1m*1m 以内，路用探地雷达电磁波穿透深度一般不超过 2m。同时考虑到计算机运算能力，设置整个建模空间尺寸为沿测线方向长度 2.7m，垂直于测线宽度 0.003m，模型高度 1.6m，该空间外的微弱电磁波可以忽略不计。建立右手坐标系，x 轴为探地雷达移动方向，探地雷达的扫描断面与 x 轴、y 轴构成的平面平行。原点 O 为仿真空间最靠左、靠下、靠后的顶点。整个仿真空间在三维直角坐标系中的位置及尺寸如图 3.5 所示。

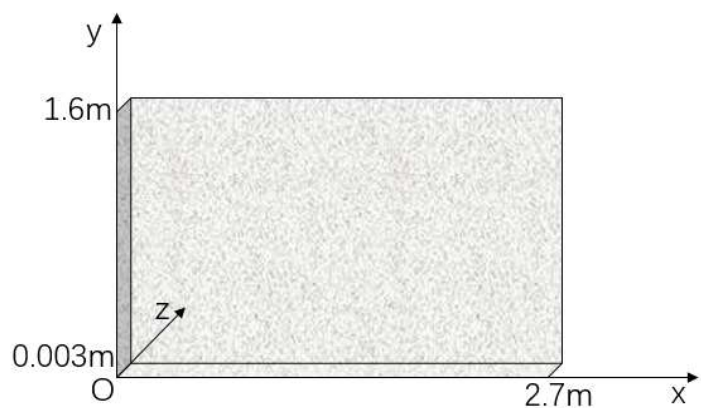

图 3.5 仿真空间范围

（2）空间精度。空间精度越高，即网格划分越细，则仿真结果越精确；但另一方面，网格划分越细，计算机处理速度越慢。根据经验，模型中网格精度应该达到最短波长的十分之一[61]。波长为波速与频率的比值，因此最短波长需要根据最小波速和最高频率计算。在本模型中波速最小的介质为水，其介电常数约为81。最高频天线选用中心频率为 1.2GHz 的天线。脉冲波为多种频率的叠加，根

据 Ricker 波的频谱,中心频率的两到三倍为较显著频率范围的上限。因此这里取 3GHz 作为最高频率。

$$\lambda = \frac{c}{f\sqrt{\varepsilon_r}} = \frac{299792458}{3\times 10^9 \times \sqrt{81}} \approx 0.0111 \ （m） \tag{3.3}$$

根据上式,同时考虑设置的最小脱空为 20mm,可以设置空间划分精度为 $dx = dy = dz = 3$（mm）。在下一章基于 A 扫描识别脱空部分,考虑到识别过程对精度要求高,且每个模型只需模拟一次扫描,因此仅在该部分采用 $dx = dy = dz = 1$（mm）空间划分精度。

（3）道路几何参数。为研究不同的脱空尺寸对脱空检测的影响,道路结构模型将选用若干组几何参数。因面层内部分层对基底脱空的检测几乎无影响,因此可以将整个沥青面层视作同一种材料。道路结构采用面层、基层、路基三层结构,尺寸固定。在路基顶部中心位置设置脱空,脱空面积、脱空高度分别选用不同尺寸值。模型选用的道路结构见表 3.1。

<p align="center">表 3.1 道路模型几何参数</p>

| 道路结构层/脱空 | 水平尺寸（m*m） | 厚度（m） |
|---|---|---|
| 沥青面层 | 3.1*0.003 | 0.15 |
| 半刚性基层 | 3.1*0.003 | 0.35 |
| 土基 | 3.1*0.003 | 0.5 |
| 脱空 | 0.2*0.003 或 0.5*0.003 | 0.02,0.05,0.1,或 0.2 |

（4）材料电磁特性。gprMax 模型需要设置材料的相对介电常数、电导率、相对磁导率、磁损耗四种电磁特性参数。非磁性材料的相对磁导率为 1,磁损耗为 0。其余两参数根据现有文献[62,43]选取。由于脱空位置可能充气或充水,注浆修复后还会被浆液填充,因此分别考虑三种不同材料的电磁性质。各种材料的介电常数见表 3.2。

<p align="center">表 3.2 各种材料电磁特性</p>

| 道路结构层/脱空 | 相对介电常数 | 电导率 | 相对磁导率 | 磁损耗 |
|---|---|---|---|---|
| 沥青面层 | 6 | 0.005 | 1 | 0 |
| 半刚性基层 | 7.5（/9.5） | 0.01 | 1 | 0 |
| 土基 | 18 | 0.2 | 1 | 0 |
| 充气脱空 | 1 | 0 | 1 | 0 |
| 充水脱空 | 81 | 1 | 1 | 0 |
| 注浆材料 | 28 | 0.01 | 1 | 0 |

（5）边界条件。在真实情形中，探地雷达的电磁波在无限大的空间内传播。但由于计算机的计算能力限制，仿真计算需要设置一个合理大小的范围，只对范围内的空间进行计算。本模型仿真的空间范围大小上文已阐述，由于范围外的电磁波非常微弱，在仿真中应当忽略不计。PML（Perfectly Matched Layer）吸收边界条件是指能够使进入的电磁波迅速衰减，从而不发生反射的边界条件。gprMax默认在仿真空间的六个面上设置 10 个单元网格厚度的 PML 吸收边界条件，本模型采用该默认值。

### 3.3.2　扫描参数设置

（1）发射波参数。gprMax 内置的激励源类型包括高斯波、正弦波、余弦波、瑞雷波（Ricker wave）等，目前各厂商的脉冲式雷达大多采用瑞雷波，因此本模型也采用瑞雷波。图 3.6（a）为瑞雷波的时域上的波幅图，图 3.6（b）为其频域上的信号强度图[63]。瑞雷波的表达式为：

$$W(t) = -\left(2\zeta(t-\chi)^2 - 1\right)e^{-\zeta(t-\chi)^2} \tag{3.4}$$

式中 $f$ 为频率，$\zeta = \pi^2 f^2$，$\chi = \dfrac{\sqrt{2}}{f}$。

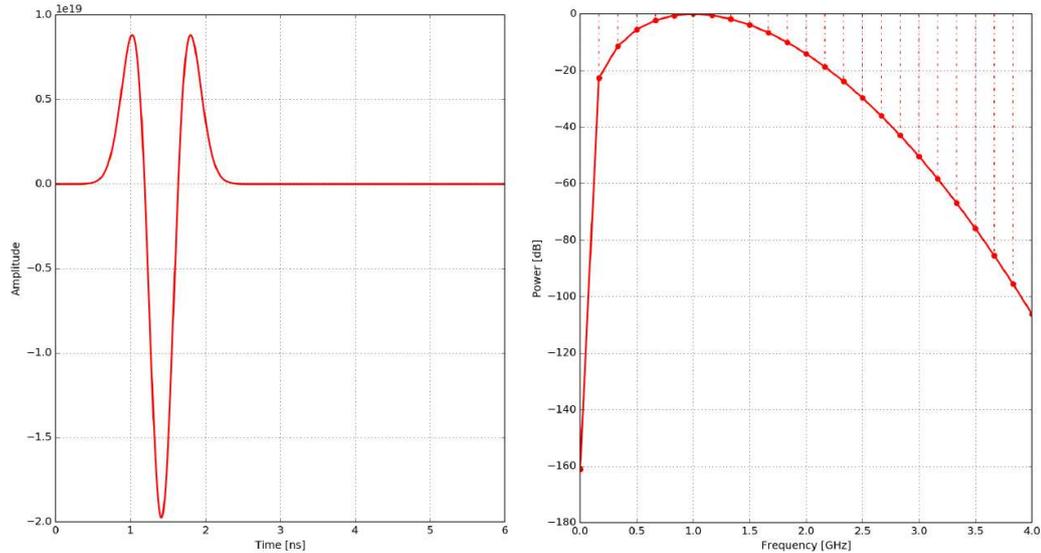

（a）时域波形图　　　　　　（b）瑞雷波频谱图

图 3.6　瑞雷波时域及频域分布[63]

（2）收发天线参数。设置一个偶极子天线（Hertz dipole）作为发射源，极化方向为 z 轴方向，即平行于水平面且垂直于雷达移动方向。发射天线起始位置坐标为（0.45，1.5，0），接收天线起始位置坐标为（0.49,1.5,0），收、发天线距离 4cm。收、发天线沿 x 轴正方向移动，移动步距 0.02m，即每移动 2cm 雷达进

行一次发射、接收。探地雷达总共移动 1.8m 距离，完成 90 次扫描，发射天线终止位置坐标为（2.25,1.5,0），接收天线终止位置坐标为（2.29,1.5,0）。由于道路结构模型的轴对称性，模拟结果图像应该关于模型中线对称。

（3）时窗。时窗为接收天线收集信号的时间长短，可以根据道路各层结构厚度及雷达波在各层中的传播速度决定。若不计各种脱空情况，根据表 3.1 和表 3.2 估算可得，在本模型的完整道路结构中电磁波的往返传播时间为 22.98ns。由于接下来将模拟多种情况，所以统一采用经验值 25ns 时窗。

$$t = \sum \frac{2h}{v} = \sum \frac{2h\sqrt{\varepsilon_r}}{c} \approx 22.98 \text{（ns）} \tag{3.5}$$

根据以上扫描对象参数设置和扫描参数设置，按照 gprMax 软件要求写成.in 文件。执行 python -m gprMax user_models/ "文件名" .in -n 90 命令，即开始建模仿真过程。

## 3.4 仿真参数及几何模型

本研究将根据研究目标建立一定数量的模型，在各模型中取固定值的参数已经在上一节"模型的建立过程"中讨论。为了研究使用不同频率天线对不同尺寸、填充情况的脱空的检测，在建立模型过程中选取 4 个指标作为变量，如表 3.3 所示。

表 3.3 仿真模型非固定指标取值

| 指标 | 取值 |
|---|---|
| 脱空高度 | 0.02，0.05， 0.1，或 0.2 （m） |
| 脱空水平尺寸 | 0.2*0.003 或 0.5*0.003 （m*m） |
| 脱空填充情况 | 充气，充水或注浆 |
| 天线中心频率 | 400，800 或 1200 （MHz） |

为了检查构建的道路结构几何模型是否正确,本研究使用 gprMax 生成.vti 文件，再在开源的数据分析和可视化软件 Paraview 中生成几何模型示意图。限于篇幅，这里仅展示两个几何模型：图 3.7 为脱空高度 0.2m、脱空水平尺寸 0.5m*0.5m 的充气脱空道路结构，图 3.8 为脱空高度 0.02m、脱空水平尺寸 0.2m*0.2m 的充水脱空道路结构。其余模型的几何形式类似，仅是脱空位置的高度、水平尺寸、填充情况发生变化，这里不再一一展示。

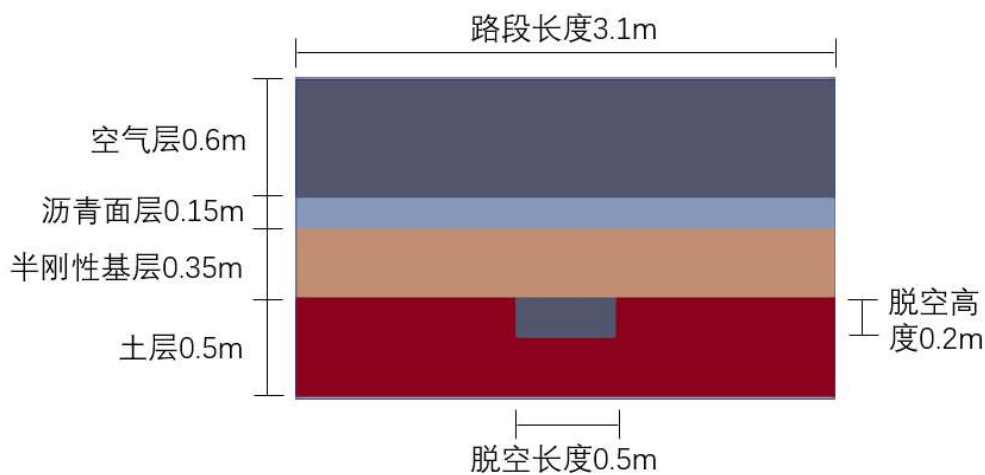

图 3.7 几何模型示意图一

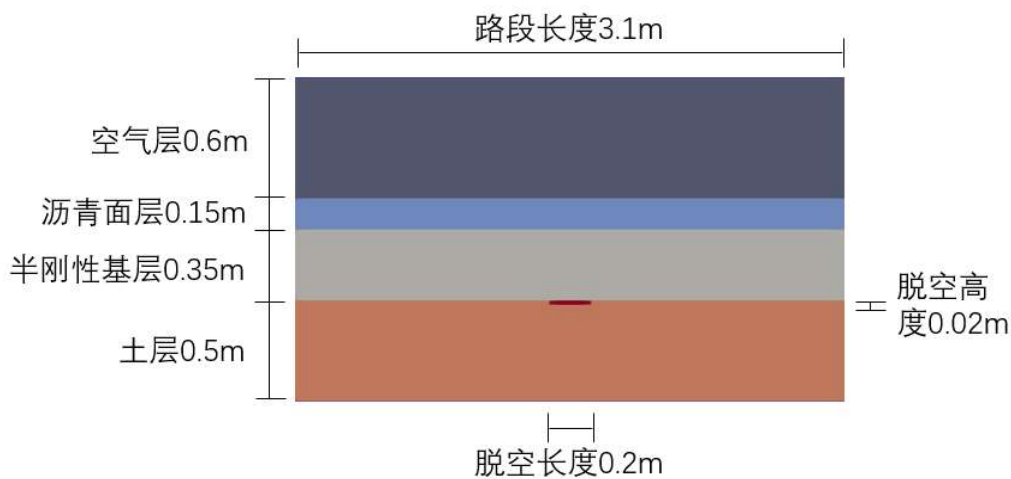

图 3.8 几何模型示意图二

## 3.5 仿真结果分析

按照上述的不同指标，仿真得到 2（脱空水平尺寸）*5（脱空竖向尺寸）*3（天线频率）*3（脱空内部物质）共 90 个道路模型的 B 扫描和中心位置 A 扫描；此外，为了下文内容中的脱空高度估算及水平尺寸估算，额外仿真生成数十个模型。

本研究生成的 A 扫描是在脱空位置中心正上方位置扫描得到的，如图 3.9 所示，包含直达波、路表反射、面层-基层界面反射、基层-脱空界面反射和脱空-土基界面反射。

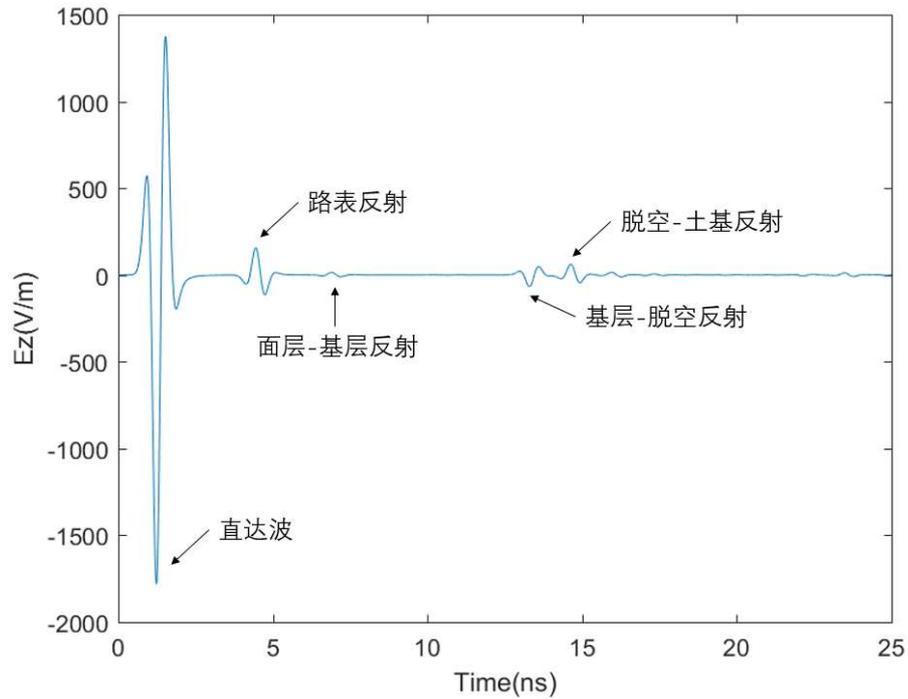

图 3.9 A 扫描示例（0.2m 充气脱空，1200MHz 天线）

本研究生成的 B 扫描是收发天线在 2.7m 长度的路段模型中间 1.8m 范围内移动扫描得到的，如图 3.10 所示。两侧分别留 0.45m 的空间，避免边界效应对 B 扫描的影响。由于直达波的波幅远大于道路内部界面的反射波，如果保留直达波会使得道路内部反射非常微弱。因此，本研究中的 B 扫描都经过了去直达波的处理。

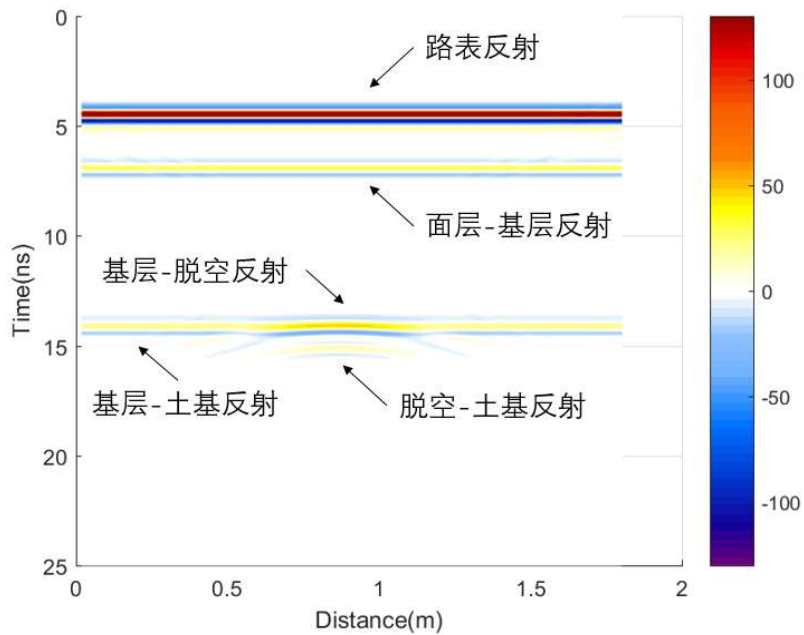

图 3.10 B 扫描示例（0.01m 高度 0.2m 长度充水脱空，1200MHz 天线）

脱空竖向尺寸、脱空内部物质等信息可以由 A 扫描获得，脱空水平尺寸可以通过 B 扫描得到，这两方面内容主要在第四章讨论。三种不同频率天线影响雷达图像的精度，如图 3.10-图 3.12 所示。400MHz 天线的反射波叠加明显，不利于分析；800MHz 和 1200MHz 天线的精度较高，考虑到 800MHz 的探测深度更适合基层底部的探测，因此下文的分析主要基于 800MHz 天线的仿真扫描数据，兼顾 1200MHz 天线的仿真扫描数据。

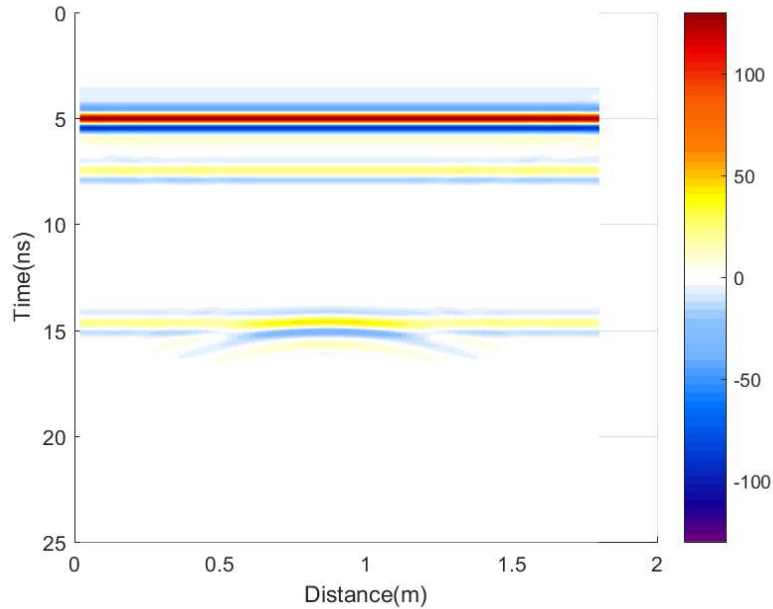

图 3.11 B 扫描示例（0.01m 高度 0.2m 长度充水脱空，800MHz 天线）

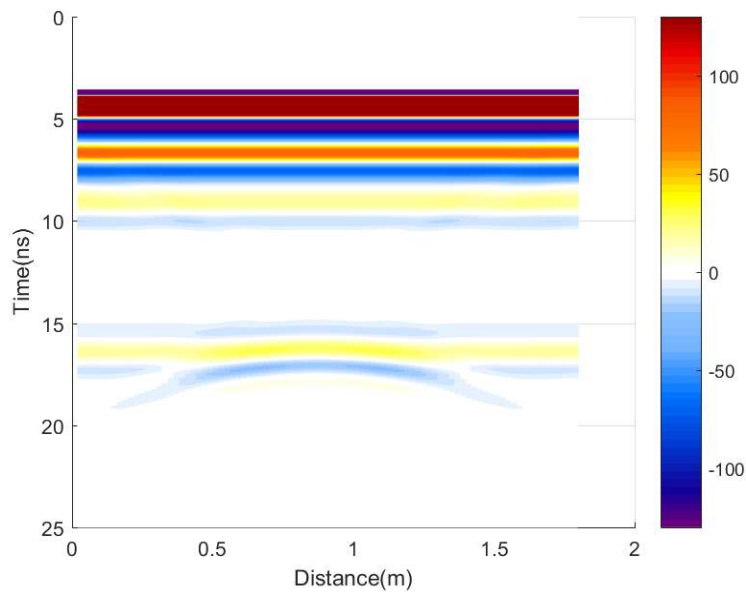

图 3.12 B 扫描示例（0.01m 高度 0.2m 长度充水脱空，400MHz 天线）

根据仿真生成的 B 扫描，可以看出基底脱空在雷达图像上的两方面特征：

首先，雷达图像上的脱空特征曲线并不是严格的等轴双曲线。依据推导得到的脱空特征曲线的方程式（3.2），若纵轴表示深度，且刻度长短与表示水平距离的水平轴的刻度长短一致，则脱空位置的特征线应该为等轴双曲线的一支，其渐近线为与坐标轴成 45° 角的直线。但是，实际的雷达图像的纵轴表示时间，若要使图像呈现标准等轴双曲线，需要先将纵轴的时间坐标乘以电磁波传播速度换算成深度坐标，且调整横轴和纵轴的刻度长短使其一致。然而，由于每个结构层内电磁波的传播速度不同，且实际检测中各层内传播速度未知，因此很难将时间换算成深度，脱空在图像上的特征也就无法被调整为严格的等轴双曲线。

其次，雷达图像上的脱空反射特征实际上叠加了周围的结构层界面反射。由于探地雷达的发射波呈发射状而非一条射线，因此雷达在某测点得到的回波信号以及图像特征，实际上反映该测点附近一定水平范围内的界面反射，而不仅仅是该测点正下方的界面反射；也就是说，同一水平高度上相邻的反射物的反射特征在雷达图像上会相互叠加。对于面层或基层内部的空洞来说，空洞附近同一水平高度上为不发生反射的均匀结构，因此空洞的图像特征不会叠加其他的反射；但是基底脱空的周围同一水平高度上是基层-土基分界面，基层-土基分界面也会发生反射，两种反射会在探地雷达扫描图像上叠加。

## 3.6 本章小结

（1）本章首先从理论上分析了基底脱空的图像特征。A 扫描上，充气脱空的特征为一个负波峰后跟着一个正波峰，充水脱空和注浆修复脱空的特征为一个正波峰后跟着一个负波峰，脱空高度较小时两波峰重叠。B 扫描上，较小范围的脱空的图像特征为等轴双曲线的一支，实际的脱空位置在双曲线的焦点上；较大范围的脱空相当于一定范围内增加的一个材料层，图像特征类似于道路结构层分界面反射。

（2）为研究脱空在探地雷达图像上的特征，并提出指标用于计算脱空的高度和水平尺寸，使用基于 FDTD 方法的 gprMax 软件数值模拟探地雷达发射波在道路中的传播。仿真模型的建立过程包括：规定空间范围、空间精度、边界条件，设定道路的几何及材料参数，选定发射波类型、收发天线的移动方式及采集时窗。

（3）仿真得到 2（脱空水平尺寸）*5（脱空竖向尺寸）*3（天线频率）*3（脱空内部物质）共 90 个道路模型的 B 扫描和中心位置 A 扫描；此外，为了下文内容中的脱空高度估算及水平尺寸估算，额外仿真生成数十个模型。根据仿真图像可知 400MHz 天线精度不能满足脱空检测要求，800MHz 为最优的天线频率。生

成的图像验证了本章第一部分对于基底脱空的图像特征的理论分析；需要注意的是水平坐标和竖向坐标的比例关系会影响双曲线特征的形状，且基层-土基界面的反射会与脱空反射重叠。

# 第四章 基底脱空及注浆识别算法

和其他道路病害相比，半刚性基层层底脱空与地面距离较大，而竖向尺寸较小，而且内部情况复杂。因此，基底脱空以及其注浆修复效果的评价一直是探地雷达在道路检测中的应用难点。现有文献对于基底脱空的检测及注浆效果的评价主要依靠经验，根据雷达色谱图上的异常位置的颜色（根据颜色判断充气或是充水）和形状（"倒 V 型"）做出判断[43,50]。但这种方法缺乏合理的理论依据，无法识别较小脱空，且容易发生误判。因此，为了实现道路基底脱空和注浆扩散范围的精确识别，需要系统研究有效的信号处理方法。道路薄层检测及隧道壁后注浆的相关研究提供了一些可以参考的识别算法，但由于工程特点的显著差异，有必要研究这些算法在道路基底脱空检测中的适用性及改进方式。

针对基底脱空的检测和注浆扩散的评价，本章首先分析基底脱空的检测特点和信号处理的需求，为信号处理技术的选用奠定基础；然后，选用三种信号处理算法，应用于上一章建立的脱空路面正演模型，分析其脱空识别及高度估算的效果；最后，研究脱空水平尺寸的估算方法。

## 4.1 基底脱空检测的相关研究

尽管几十年来探地雷达在道路多种病害的检测中得到了显著的进展，半刚性基层层底脱空的检测依旧是一个尚未得到理想解决的问题。其难度主要来自于三方面。首先，基层层底脱空与地面距离较大，雷达信号经过路表、道路内部结构层分界面的多次反射以及有耗介质材料造成的能量耗散，传递至脱空位置时信号已大大减弱，微弱的反射信号很容易被噪声掩盖。其次，层底脱空尺寸相对于其深度来说较小，按照探地雷达使用的经验，雷达精度大致为深度的十分之一尺寸，基底脱空的竖向尺寸有时小于这一标准。在层底脱空检测雷达天线频率的选择上，由于基底脱空的深度较大，为了达到其探测深度，需要选用较低频率的天线；但天线频率越低，探测精度也就越低，而基底脱空相对较小的尺寸又要求天线具有较高精度，因此层底脱空的较大深度和较小竖向尺寸这对"相互矛盾"的要求大大增加了检测难度。再次，由于基底脱空与地表隔着多层道路结构层，无法用地表的干湿情况判断基底脱空内的干湿情况，地表干燥时基底脱空内部可能有滞留水分，而地表潮湿时，水分亦有可能并没有渗透到基层底部。因此，基底脱空内部填充情况的不确定，也增加了检测的难度。

现有文献[43,50]以及工程实践中，常用的判断方法为：先在探地雷达 B 扫描图像上寻找若干近似水平的连续条带，根据钻芯或设计资料确定这些连续条带所对

应的道路结构层分界面；确定扫描图像上的道路分层情况后，寻找除结构层界面条带以外的颜色异常区域，并根据异常位置的颜色所对应的反射波的正负，判断该位置为充水或是充气脱空，特别的是，如果发现有"倒 V 形"特征位置，判断为脱空。注浆修复后，在原来的脱空区域的路表再次使用探地雷达检测，寻找结构层分界面以外的异常位置，分析其反射波幅值大小，并与注浆前的探地雷达图像对比。这种方法可以用于粗略的脱空检测及注浆评价，但是存在明显的问题。一方面，仅根据 GPR 图像上的颜色异常判断脱空，有可能将局部含水率高、压实度不均匀等其他病害误判为脱空；另一方面，由于脱空的竖向尺寸较小，脱空上下界面反射波发生叠加，而经验法判断充气或充水是基于脱空上表面的反射波，即根据单个反射波考虑，因此有可能造成误判。

目前也有一些关于路基脱空仿真的研究[64]，但路基脱空的一个特点往往被忽略：对于公路路基来说，脱空多发生于基层底部，由行车荷载下路基土永久变形、路基刚度差异以及基基中水的作用引起[2]。而这些现有研究建立的模型中脱空位于路基内部，与实际情况不同。从探地雷达检测的角度考虑，基底脱空位于半刚性基层底部，意味着脱空位置的脱空上边界，和周围无脱空区域的基层与路基分界面，处于同一深度。反映在探地雷达图像上，脱空顶部的反射波与周围无脱空区域土基顶面的反射波处于同一高度，因此要通过二者反射波幅的区别判断，以免把脱空顶部反射误当作基层和土基的层间界面反射。当然，现实中基底脱空也可能不只是路基顶面下降，基层的底面也有一定程度的冲刷损耗，但作为数值模拟的典型结构模型，本研究把脱空设置在与基底接触的土基上层。

从信号分析的角度考虑，基层底部脱空识别或是注浆效果评价，本质上是一个薄层识别的问题。对于这一问题，目前有一些相近的研究。盾构法施工的隧道中，隧道壁由拼接的衬砌组成，衬砌后的空隙是衬砌开裂的重要原因，需要及时注浆修复。同济大学的 Xiongyao Xie 和 Tohoku University 的 Rai Liu 等人应用带通滤波器、K-L 滤波器、CIRM 方法和 CMP 方法等，在基于探地雷达的衬砌后空洞的探测和注浆评价方面做了相关研究[53,65]。道路在运营过程中使用性能下降，薄层罩面是较常用的一种修复方法，探地雷达可以用于薄加铺层的厚度检验。Institut Supe´rieur des Sciences Applique´es et de Technologie de Sousse 的 Samer Lahouar 等人使用"匹配去除"的方法循环去除识别出的反射波从而避免其掩盖后面的较弱反射波[17]，效果较好但耗时较长；University of Illinois at Urbana-Champaign 的 Shan Zhao 等人[66]将解卷积、L 曲线方法用于薄层上下界面反射波的分离，取得较好的效果。然而，以上两方面研究的工程结构与基底脱空存在区别：道路沥青薄层往往作为表面加铺层，其上部没有其他结构的遮挡，薄层上下界面的反射波较明显，不存在基底脱空反射波信号强度弱的困难；隧道壁厚注浆

的检测虽然隔着隧道壁,但隧道壁一般比道路基底脱空上方的道路结构的厚度更薄,而壁后空隙和注浆的厚度又比道路基底脱空厚,因此检测难度小于道路基底脱空。因此,本章其余部分研究各种信号处理方法在道路基底脱空检测中的适应性,以及可能的改进方法。

## 4.2 脱空的识别及脱空高度的计算

脱空的介电常数异于周围的道路材料,可以视作道路结构中的一个"薄层"。但是,与沥青路面表面薄层罩面的识别及厚度测量不同的是,半刚性基层底部脱空的深度较大,由于信号的衰减,反射信号强度较弱,而且脱空内部情况复杂,这些给脱空的识别和高度计算造成很大困难。针对半刚性基层底部脱空的特点,以下研究相应的识别和高度计算方法。

针对不同填充情况的脱空,本研究采用不同的识别方法。对于充气脱空,采用最小二乘系统辨识法和解卷积法实现脱空的识别和高度估算,在4.2.2和4.2.3中讨论。对于充水脱空及注浆修复脱空,采用介电常数法实现脱空的识别,在4.2.4中讨论。

### 4.2.1 不同高度脱空的图像特征

本小节分析不同高度和不同填充情况的脱空在A扫描图像上的特征,作为下文研究脱空识别及尺寸估计算法的基础。为避免脱空水平尺寸对高度的计算产生影响,本节脱空的水平尺寸均选取足够大,即与道路结构的水平尺寸一致。根据第三章数值仿真的结果,400MHz天线的精度较低,信号重叠严重,而800MHz和1200MHz的精度较高,且反射波强度能够被识别,因此,本章采用800MHz和1200MHz两种频率的天线。

（1）充气脱空

雷达波在穿过充气脱空的过程中,先在基层-脱空界面上产生一个负的反射波,再在脱空-土基界面上产生一个正的反射波;脱空高度较大时,两个反射波保持一定间隔,脱空高度较小时,两个反射波重叠。需要说明的是,由于直达波的最大幅值为负值,因此与直达波电场方向一致的反射波是负的,与直达波电场方向相反的反射波是正的。

先用800MHz天线进行数值模拟,得到不同脱空高度的A扫描图像。按照0.01m的间距,生成0.01m到0.3m高度的脱空道路结构A扫描图像。由于篇幅所限,这里展示0.3m、0.25m、0.2m、0.15m、0.1m、0.05m、0.01m高度脱空的图像,如图4.1-图4.7所示。

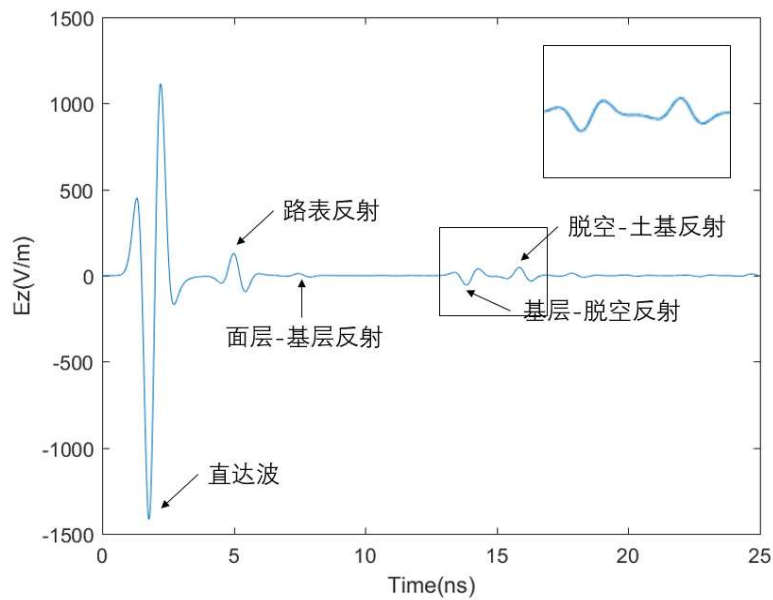

图 4.1 A 扫描图像（0.3m 充气脱空，800MHz 天线）

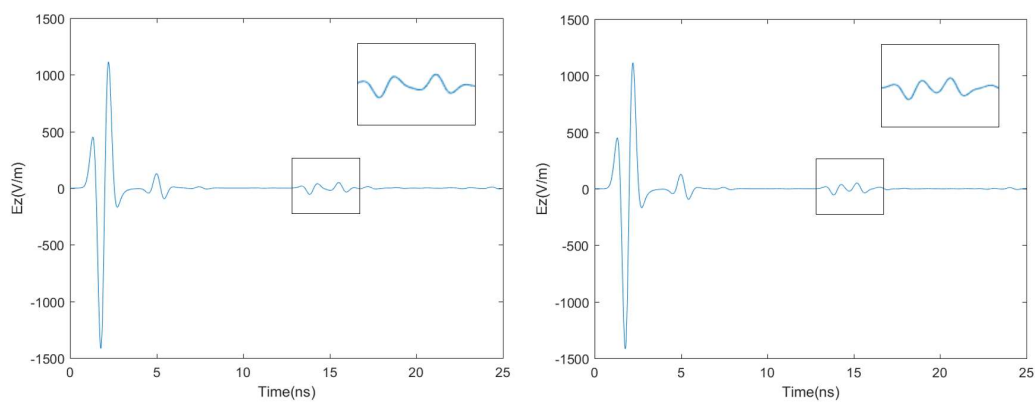

图 4.2 A 扫描图像（0.25m 充气脱空，800MHz 天线）

图 4.3 A 扫描图像（0.2m 充气脱空，800MHz 天线）

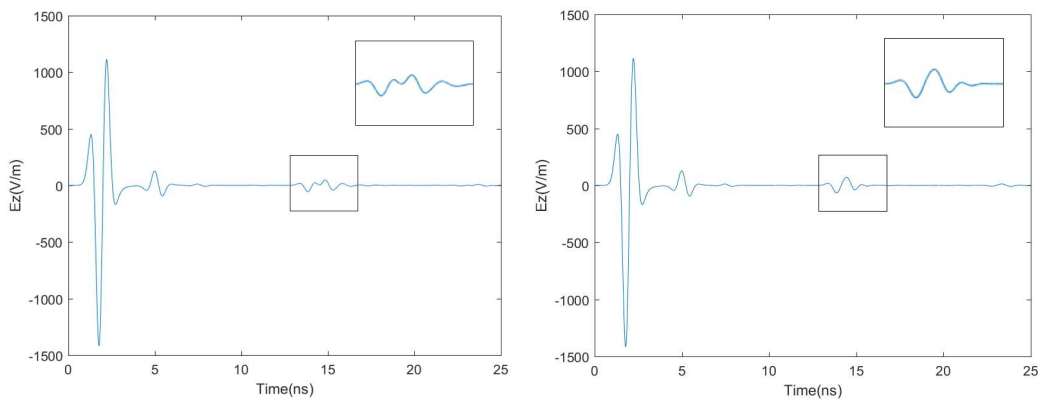

图 4.4 A 扫描图像（0.15m 充气脱空，800MHz 天线）

图 4.5 A 扫描图像（0.1m 充气脱空，800MHz 天线）

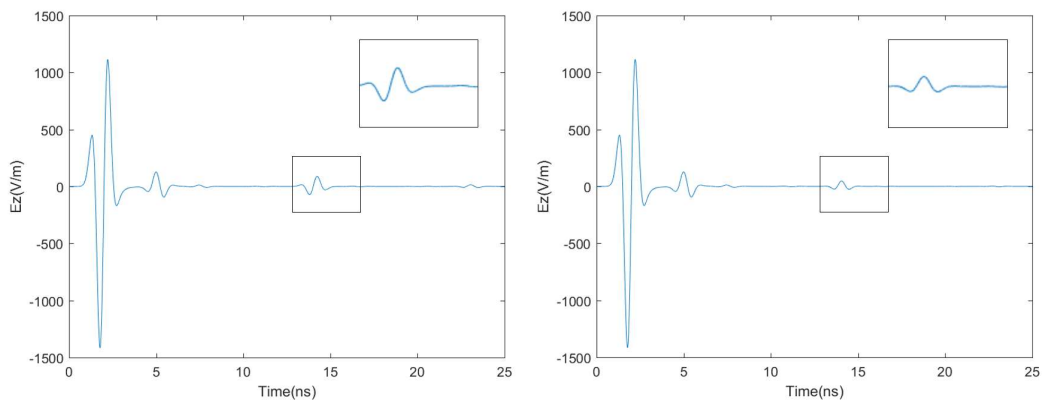

图 4.6 A 扫描图像（0.05m 充气脱空，800MHz 天线）

图 4.7 A 扫描图像（0.01m 充气脱空，800MHz 天线）

从以上一系列图像可以看出，充气脱空的图像特征是前后两个方向相反的反射波。第一个反射子波，基层-充气脱空界面反射，与发射波方向相同；第二个反射子波，充气脱空-土基界面反射，与反射波方向相反。当脱空高度较大时，上下两个界面间距较大，两个子波也不发生重叠，两子波中间有一段幅值近似等于零的范围。随着脱空高度的减小，两个反射子波间距变小，发生部分重叠。图 4.1 到图 4.7 展示了随着脱空高度变小 A 扫描图像上的变化规律。可以将脱空图像的特征随脱空高度减小的变化分成三个阶段。

第一个阶段特征为两个反射子波不发生重叠。如图 4.1 所示（脱空高度 0.3m），两个子波不发生重叠，各自的波形不受对方影响，两子波中间有一段幅值近似等于零的范围。随着脱空高度的减小，两个子波之间近似等于零的范围的长度不断减小以至于消失（如图 4.2 脱空高度 0.25m）。

第二个阶段特征为两个子波中间的"山谷"。脱空高度减小到某值后，前一个子波的"尾部"（幅值为正）和后一个子波的"头部"（幅值为正）发生加，两个子波各自的正的波峰之间出现一个幅值为正的"山谷"，且该"山谷"的幅值随脱空高度减小而增大（如图 4.3 脱空高度 0.2m 和图 4.4 脱空高度 0.15m）。与单个波的波谷不同的是，单个波的波谷幅值为负，而这里的"山谷"幅值为正。

第三个阶段特征为波峰唯一。随着脱空高度的减小，"山谷"的幅值不断增加，最后变成一个波峰（如图 4.5 脱空高度 0.1m，图 4.6 脱空高度 0.05m 和图 4.7 脱空高度 0.01m）。这个阶段的唯一波峰实际上是前一个子波的后半部分和后一个子波的前半部分的叠加。随着脱空高度的减小，叠加后的波峰先增大后减小，叠加波峰达到最大值时，前后两个子波各自的最大幅值恰好重叠。

接下来，再使用 1200MHz 天线进行数值模拟，得到不同脱空高度的 A 扫描图像。由于电磁波的频率和波长成反比，因此 1200MHz 天线具有更高的精度，脱空上下反射面的反射波也更不容易重叠。按照 0.01m 的间距，生成 0.01m 到

0.3m 高度的脱空道路结构 A 扫描图像。由于篇幅所限，这里展示 0.3m、0.25m、0.2m、0.15m、0.1m、0.05m、0.01m 高度充水脱空的图像，如图 8-图 14 所示。

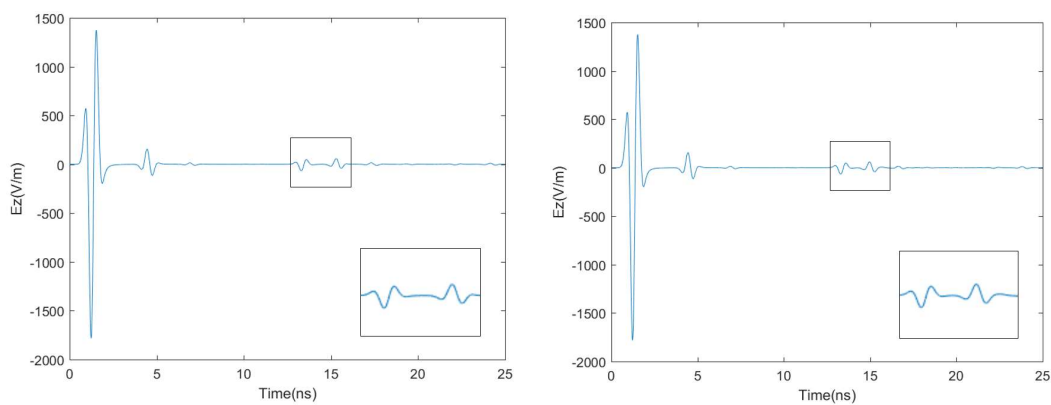

图 4.8 A 扫描图像（0.3m 充气脱空，1200MHz 天线）

图 4.9 A 扫描图像（0.25m 充气脱空，1200MHz 天线）

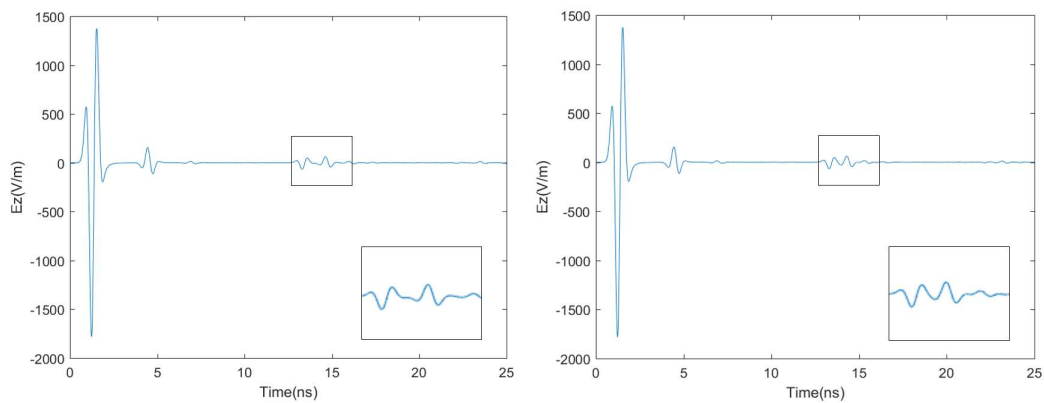

图 4.10 A 扫描图像（0.2m 充气脱空，1200MHz 天线）

图 4.11 A 扫描图像（0.15m 充气脱空，1200MHz 天线）

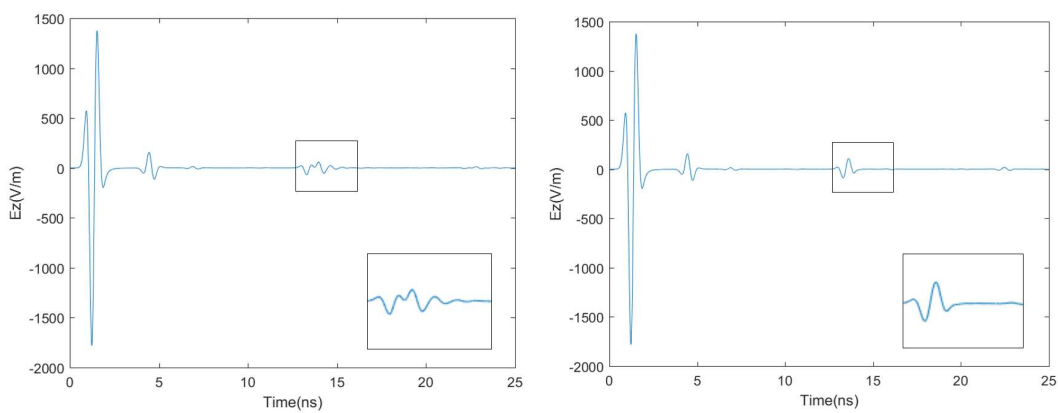

图 4.12 A 扫描图像（0.1m 充气脱空，1200MHz 天线）

图 4.13 A 扫描图像（0.05m 充气脱空，1200MHz 天线）

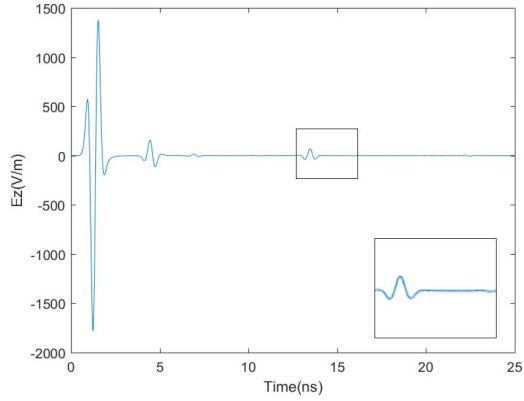

图 4.14 A 扫描图像（0.01m 充气脱空，1200MHz 天线）

从以上一系列图像可以看出，由于 1200MHz 天线的波长比 800MHz 天线的波长短，在相同的基层脱空高度下 1200MHz 天线获取的信号中两个子波更不容易重叠。对于 1200MHz 频率的天线，0.3m-0.17m 高度的脱空属于第一个阶段，特征为两个反射子波不发生重叠。0.16m-0.09m 高度的脱空属于第二个阶段，特征为两个子波中间的幅值为正的"山谷"。0.08m 及以下属于第三个阶段，特征为波峰唯一，随着脱空高度的减小，叠加后的波峰先增大后减小，叠加波达到最大值时，前后两个子波各自的最大幅值恰好重叠。

（2）充水脱空

理论上，雷达波在穿过充水脱空的过程中，先在基层-水层界面上产生一个正的反射波，再在水层-土基界面上产生一个负的反射波；充水脱空高度较大时，两个反射波保持一定间隔，充水脱空高度较小时，两个反射波重叠。

用 800MHz 天线进行数值模拟，得到不同充水脱空高度的 A 扫描图像。按照 0.01m 的间距，生成 0.01m 到 0.3m 高度的充水脱空道路结构 A 扫描图像。由于篇幅所限，这里展示 0.3m、0.2m、0.1m、0.04m、0.03m、0.02m、0.01m 高度充水脱空的图像，如图 4.15-图 4.21 所示。

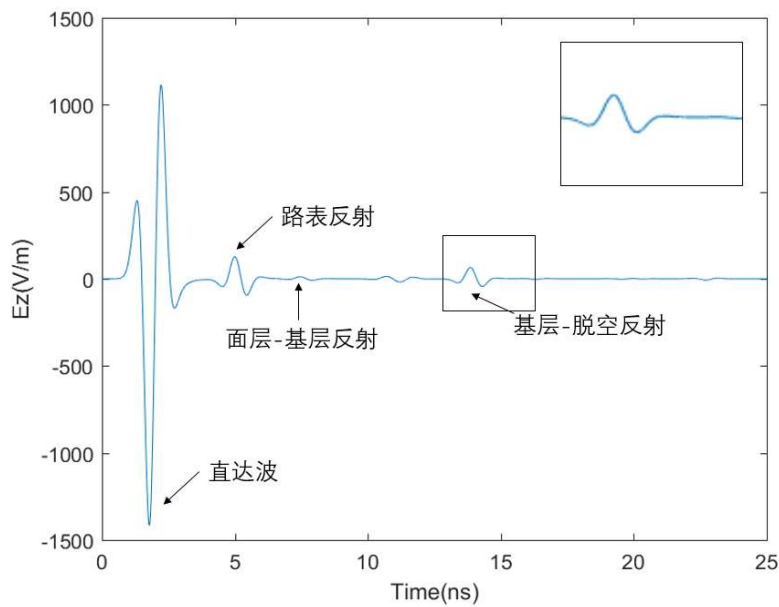

图 4.15 A 扫描图像（0.3m 充水脱空，800MHz 天线）

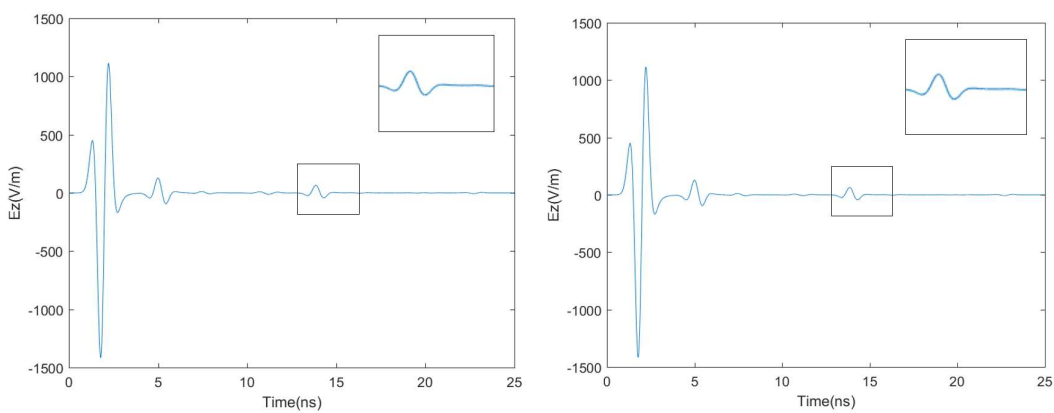

图 4.16 A 扫描图像（0.2m 充水脱空，800MHz 天线）

图 4.17 A 扫描图像（0.1m 充水脱空，800MHz 天线）

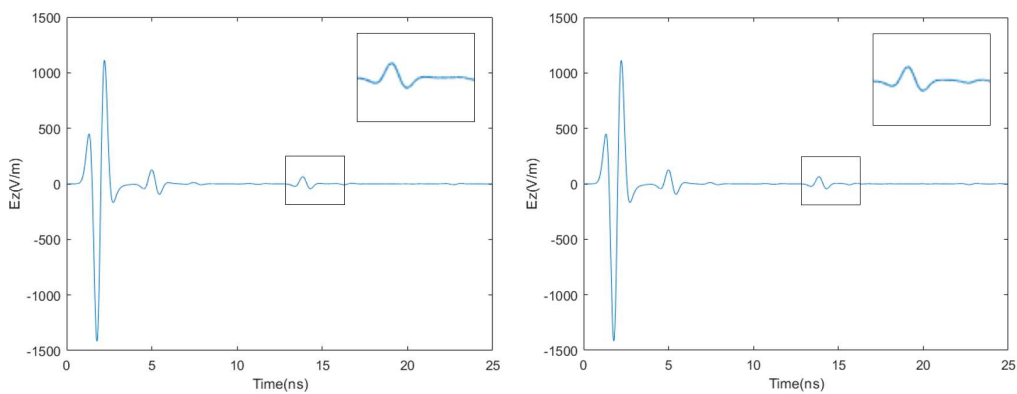

图 4.18 A 扫描图像（0.04m 充水脱空，800MHz 天线）

图 4.19 A 扫描图像（0.03m 充水脱空，800MHz 天线）

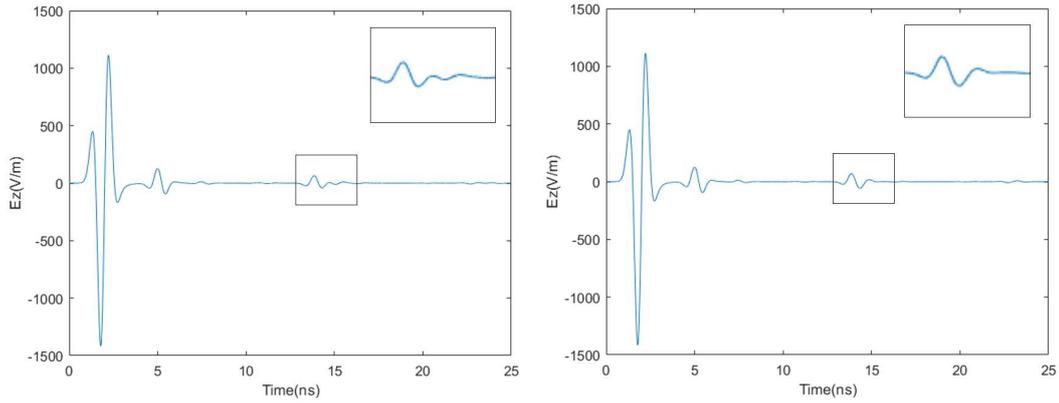

图 4.20 A 扫描图像（0.02m 充水脱空，800MHz 天线）

图 4.21 A 扫描图像（0.01m 充水脱空，800MHz 天线）

由以上一系列图像可看出，不同于充气脱空，充水脱空的脱空-土基界面反射非常微弱，当充水脱空高度大于 0.04m 时，该反射难以被识别，当充水脱空高度小于 0.04m 时，该反射随着水层高度的减小而幅度增加，但反射波始终较弱。这是因为水具有导电性，电磁波在水中衰减迅速，而且探地雷达采用的是高频天线，雷达信号在水中只能传播很短的距离。因此，充水脱空的识别主要依靠基层-充水脱空界面的反射波，而且充水脱空的高度难以估测。

（3）注浆修复脱空

理论上，雷达波在穿过注浆修复脱空的过程中，先在基层-注浆材料界面上产生一个正的反射波，再在注浆材料-土基界面上产生一个负的反射波；注浆修复脱空高度较大时，两个反射波保持一定间隔，注浆修复脱空高度较小时，两个反射波重叠。

用 800MHz 天线进行数值模拟，得到不同注浆修复脱空高度的 A 扫描图像。按照 0.01m 的间距，生成 0.01m 到 0.3m 高度的注浆修复脱空道路结构 A 扫描图像。由于篇幅所限，这里展示 0.3m、0.25m、0.2m、0.15m、0.1m、0.05m、0.01m 高度注浆修复脱空的图像，如图 4.22-图 4.28 所示。

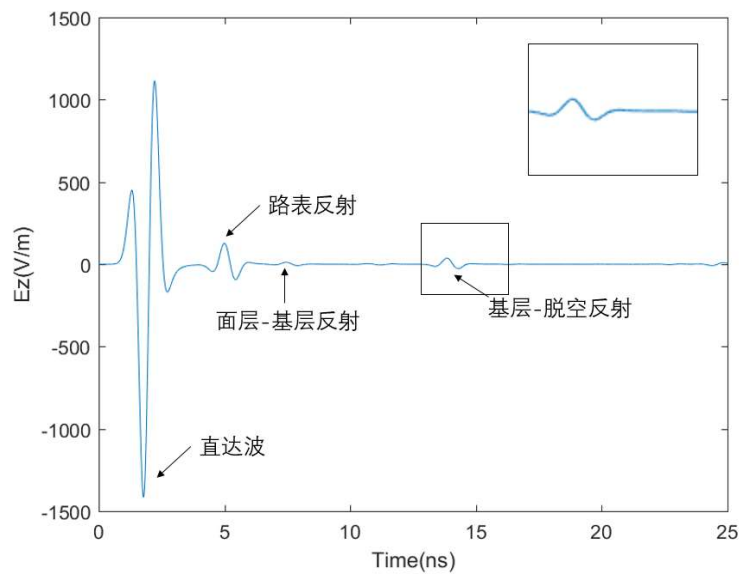

图 4.22 A 扫描图像（0.3m 注浆修复脱空，800MHz 天线）

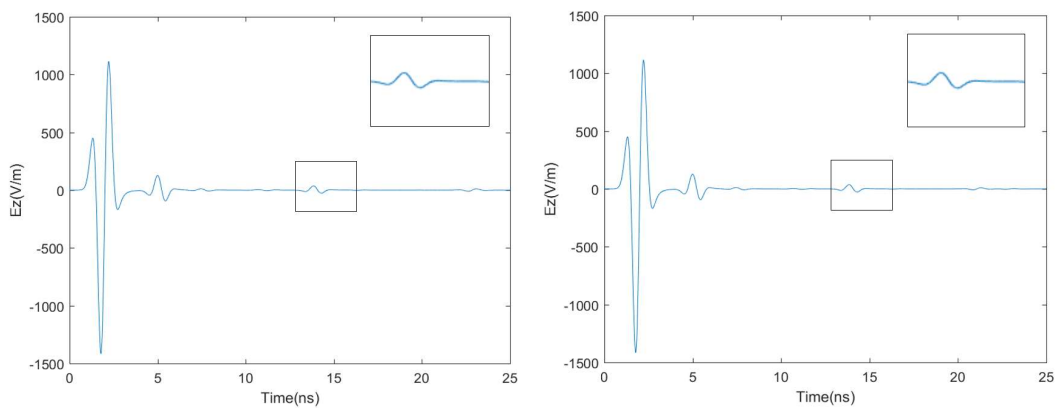

图 4.23 A 扫描图像（0.25m 注浆修复脱空，800MHz 天线）

图 4.24 A 扫描图像（0.2m 注浆修复脱空，800MHz 天线）

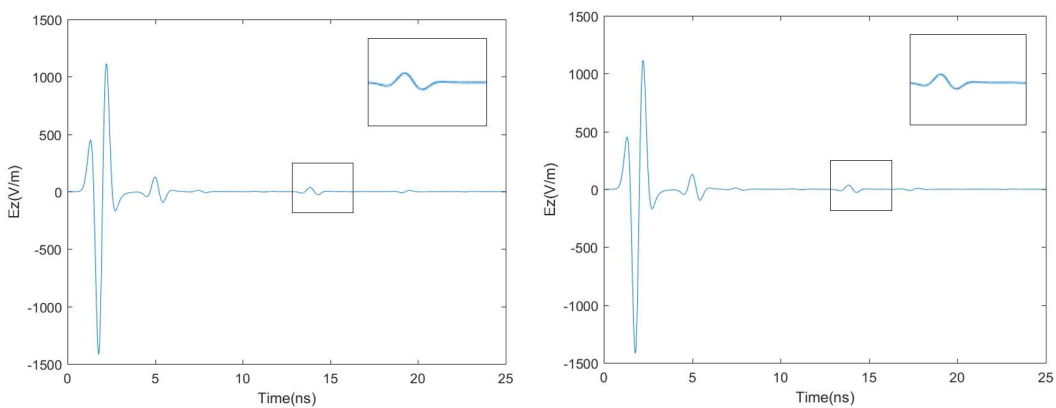

图 4.25 A 扫描图像（0.15m 注浆修复脱空，800MHz 天线）

图 4.26 A 扫描图像（0.1m 注浆修复脱空，800MHz 天线）

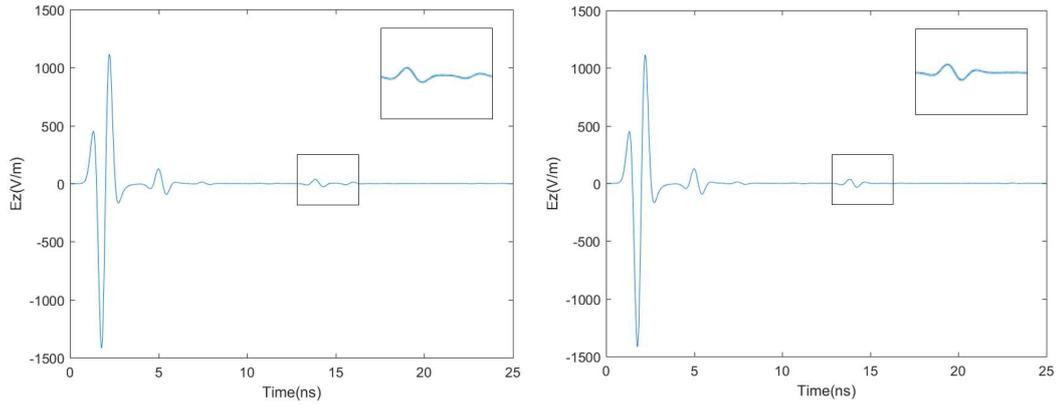

图 4.27 A 扫描图像（0.05m 注浆修复脱空，800MHz 天线）

图 4.28 A 扫描图像（0.01m 注浆修复脱空，800MHz 天线）

由以上一系列图可知，类似于充气脱空，注浆修复脱空的下界面的反射波也可见，而且波幅不随脱空高度变化，这说明电磁波在理想的注浆材料内部传播不发生损耗。然而，注浆脱空-土基界面的反射波幅度很小，尤其是在有噪声存在的情况下很难被识别，因此在注浆修复脱空的识别过程中基层-注浆脱空界面的反射波更为重要。

### 4.2.2 最小二乘系统辨识

系统辨识是指确定系统的数学模型及参数的一类问题，很多工程领域都需要用到辨识技术。很多系统的机理未知或部分未知，难以用理论分析法建立数学模型；但是，大多数系统的输入值、输出值可以测量，利于系统输入值、输出值提供的信息建立系统的数学模型的方法，就是系统辨识技术。本节研究最小二乘系统辨识法在充气脱空的识别及高度估算中的应用。

对于一个各层尺寸确定、材料确定的道路结构来说，其探地雷达回波应当是确定的（除去随机噪声部分）。而探地雷达回波的单条信号，可以被看作是因变量 $Ez$ 关于时间 $t$ 的模型或函数，其图像即为探地雷达 A 扫描。从这个角度看，脱空尺寸的计算实际上是一个系统辨别问题：即根据系统的输入数据（时间）、输出数据（z 方向电场强度 $Ez$），求解待定参数（脱空层高度）。因此，可以采用系统辨识理论中的方法，解决基于探地雷达的道路脱空高度估算问题。

系统辨识有三个要素，即数据、模型类和准则。数据是指建立模型所基于的系统输入值和输出值，在本研究中为探地雷达在道路上的某条实测回波信号的幅值（输出值）和时间（输入值）。模型类是指根据已知条件确定的模型的辨识范围，在本研究中道路结构确定，道路各层材料确定，脱空高度不确定，则模型类就是不同脱空高度下已知的结构和材料的道路的反射信号的集合。准则是用于选

择最优模型的标准，常被表示为误差的泛函数，本研究采用输出误差准则，把各采样点的实测波幅与模型输出值之间的误差的平方和作为选择最优模型的标准。

系统辨识主要包括以下步骤[67]：

（1）辨识目的。决定模型的类型、精度要求和辨识方法。这里的模型类型是各脱空高度的道路结构的雷达回波所代表的函数，该函数无法用公式表达，通过雷达波正演模拟得到各模型，而且该函数是一个非连续函数（因为探地雷达用等间隔的采样点记录数字信号）。辨识方法采用经典的最小二乘法。

（2）实验设计；输入、输出数据的采集、处理。本研究中即为探地雷达回波信号的采集过程。

（3）模型结构选取和辨识。

（4）模型参数辨识。在本研究中为找出误差平方和最小的模型，得到其脱空高度。

（5）模型检验。检验选取的模型是否恰当地表示了被辨识的系统。

接下来将参考以上的流程研究脱空高度参数的辨识方法。

前面已经仿真生成了 800MHz 天线在存在 0.01m-0.3m 高度脱空的路面上的 A 扫描信号。0.01m 到 0.3m 共 30 种脱空高度可以看做是脱空高度这一未知参数的 30 种取值，相应地，这 30 条信号可以看做是在各脱空高度参数取值下的信号幅值关于时间的函数。如果给定一个该结构道路上实测的信号，求基层脱空高度，则可以比较该信号与上述 30 条信号之间的误差平方和，与实测信号误差平方和最小的已知信号的脱空高度值就被认为是实测信号对应的脱空高度值。

本研究通过给仿真信号加高斯白噪声来模拟实测信号。由于探地雷达图像处理中通常去除直达波，因此以下研究中的信号都经过去直达波处理。

首先，任取 3 个脱空高度值 0.27m、0.17m、0.02m，在相应的三个回波信号上分别添加不同信噪比的高斯白噪声，然后，求模拟的实测信号与已知的信号之间的误差平方和，找出误差平方和最小值从而估计脱空高度。

（1）0.27m 脱空高度

图 4.29、图 4.30 为存在 0.27m 高度脱空的道路结构用 800MHz 天线扫描得到的 A 扫描图像，信噪比分别为 10、5。红线为仿真得到的回波信号，蓝线为在仿真回波信号基础上加噪声得到的回波信号。

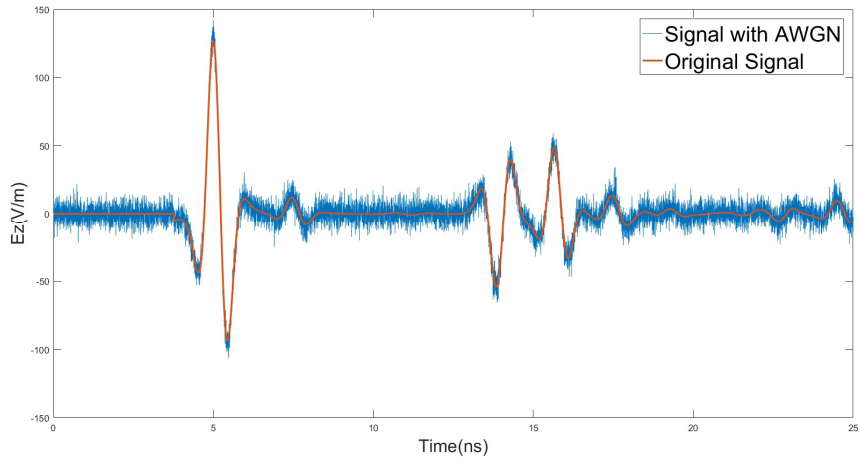

图 4.29 模拟实测 A 扫描（0.27m 充气脱空，800MHz 天线，信噪比 10）

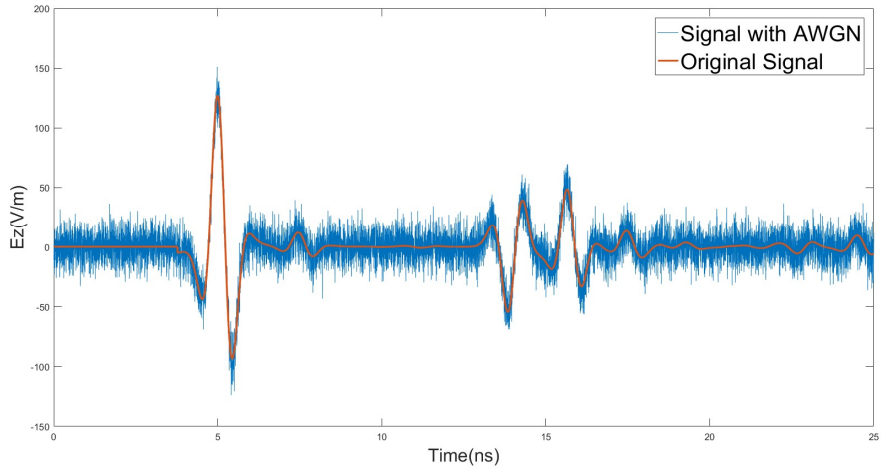

图 4.30 模拟实测 A 扫描（0.27m 充气脱空，800MHz 天线，信噪比 5）

之后，计算模拟实测信号与之前仿真得到的 30 种脱空高度的道路结构的回波信号之间的差值，把各采样点的信号差值求平方和，得到 30 种脱空高度一一对应的 30 个差值平方和，并作图（图 4.31 和图 4.32）。

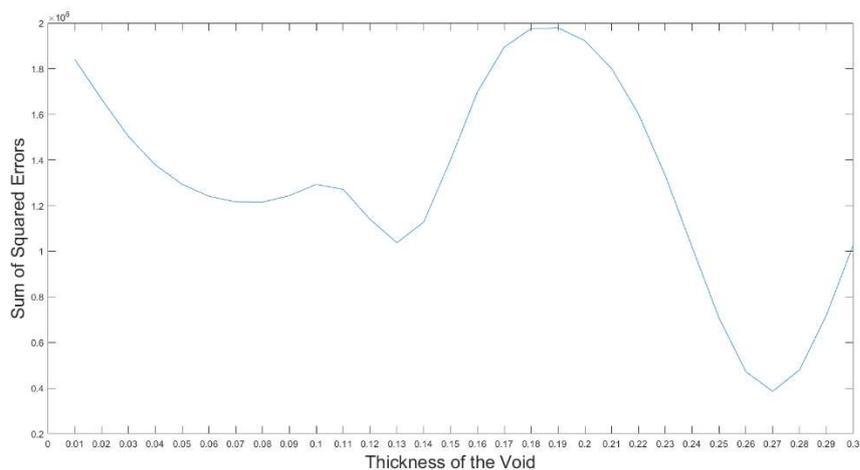

图 4.31 各脱空高度道路信号与待测信号的差值平方和（待测 0.27m 充气脱空，
800MHz 天线，信噪比 10）

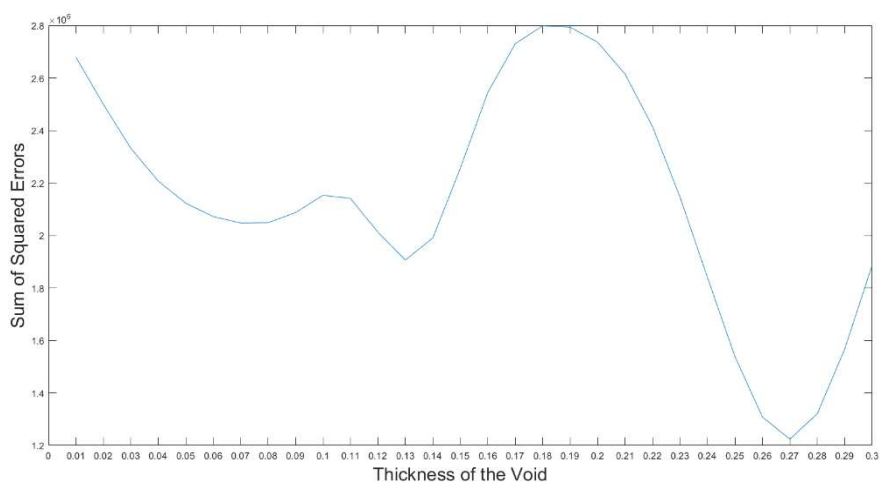

图 4.32 各脱空高度道路信号与待测信号的差值平方和（待测 0.27m 充气脱空，
800MHz 天线，信噪比 5）

由图可看出，不论是信噪比是 10 还是 5，已知的仿真信号与待测信号的差值平方和均在脱空高度为 0.27m 处取到最小值，可以判断待测信号的检测道路的脱空高度为 0.27m，结果准确。

（2）0.17m 脱空高度

类似地，给 0.17m 的仿真信号添加噪声（信噪比为 10、3），并求与各脱空高度信号差值的平方和，如图 4.33-图 4.36。

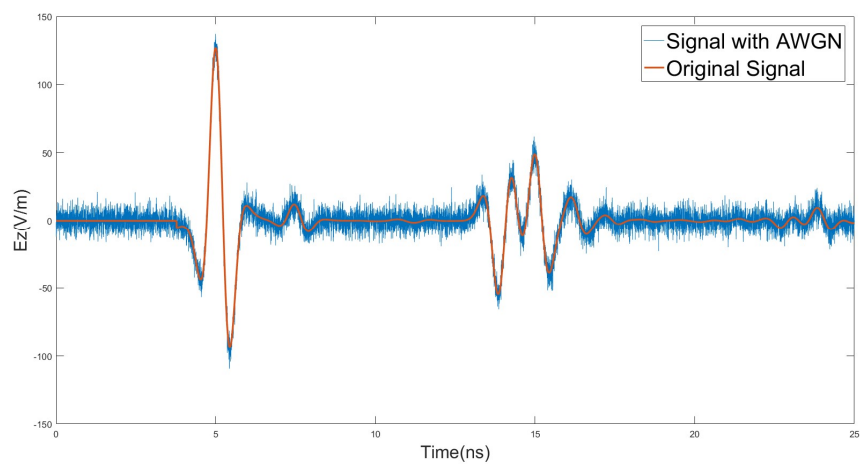

图 4.33 模拟实测 A 扫描（0.17m 充气脱空，800MHz 天线，信噪比 10）

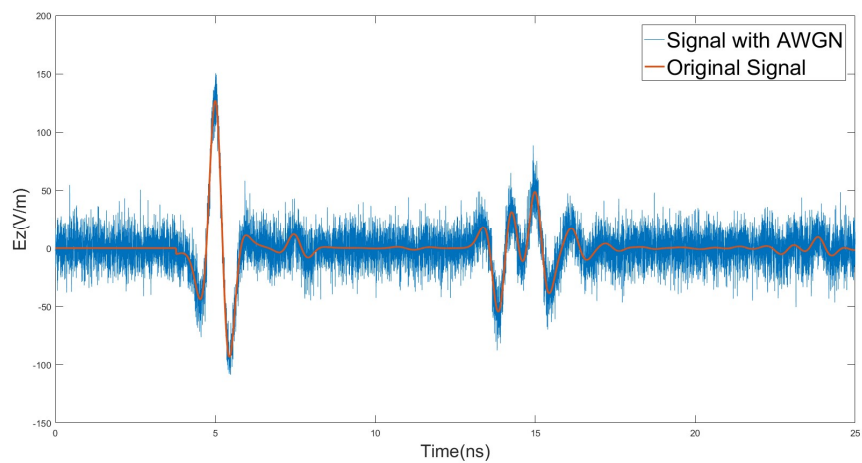

图 4.34 模拟实测 A 扫描（0.17m 充气脱空，800MHz 天线，信噪比 3）

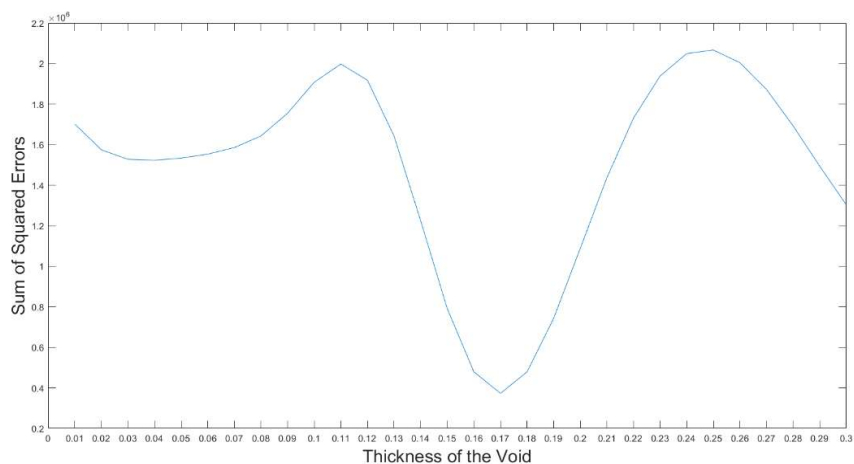

图 4.35 各脱空高度道路信号与待测信号的差值平方和（待测 0.17m 充气脱空，
800MHz 天线，信噪比 10）

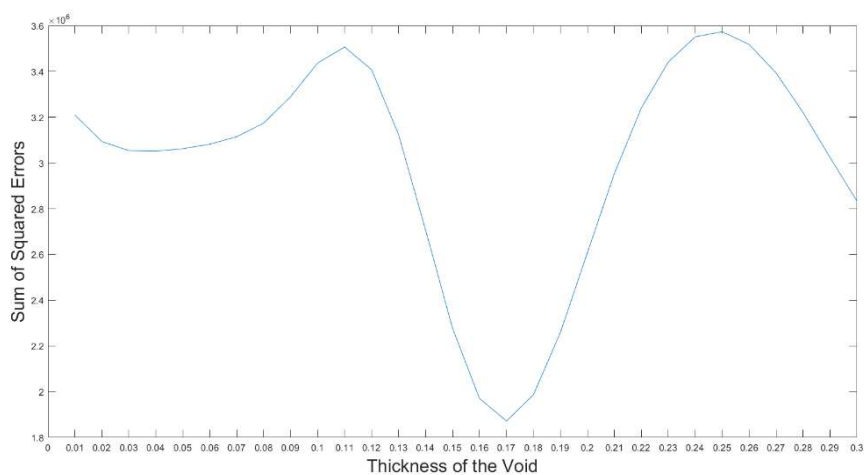

图 4.36 各脱空高度道路信号与待测信号的差值平方和（待测 0.17m 充气脱空，
800MHz 天线，信噪比 3）

由图可知，对于 0.17m 高度脱空的路面，最小二乘辨识法研究可以准确识别脱空高度。即使信噪比降低至 3，也不影响识别效果。

（3）0.02m 脱空高度

类似地，给 0.02m 的仿真信号添加噪声（信噪比为 10、2），并求与各脱空高度信号差值的平方和，如图 4.37-图 4.40。

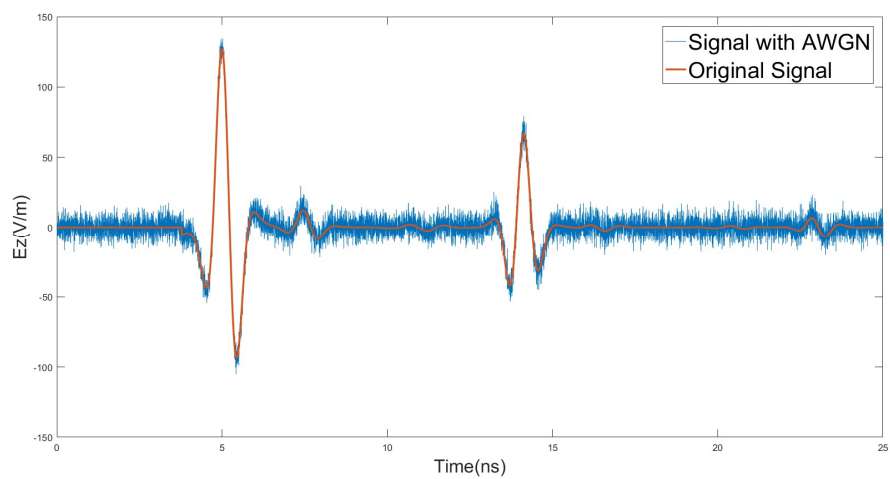

图 4.37 模拟实测 A 扫描（0.02m 充气脱空，800MHz 天线，信噪比 10）

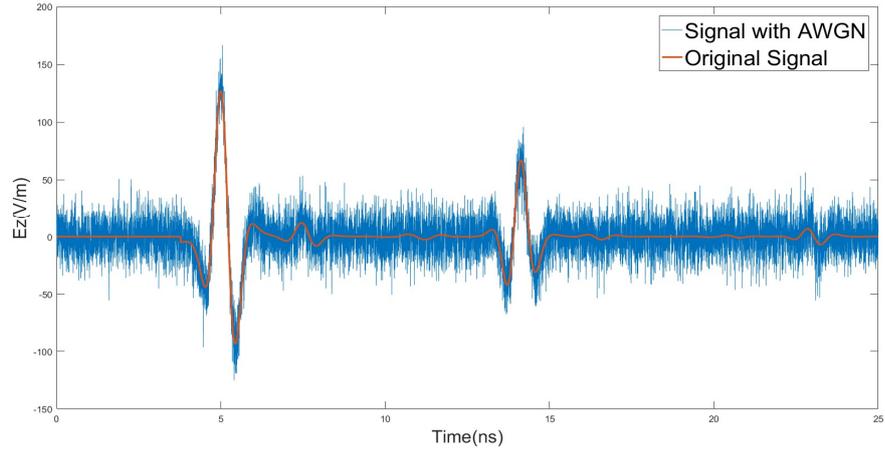

图 4.38 模拟实测 A 扫描（0.02m 充气脱空，800MHz 天线，信噪比 2）

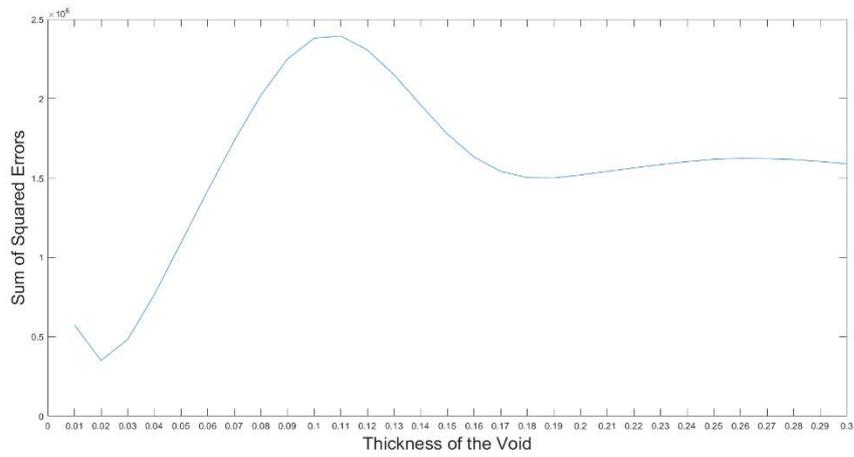

图 4.39 各脱空高度道路信号与待测信号的差值平方和（待测 0.02m 充气脱空，
800MHz 天线，信噪比 10）

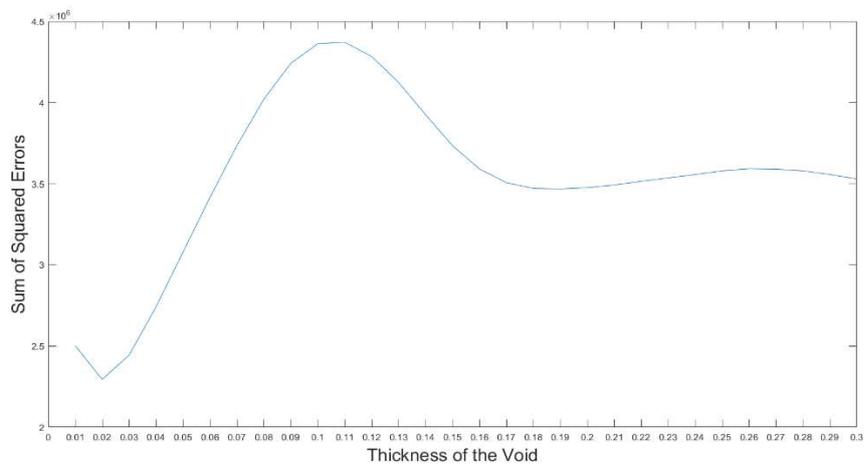

图 4.40 各脱空高度道路信号与待测信号的差值平方和（待测 0.02m 充气脱空，
800MHz 天线，信噪比 2）

由图可知，对于 0.02m 高度脱空的路面，最小二乘辨识法研究可以准确识别脱空高度。当信噪比降低至 2 时，由于噪声的影响，差值平方和的绝对值增大，但随脱空高度的变化趋势不变，依旧可以根据差值平方和最小值识别出脱空高度。

以上讨论了 800MHz 天线的情况，结果表明，对于三个阶段的脱空高度，信噪比从 10 降至 2，最小二乘系统辨识法均可正确识别脱空高度。

接下来讨论 1200MHz 天线的情况。真实道路结构的脱空高度可能介于事先仿真的脱空高度，比如真实脱空高度是 0.023m，而事先只仿真了 0.02m 和 0.03m 高度脱空的信号。接下来将选取 0.193m、0.128m、0.023m 三个脱空高度值，信噪比均设定为 2，检验最小二乘辨识法能否准确判断其近似值。作图 4.41-4.46。

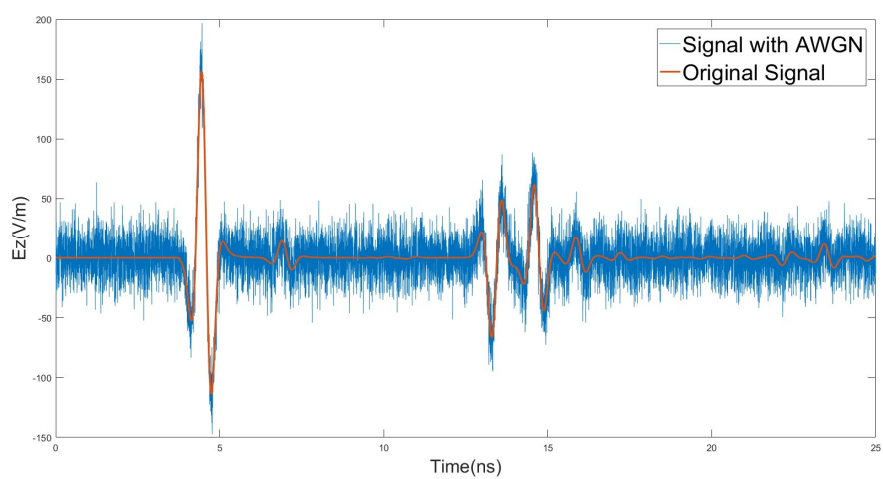

图 4.41 模拟实测 A 扫描（0.193m 充气脱空，1200MHz 天线，信噪比 2）

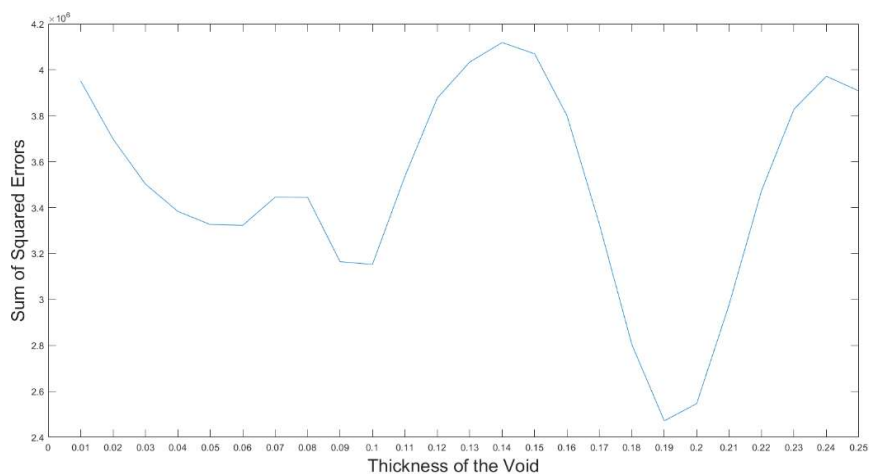

图 4.42 各脱空高度道路信号与待测信号的差值平方和（待测 0.193m 充气脱空，1200MHz 天线，信噪比 2）

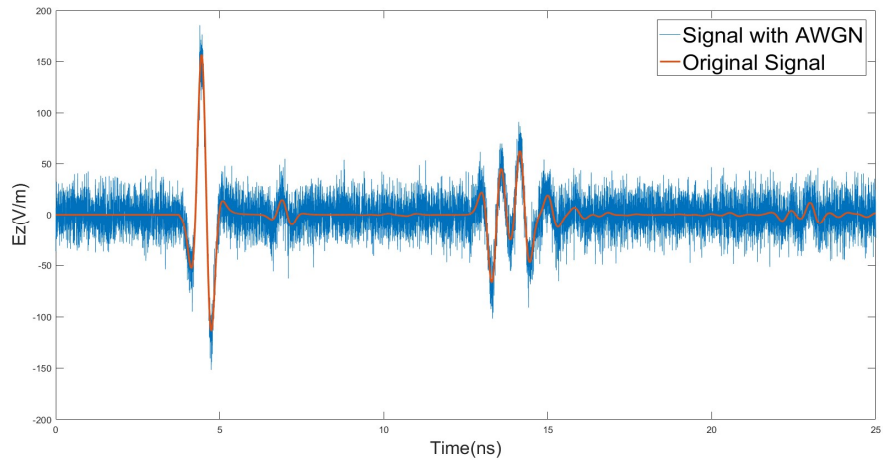

图 4.43 模拟实测 A 扫描（0.128m 充气脱空，1200MHz 天线，信噪比 2）

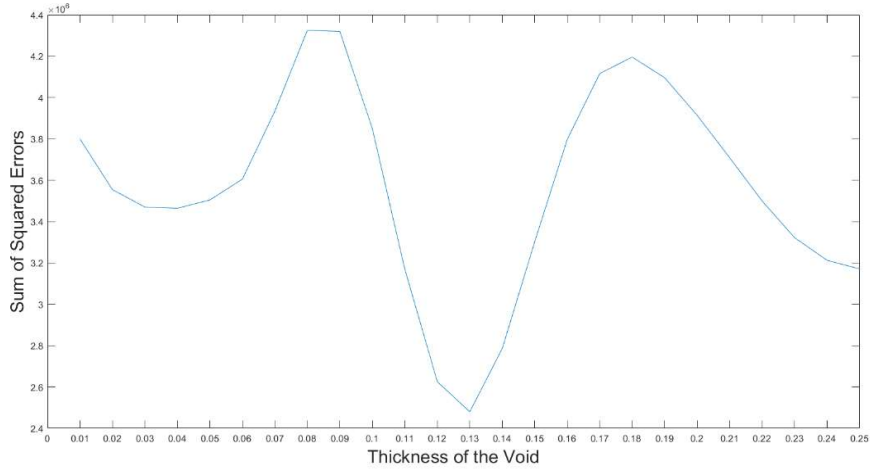

图 4.44 各脱空高度道路信号与待测信号的差值平方和（待测 0.128m 充气脱空，
1200MHz 天线，信噪比 2）

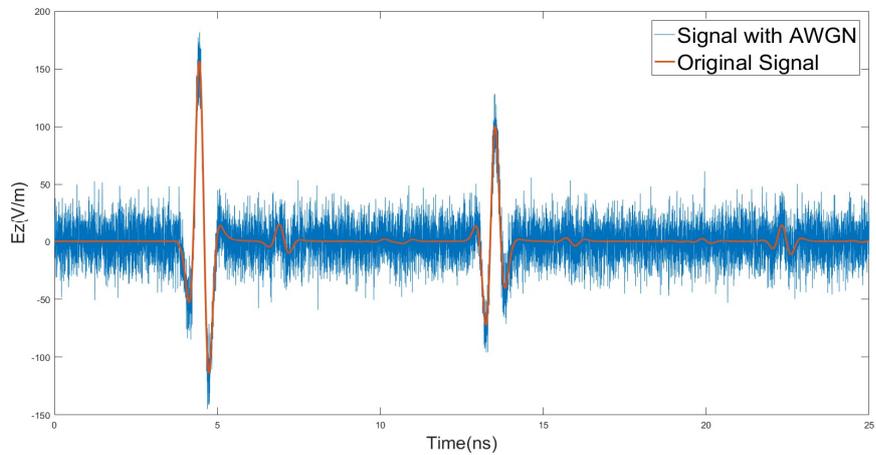

图 4.45 模拟实测 A 扫描（0.023m 充气脱空，1200MHz 天线，信噪比 2）

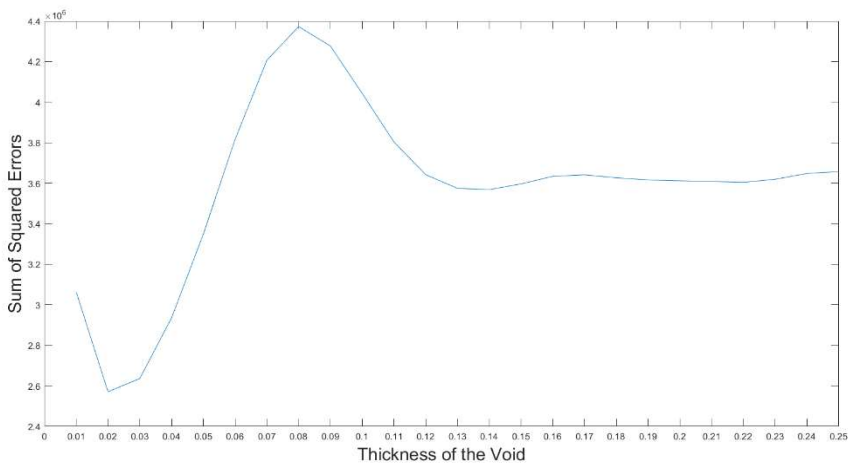

图 4.46 各脱空高度道路信号与待测信号的差值平方和（待测 0.023m 充气脱空，
1200MHz 天线，信噪比 2）

由图 4.41-图 4.46 可知，最小二乘辨识法对于 1200MHz 天线、信噪比为 2 的情况仍然适用。对于真实道路结构的脱空高度介于事先仿真的脱空高度的情况，最小二乘辨识法可以识别出最接近的近似值。在以上三个例子中，0.193m、0.128m、0.023m 三个脱空高度值由最小二乘辨识法分别判断为 0.19m、0.13m、0.02m，满足要求。

综上，最小二乘系统辨识法具有很好的辨识效果。但是，在上述辨识过程中，实际上有若干前提条件。由于系统辨识的标准是待测系统与模型类中各模型的相似性，所以这种方法在信号识别中就要求各反射子波在时间轴上的位置固定，也即要求实际道路结构层厚度均匀，道路内部反射面平整，否则无法保证辨别精度。而且，实际道路可能会存在其他病害，产生异常反射，也会影响辨别精度。此外，模型类里的信号的差异只在于脱空高度、填充情况，道路结构、材料完全一致，这一点在旧路中往往难以保证。

为保证反射子波在时间轴上位置一致，并排除其他异常回波的影响，可以将疑似脱空的信号段截出，与仿真的各种脱空高度的信号的相应位置的信号片段做最小二乘系统辨识。但这种方法耗时长。因此，最小二乘系统辨识的方法可以用于充气脱空的精确分析；而对于充水脱空和注浆修复脱空，由于脱空下方界面的反射波微弱，不适宜使用最小二乘系统辨识法，下文使用介电常数法加以判别。

### 4.2.3 解卷积

充气脱空检测的一个主要难点在于，如果选用较高频率的天线，虽然可以提高精度减少反射子波的重叠，但探测深度会减小；如果选用较低频率的天线，探测深度可以提高，但脉冲波波长增加，脱空上下两界面的反射子波容易重叠。因

此,如果能够在保证探测深度的同时,通过信号处理算法把脉冲波的波长人为"缩短",把重叠的反射子波分离,就可以提高脱空高度的识别精度。

探地雷达的发射波可以看作是一个在时间轴上的函数(系统的冲激响应);另一方面,道路结构内部各界面对于雷达信号的反射幅值与发射波幅值的比值在时间轴上的分布,可以看做是道路系统的反射函数(冲激输入序列),由各道路结构层的厚度和相对介电常数决定。道路结构反射信号的表达式为式(2.33),可以看出,接收天线采集的反射信号是若干个经延迟、衰减的发射信号的叠加,也就是冲激响应(即发射信号)与输入信号(即道路结构的反射函数)的卷积的结果。因此,由采集到的反射信号(系统的输入)求道路结构反射函数,可以使用反卷积的方法。经过反卷积,可以缩短反射信号中的子波波长,使原本重叠的一些反射子波分离开。

为进一步说明反射子波重叠的影响,仿真一个道路结构上的反射波。该道路有面层、基层、土基三层,土基顶部基层下方有注浆修复脱空,各界面反射系数如图 4.47 所示。图 4.48 为道路结构的反射函数(即反射界面对应的冲激序列)和各反射子波(冲激响应),需要注意的是,这里反射函数在绘制过程中向右平移了 2ns,从而与各反射子波的最大幅值位置重叠;数值仿真得到的反射波如图 4.49 所示,该反射波可以看做是系统输出。图 4.48 中的各衰减、时延的冲激响应叠加即得到图 4.49 中的系统响应。

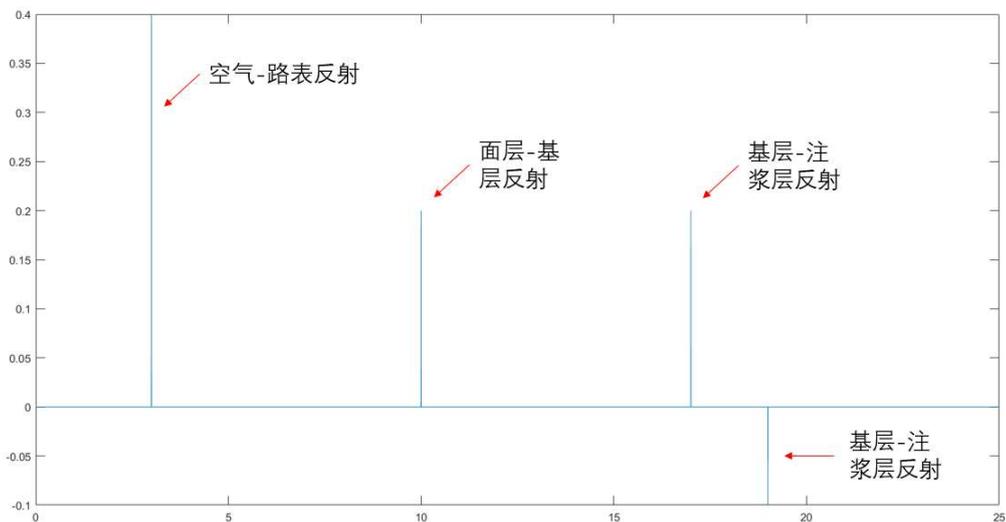

图 4.47 各界面反射系数(冲激序列)

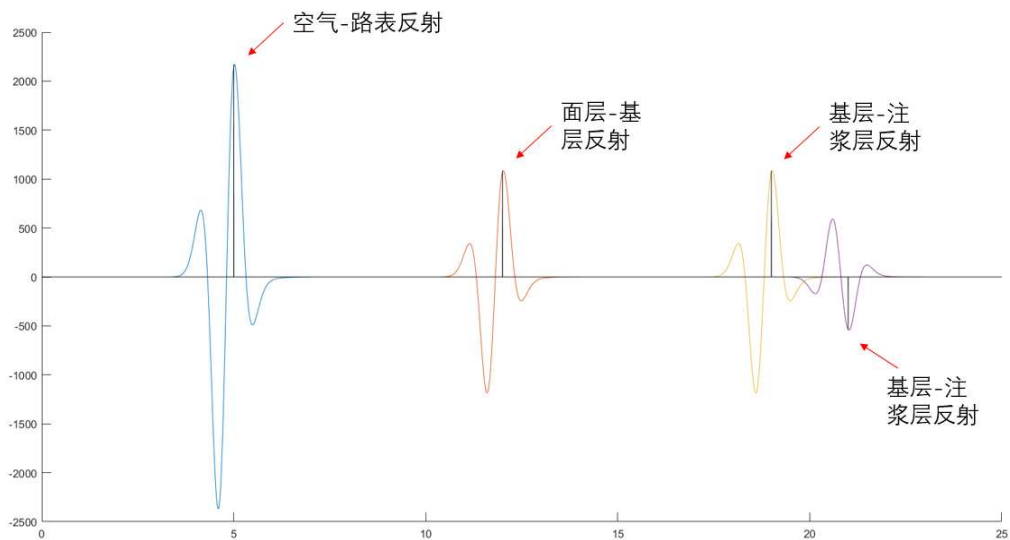

图 4.48 道路结构反射函数（冲激序列）和各反射子波（冲激响应）

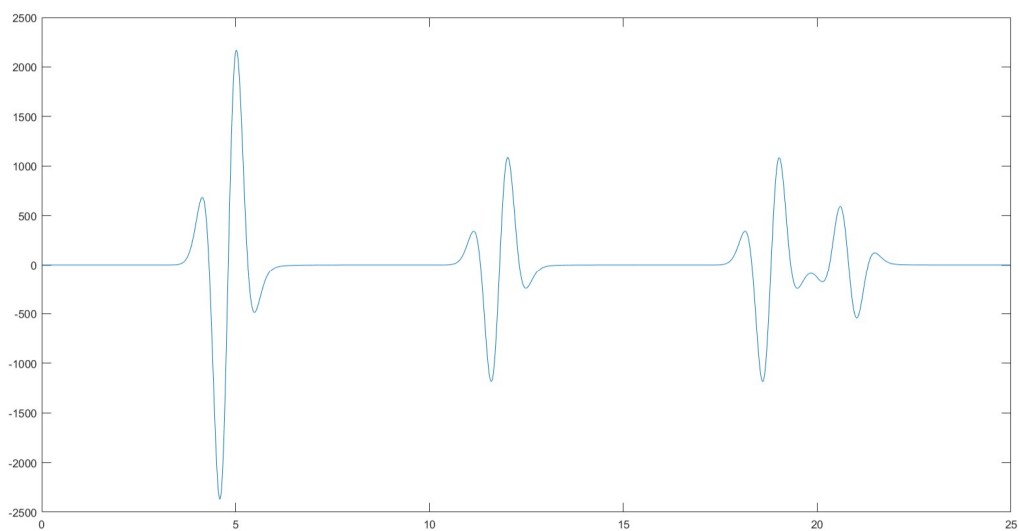

图 4.49 存在子波重叠的反射波（系统输出）

接下来讨论解卷积的实现方法。记系统的冲激响应为 $h(t)$，冲激序列（系统输入）为 $x(t)$，系统输出为 $y(t)$。由于探地雷达数据是离散的采样点，所以可设冲激序列、冲激响应和系统输出都是离散函数。则在 0 时刻的输出为

$$y(0) = h(0)x(0) \tag{4.1}$$

在 1 时刻的输出为当前输入的影响 $h(0)x(1)$ 加上 0 时刻的输入的响应持续到 1 时刻的影响 $h(1)x(0)$：

$$y(1) = h(1)x(0) + h(0)x(1) \tag{4.2}$$

类似地，得到各时刻系统输出的计算式[68]：

$$y(2) = h(2)x(0) + h(1)x(1) + h(0)x(2)$$
$$y(3) = h(3)x(0) + h(2)x(1) + h(1)x(2) + h(0)x(3)$$
$$\vdots \qquad\qquad\qquad\qquad\qquad\qquad (4.3)$$
$$y(n) = h(n)x(0) + h(n-1)x(1) + \ldots + h(0)x(n)$$

为能够在 MATLAB 中计算，将以上各式写成矩阵相乘的形式：

$$\begin{pmatrix} y(0) \\ y(1) \\ y(2) \\ \vdots \\ y(n) \end{pmatrix} = \begin{pmatrix} h(0) & 0 & 0 & \ldots & 0 \\ h(1) & h(0) & 0 & \ldots & 0 \\ h(2) & h(1) & h(0) & \ldots & 0 \\ \vdots & \vdots & \vdots & \vdots & \vdots \\ h(n) & h(n-1) & h(n-2) & \ldots & h(0) \end{pmatrix} * \begin{pmatrix} x(0) \\ x(1) \\ x(2) \\ \vdots \\ x(n) \end{pmatrix} \qquad (4.4)$$

可记作

$$\mathbf{y = Eh} \qquad\qquad (4.5)$$

其中，$\mathbf{y}$ 为接收天线采集的反射信号序列，$\mathbf{h}$ 为道路结构反射函数，$\mathbf{E}$ 为基于冲激响应（即发射波）$\mathbf{x}$ 建立的一个托普利兹矩阵（Toeplitz Matrix），矩阵主对角线上的元素相等，平行于主对角线上的线上的元素也相等。

在 $\mathbf{y}$ 和 $\mathbf{E}$ 已知的情况下，为了求解 $\mathbf{h}$，可以将 (4.5) 式变形，得

$$\mathbf{h = E^{-1}y} \qquad\qquad (4.6)$$

这样，就建立了道路结构反射函数的解卷积计算方法。根据 4.2.1 中的结果，800MHz 天线用于检测充气脱空道路，如果脱空高度小于 0.2m，脱空上下界面的反射子波就会重叠。接下来使用解卷积的方法，分离重叠的反射子波，处理结果如图 4.50-图 4.54 所示。

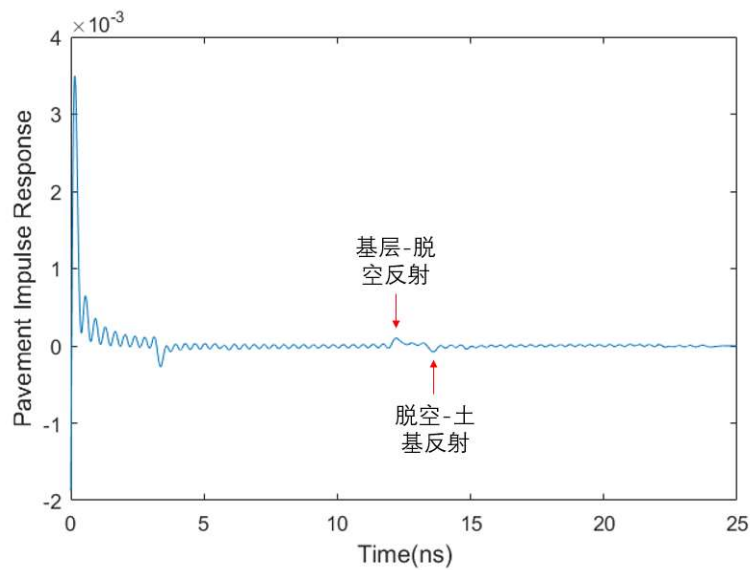

图 4.50 解卷积后的 A 扫描（0.2m 充气脱空，800MHz 天线）

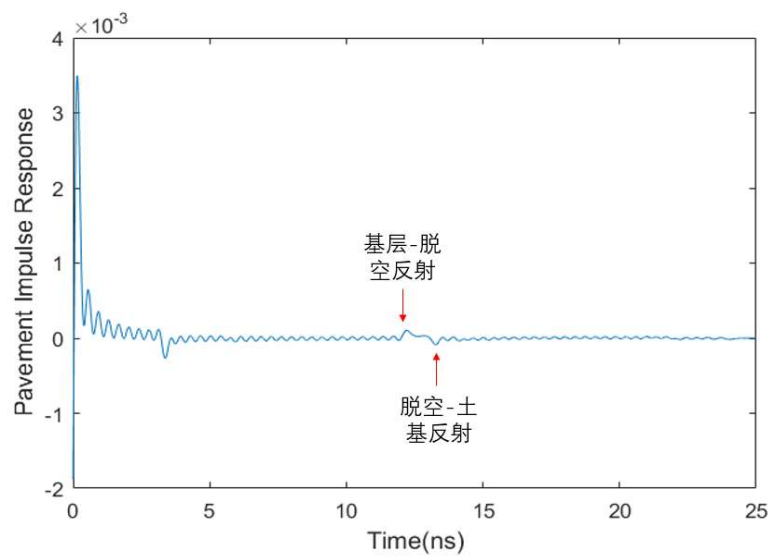

图 4.51 解卷积后的 A 扫描（0.15m 充气脱空，800MHz 天线）

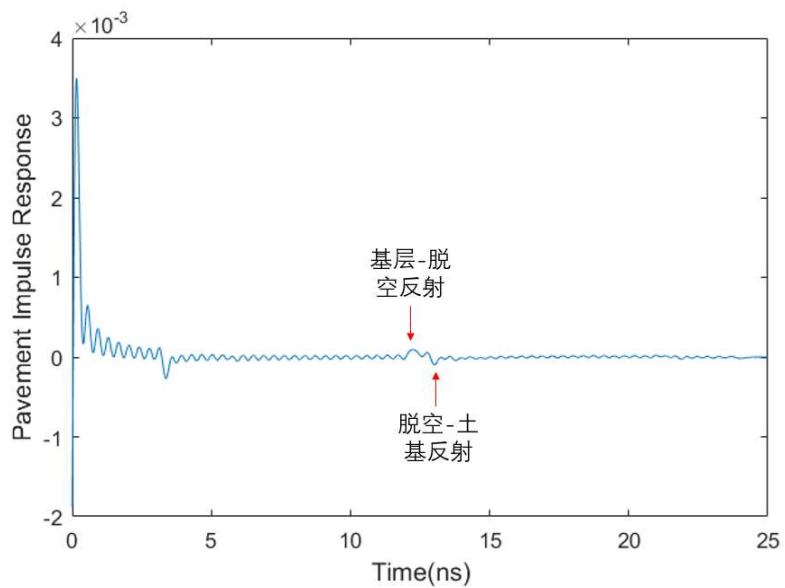

图 4.52 解卷积后的 A 扫描（0.12m 充气脱空，800MHz 天线）

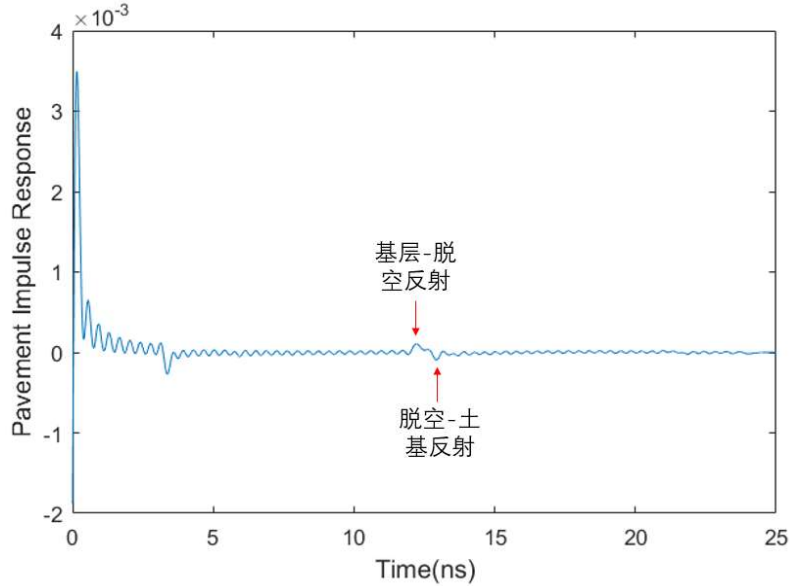

图 4.53 解卷积后的 A 扫描（0.1m 充气脱空，800MHz 天线）

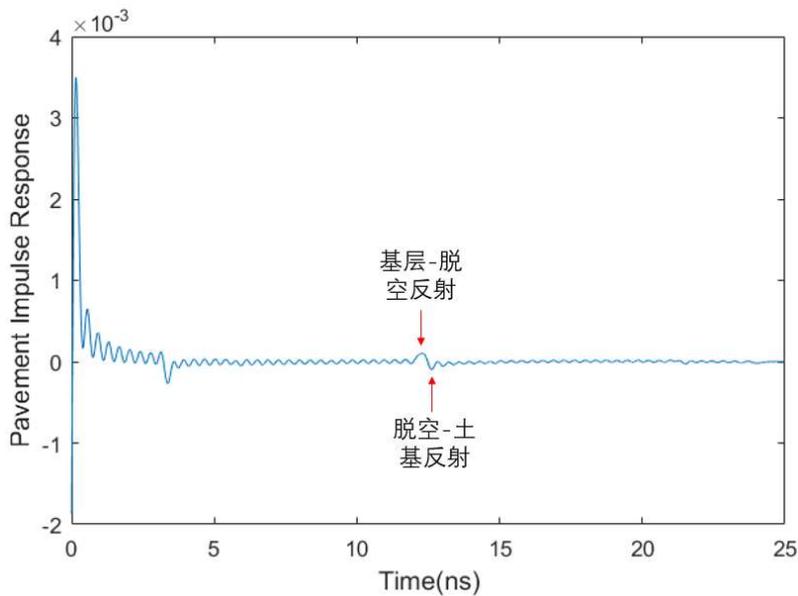

图 4.54 解卷积后的 A 扫描（0.06m 充气脱空，800MHz 天线）

由图 4.50-图 4.54 可知，解卷积对于一定范围内的反射波叠加有较好的分离效果。解卷积前，高度小于 0.2m 的脱空的上下界面反射发生重叠；解卷积后，高度 0.1m 到 0.2m 之间的脱空的上下界面反射分离，能够从图像上估测脱空高度。但是，解卷积后的子波仍然有一定宽度，当脱空高度小于 0.1m 时，上下界面的反射还是会重叠。

解卷积的一个特点是易受噪声的干扰，原始数据上很小的误差会造成解的很大改变，这在数学上被称为不适定问题[69]。图 4.55 为 0.12m 充气脱空 800MHz 天线 A 扫描加高斯白噪声后的图像，在这组加噪后的数据的基础上解卷积，得

到图 4.56，可以看出，由于噪声的影响，解卷积后的结果完全无法解读。

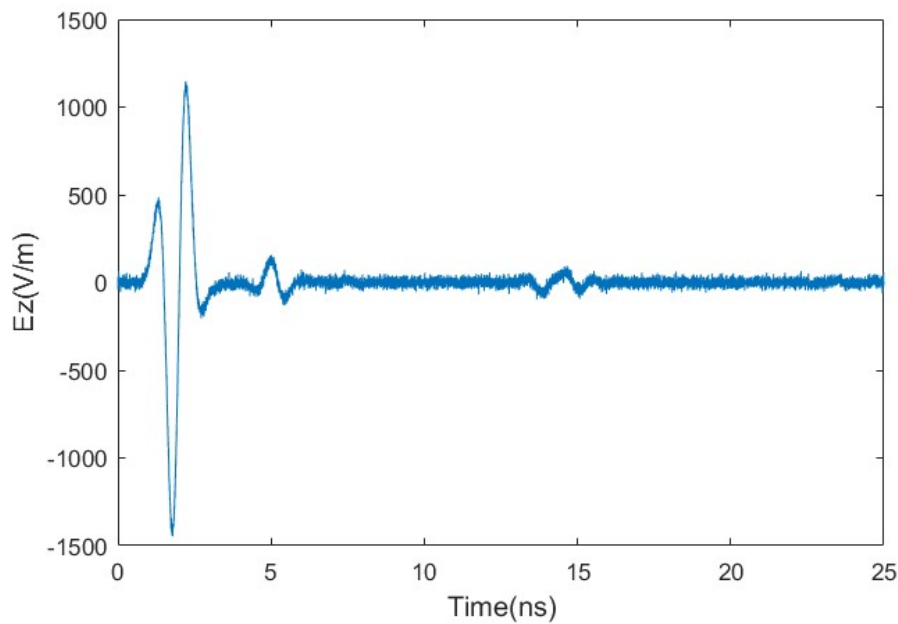

图 4.55 加噪后的 A 扫描（0.12m 充气脱空，800MHz 天线，信噪比 20）

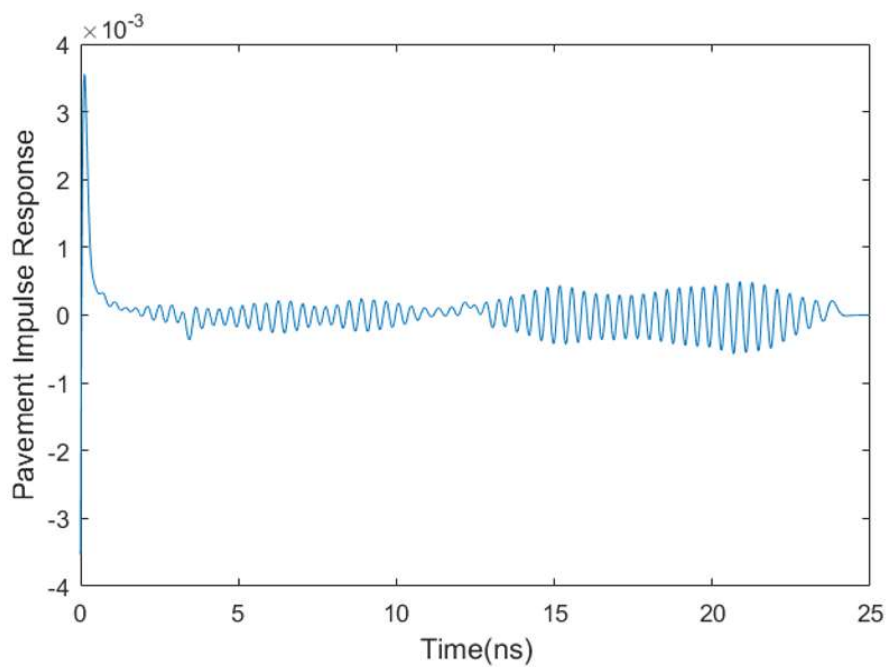

图 4.56 解卷积后的 A 扫描（0.12m 充气脱空，800MHz 天线，信噪比 20）

要解决这个问题，可以采用基于变分原理的吉洪诺夫正则化（Tikhonov Regularization）的方法，引入正则算子将不适定问题（ill posed problem）转化为适定问题（well posed problem）[70]。伊利诺伊大学的 Shan Zhao 等人[66]在 GPR 用于路面加铺层厚度测算中即使用了这种方法。

吉洪诺夫正则化后的解卷积公式变为

$$\mathbf{h} = \left(\mathbf{E}^T\mathbf{E} + \alpha\mathbf{I}\right)^{-1}\mathbf{E}^T\mathbf{y} \tag{4.7}$$

式中，$\mathbf{I}$ 为单位矩阵，$\alpha$ 为正则化参数，其取值应当与噪声水平相匹配。正则化参数的选取有偏差原理、误差极小化原理、L 曲线准则、广义交叉校验准则等方法。这里正则化参数取 $5\times10^6$。

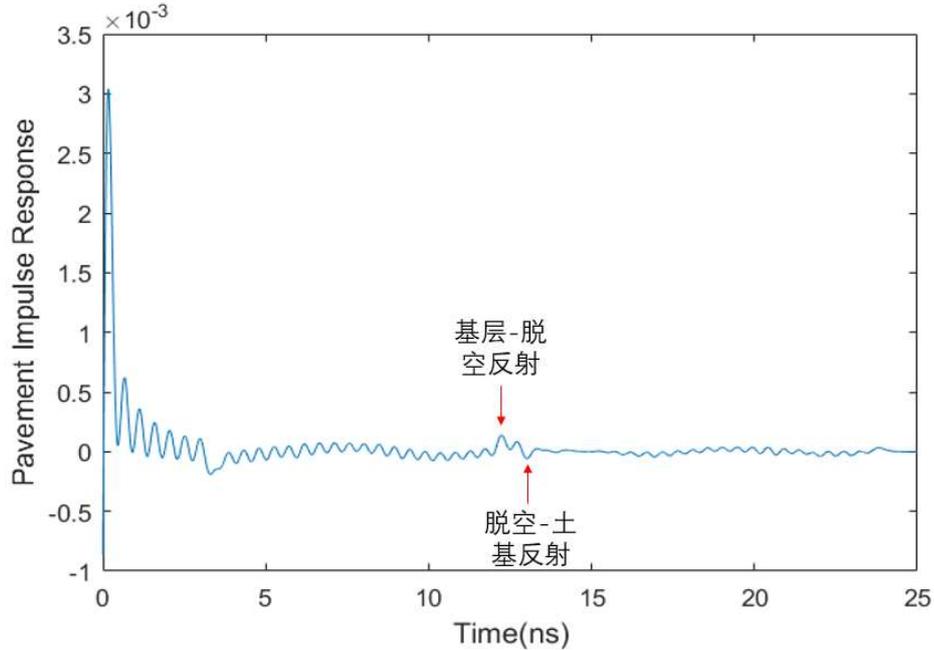

图 4.57 正则化解卷积后的 A 扫描（0.12m 充气脱空，800MHz 天线，信噪比 20）

经过正则化解卷积处理的 A 扫描如图 4.57 所示，可明确识别基层-脱空界面的正反射波幅和脱空-土基界面的负反射波幅。可以看出，正则化处理大大降低了原信号中的噪声对解卷积结果的影响。

综上，解卷积法可以较好地消除充气脱空反射子波重叠的影响，而解卷积的不适定问题可以用吉洪诺夫正则化方法解决。结果表明，解卷积的方法能够把 0.1m-0.2m 脱空高度原本重叠的信号分离，但对于小于 0.1m 的脱空高度，这种方法无法分离重叠的反射波。这种方法的优势在于对原始图像的要求低于最小二乘系统辨识法，可用于大范围的充气脱空识别及尺寸估测。

### 4.2.4 介电常数法

上文讨论了充气脱空识别和高度估算的两种方法，即系统辨识法和解卷积法，但这两种方法不适用于充水脱空和注浆修复脱空。电磁波在水中迅速衰减，充水脱空底部与路基的界面的反射很难识别；由于一般情况下注浆材料与路基土介电常数差别不大，而且注浆过程中注浆材料容易和路基表面土层混合，使两种材料

的介电常数逐渐过渡，所以注浆修复脱空底部的反射很弱，容易被噪声掩盖。

但是，如果不考虑脱空厚度的估测，而只考虑脱空的识别，则可以根据计算介电常数来判断充水脱空和注浆修复脱空的存在。水的介电常数为 81，远大于各种道路修筑材料；注浆材料虽然与道路修筑材料的介电常数差异不大，但如果注浆前已经判断出充气脱空的位置，则可以把相同位置的注浆前后的检测结果作对比，如果注浆材料填充了脱空位置，则相应位置的负反射波幅消失或变为正反射波幅。

在 2.6 章节中介绍了介电常数法，用此方法计算充水脱空的介电常数，模型中材料的电磁学特性参数按表 3.2 选取，脱空高度从 0.01m 按 0.01m 的间隔增加到 0.3m。部分结果如表 4.1 所示。

表 4.1 充水脱空道路结构介电常数估算值

| 结构层 脱空高度 | 面层 | 基层 | 充水脱空 |
|---|---|---|---|
| 0.01m | 5.9 | 7.1 | 22.1 |
| 0.02m | 5.9 | 7.1 | 20.3 |
| 0.03m | 5.9 | 7.1 | 20.3 |
| 0.05m | 5.9 | 7.1 | 20.3 |
| 0.15m | 5.9 | 7.1 | 20.3 |
| 0.3m | 5.9 | 7.1 | 20.3 |

面层的介电常数为 6，估算值为 5.9；基层的介电常数为 7.5，估算值为 7.1，这两层的估算值比较准确。但水的介电常数为 81，而估测值为 20-22，误差很大，可能是由于电磁损耗导致。

从表 4.1 中数据可以看出，如果脱空高度在 0.02m-0.3m 范围内，充水脱空的估测值为定值；如果脱空高度为 0.01m，充水脱空的估测值略微减小。这说明充水脱空与充气脱空不同，前者上下界面波形叠加效应不明显。其原因有二。一方面，电磁波在水中的传播速度是空气中传播速度的 1/9，波速较小使得充水脱空在更小的高度下，充水脱空的上下界面反射波不重叠。另一方面，由于电磁波在水中的衰减较快，所以充水脱空-土基分界面反射很微弱，当脱空高度不小于 0.02m 时不会对基层-充水脱空界面反射波形产生影响。

为研究电磁损耗对介电常数估算值的影响，将表 3.2 中道路面层、基层的电导率改为 0，使模型中的路面结构成为无电磁耗损介质。重新生成电磁传播模型，并用反射波幅计算道路材料的介电常数，结果如表 4.2 所示。

表 4.2 充水脱空道路结构介电常数估算值（无电磁损耗）

| 结构层<br>脱空高度 | 面层 | 基层 | 充水脱空 |
|---|---|---|---|
| 0.01m | 5.9 | 7.2 | 66.9 |
| 0.02m | 5.9 | 7.2 | 50.2 |
| 0.03m | 5.9 | 7.1 | 50.2 |
| 0.3m | 5.9 | 7.1 | 50.2 |

由表 4.2 中数据可看出，在设定电磁波在道路结构层内部强度不衰减的情况下，估测得到的充水脱空介电常数为 50-67，与真实值 81 明显接近。这说明道路材料的电导率对脱空位置处介电常数的估算有很大影响。

类似地，用反射波幅计算充水脱空的介电常数，模型中材料的电磁学特性参数按表 3.2 选取，脱空高度从 0.01m 按 0.01m 的间隔增加到 0.3m。部分结果如表 4.3 所示；再设定模型中的路面结构为无电磁耗损介质，重新生成电磁传播模型，并用反射波幅计算道路材料的介电常数，结果如表 4.4 所示。

表 4.3 注浆修复脱空道路结构介电常数估算值

| 结构层<br>脱空高度 | 面层 | 基层 | 注浆修复脱空 |
|---|---|---|---|
| 0.01m | 5.9 | 7.2 | 12.8 |
| 0.02m | 5.9 | 7.2 | 13 |
| 0.03m | 5.9 | 7.1 | 12.8 |
| 0.3m | 5.9 | 7.1 | 12.8 |

表 4.4 注浆修复脱空道路结构介电常数估算值（无电磁损耗）

| 结构层<br>脱空高度 | 面层 | 基层 | 注浆修复脱空 |
|---|---|---|---|
| 0.01m | 5.9 | 7.2 | 23.6 |
| 0.02m | 5.9 | 7.2 | 22.8 |
| 0.03m | 5.9 | 7.2 | 21.6 |
| 0.3m | 5.9 | 7.2 | 21.6 |

仿真模型中注浆材料的介电常数为 28，存在电磁损耗得到的估算值为 12-13，消除电磁损耗得到的估算值为 21-24。

这种方法的介电常数估算公式中没有考虑电磁耗损，而实际道路材料（尤其是当湿度较大时）的电导率不为零。因此，如果通过这种方法估算的介电常数来判断脱空类型，需要现场校对，通过若干校对点探地雷达扫描和钻芯取样对比，建立基层底部材料介电常数和反射波幅的关系，然后再进行整个路段雷达图像的分析判断。

## 4.3 充气脱空水平尺寸的计算

在实际道路检测中，识别出脱空并估算脱空高度后，往往还需要估算脱空的水平尺寸，以便估计病害的严重程度，并为脱空修复所需的注浆量提供参考。脱空的填充状态会影响脱空水平尺寸的估计：若脱空处于充水状态，脱空附近的道路结构也处于浸湿状态，二者介电常数相近，不易从雷达图像上分辨；若脱空处于充气状态，脱空与附近的道路结构介电常数差异较大，方便脱空水平尺寸的估算。因此，本节以充气脱空为研究对象，相应地，实际工程检测应当选择在少雨季节。

### 4.3.1 充气脱空水平尺寸计算指标的选取

由于探地雷达天线有一个发散型的扫描范围，某测点的探地雷达图像反映的是测点附近一定范围内的情况。因此，在探地雷达图像上由脱空引起的异常位置的水平尺寸，大于脱空本身的水平尺寸。为了根据探地雷达图像上的特征估算脱空的水平尺寸，就需要研究图像上异常位置水平尺寸与脱空实际水平尺寸之间的关系。

除了脱空的水平尺寸，脱空高度也会对图像特征产生影响。为了使脱空水平尺寸的估算不受脱空高度的影响，需要选取合理的估算指标。通过对 800MHz 天线、0.5m 水平长度、不同高度（0.01m、0.02m、0.05m、0.1m、0.2m）脱空的道路结构的图像分析，发现位于脱空位置两侧的基层-土基界面反射的边缘间距不随脱空高度而变化，如图 4.58、图 4.59 和表 4.5 所示，故将该特征长度选为估算脱空水平尺寸的指标，记作 d。

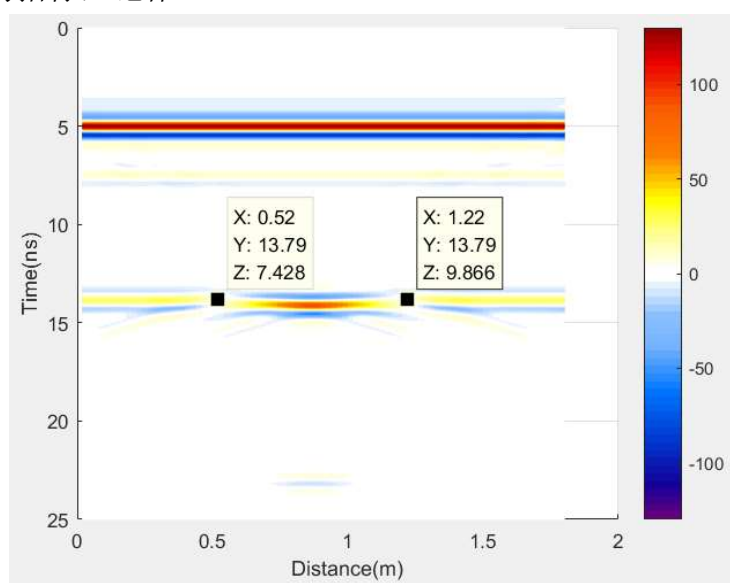

图 4.58 B 扫描脱空水平尺寸指标（0.02m 高度 0.5m 长度充气脱空，800MHz 天线）

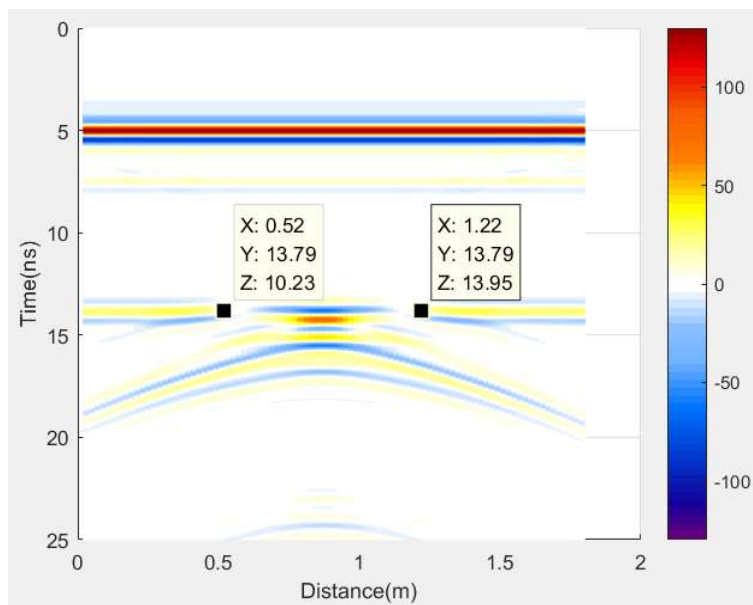

图 4.59 B 扫描脱空水平尺寸指标（0.2m 高度 0.5m 长度充气脱空，800MHz 天线）

表 4.5 不同高度 0.5m 长度的充气脱空的特征长度 d

| 脱空高度（m） | 0.01 | 0.02 | 0.05 | 0.1 | 0.2 |
|---|---|---|---|---|---|
| 左侧边界（m） | 0.52 | 0.52 | 0.52 | 0.52 | 0.52 |
| 右侧边界（m） | 1.22 | 1.22 | 1.22 | 1.22 | 1.22 |
| 特征长度 d（m） | 0.70 | 0.70 | 0.70 | 0.70 | 0.70 |

接下来研究该特征长度是否受道路结构层厚度及材料电磁特性的影响。研究方法为：选取具有不同结构层厚度和材料电磁特性参数的 5 种结构，充气脱空的高度和水平尺寸统一取 0.1m 和 0.5m*0.003m，分别仿真生成 B 扫描图像，读取各图像上的特征长度并比较。结构 1 的道路模型几何参数及各层材料电磁特性按表 3.1 和表 3.2 选取，汇总为表 4.6；结构 2、结构 3、结构 4、结构 5 的层厚及参数见表 4.7-4.10。这里各种材料均为非磁性材料，相对磁导率为 1，磁损耗为 0。基于结构 1-结构 5 生成的 B 扫描图像分别为图 4.60-图 4.64。

表 4.6 结构 1 的道路模型几何参数及各层材料电磁特性

| 道路结构层 | 水平尺寸（m*m） | 厚度（m） | 相对介电常数 | 电导率 |
|---|---|---|---|---|
| 沥青面层 | 3.1*0.003 | 0.15 | 6 | 0.005 |
| 半刚性基层 | 3.1*0.003 | 0.35 | 7.5 | 0.01 |
| 土基 | 3.1*0.003 | 0.5 | 18 | 0.2 |
| 脱空 | 0.5*0.003 | 0.1 | 1 | 0 |

表 4.7 结构 2 的道路模型几何参数及各层材料电磁特性

| 道路结构层 | 水平尺寸（m*m） | 厚度（m） | 相对介电常数 | 电导率 |
|---|---|---|---|---|
| 沥青面层 | 3.1*0.003 | 0.1 | 6 | 0.005 |
| 半刚性基层 | 3.1*0.003 | 0.25 | 7.5 | 0.01 |
| 土基 | 3.1*0.003 | 0.5 | 18 | 0.2 |
| 脱空 | 0.5*0.003 | 0.1 | 1 | 0 |

表 4.8 结构 3 的道路模型几何参数及各层材料电磁特性

| 道路结构层 | 水平尺寸（m*m） | 厚度（m） | 相对介电常数 | 电导率 |
|---|---|---|---|---|
| 沥青面层 | 3.1*0.003 | 0.2 | 6 | 0.005 |
| 半刚性基层 | 3.1*0.003 | 0.45 | 7.5 | 0.01 |
| 土基 | 3.1*0.003 | 0.5 | 18 | 0.2 |
| 脱空 | 0.5*0.003 | 0.1 | 1 | 0 |

表 4.9 结构 4 的道路模型几何参数及各层材料电磁特性

| 道路结构层 | 水平尺寸（m*m） | 厚度（m） | 相对介电常数 | 电导率 |
|---|---|---|---|---|
| 沥青面层 | 3.1*0.003 | 0.15 | 9 | 0.005 |
| 半刚性基层 | 3.1*0.003 | 0.35 | 12 | 0.01 |
| 土基 | 3.1*0.003 | 0.5 | 22 | 0.2 |
| 脱空 | 0.5*0.003 | 0.1 | 1 | 0 |

表 4.10 结构 5 的道路模型几何参数及各层材料电磁特性

| 道路结构层 | 水平尺寸（m*m） | 厚度（m） | 相对介电常数 | 电导率 |
|---|---|---|---|---|
| 沥青面层 | 3.1*0.003 | 0.15 | 4 | 0.005 |
| 半刚性基层 | 3.1*0.003 | 0.35 | 7 | 0.01 |
| 土基 | 3.1*0.003 | 0.5 | 15 | 0.2 |
| 脱空 | 0.5*0.003 | 0.1 | 1 | 0 |

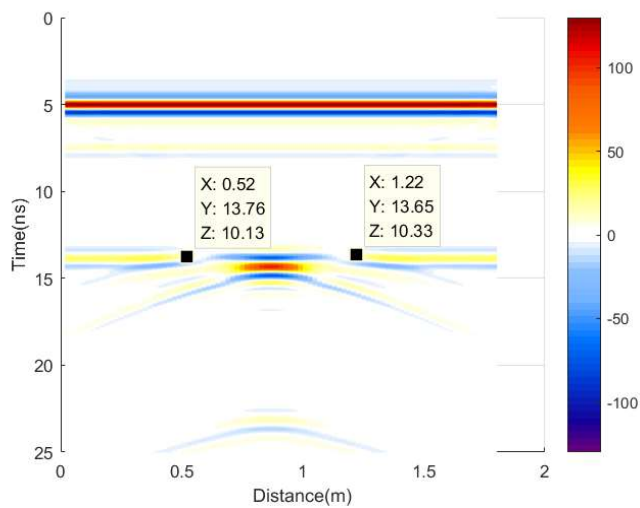

图 4.60 B 扫描脱空水平尺寸指标（结构 1，800MHz 天线）

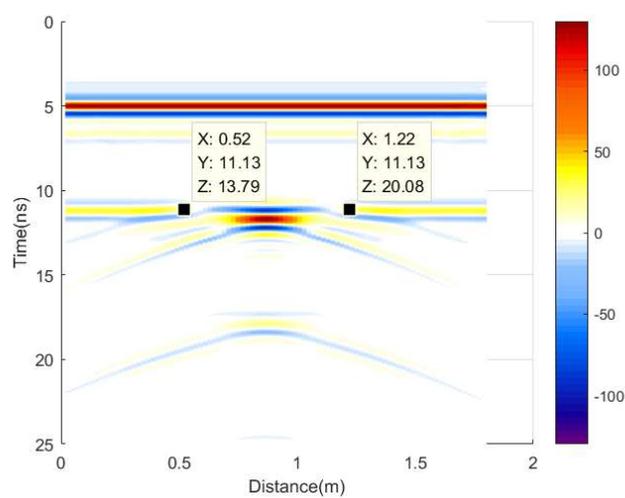

图 4.61 B 扫描脱空水平尺寸指标（结构 2，800MHz 天线）

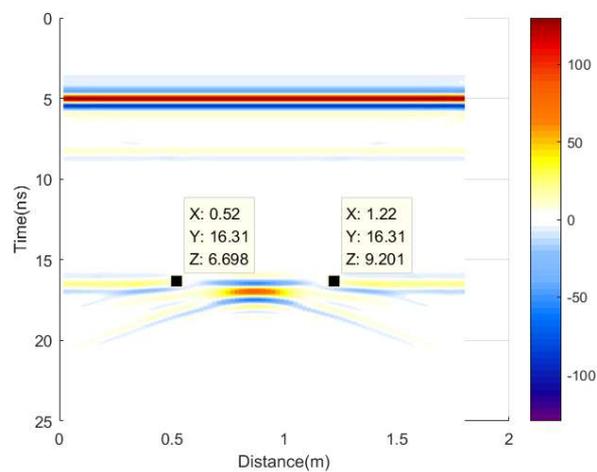

图 4.62 B 扫描脱空水平尺寸指标（结构 3，800MHz 天线）

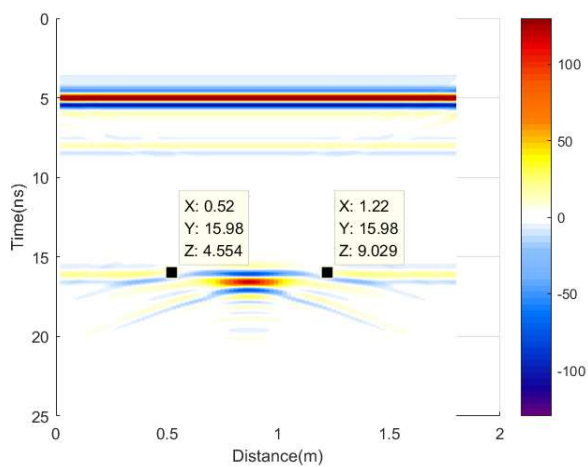

图 4.63 B 扫描脱空水平尺寸指标（结构 4，800MHz 天线）

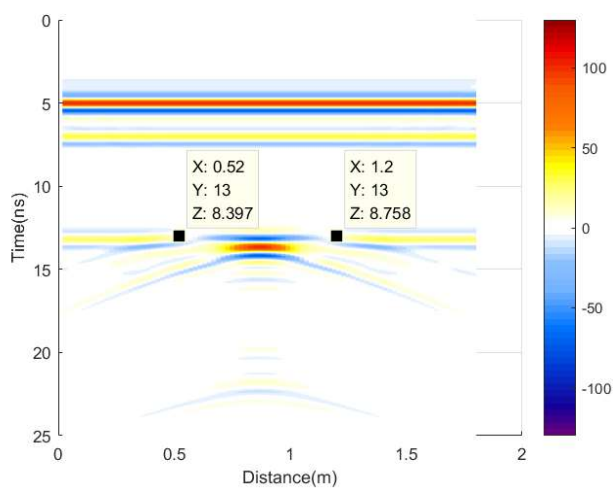

图 4.64 B 扫描脱空水平尺寸指标（结构 5，800MHz 天线）

结构 2、3 的道路结构层厚异于结构 1，结构 4、5 的结构层电磁特性参数异于结构 1，但是从图 4.60-图 4.64 可以看出，脱空的水平尺寸指标均为 0.7m，如表 4.11 所示。因此，可以初步认为脱空的水平尺寸指标不受道路结构层厚度及材料的电磁特性参数的影响。

表 4.11 不同道路结构的充气脱空的特征长度 d

| 道路结构 | 结构 1 | 结构 2 | 结构 3 | 结构 4 | 结构 5 |
|---|---|---|---|---|---|
| 左侧边界（m） | 0.52 | 0.52 | 0.52 | 0.52 | 0.52 |
| 右侧边界（m） | 1.22 | 1.22 | 1.22 | 1.22 | 1.22 |
| 特征长度 d（m） | 0.70 | 0.70 | 0.70 | 0.70 | 0.70 |

综上，本节选取的特征长度，即脱空位置两侧的基层-土基界面反射的边缘间距，不随脱空高度、路面结构层厚度及路面结构电磁特性参数的影响，只与实

际的脱空水平尺寸有关。因此，该特征长度可以作为估算脱空水平尺寸的指标，用于下文回归得到脱空水平尺寸的估算公式。

### 4.3.2 计算方法

选定脱空长度的计算指标后，接下来研究脱空长度关于计算指标的表达式。首先，仿真生成一组脱空长度不同、其他参数相同的道路模型的 B 扫描，读取每个 B 扫描上的特征长度 d，如表 4.12 所示。

表 4.12 不同长度脱空的特征长度 d（0.1m 高度充气脱空 800MHz 天线）

| 脱空长度（m） | 左侧边界（m） | 右侧边界（m） | 特征长度 d（m） |
|---|---|---|---|
| 0.04 | —— | —— | —— |
| 0.08 | 0.70 | 1.04 | 0.34 |
| 0.12 | 0.68 | 1.06 | 0.38 |
| 0.16 | 0.66 | 1.08 | 0.42 |
| 0.20 | 0.64 | 1.10 | 0.46 |
| 0.24 | 0.62 | 1.12 | 0.50 |
| 0.28 | 0.62 | 1.12 | 0.50 |
| 0.32 | 0.60 | 1.14 | 0.54 |
| 0.36 | 0.58 | 1.16 | 0.58 |
| 0.40 | 0.56 | 1.18 | 0.62 |
| 0.44 | 0.54 | 1.20 | 0.66 |
| 0.48 | 0.54 | 1.20 | 0.66 |
| 0.52 | 0.52 | 1.22 | 0.70 |

当脱空长度为 0.04m 时，B 扫描图像如图 4.65 所示，由于脱空长度过短，难以寻找特征长度。当脱空长度不小于 0.08m 时，可以依据 B 扫描读取特征长度，如图 4.66 和图 4.67 所示。

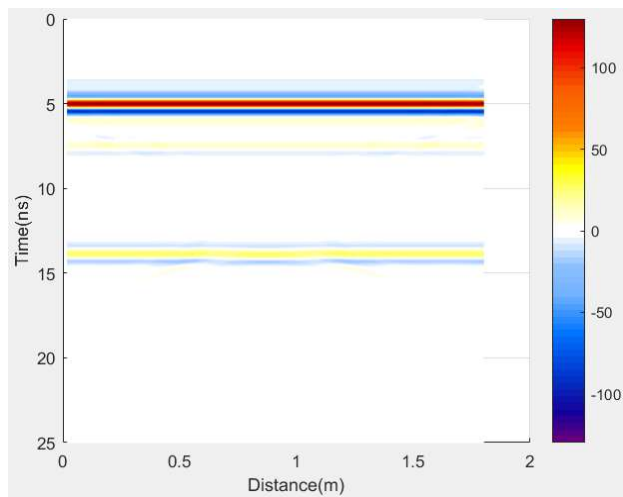

图 4.65 B 扫描脱空水平尺寸（0.04m 长度 0.1m 高度充气脱空，800MHz 天线）

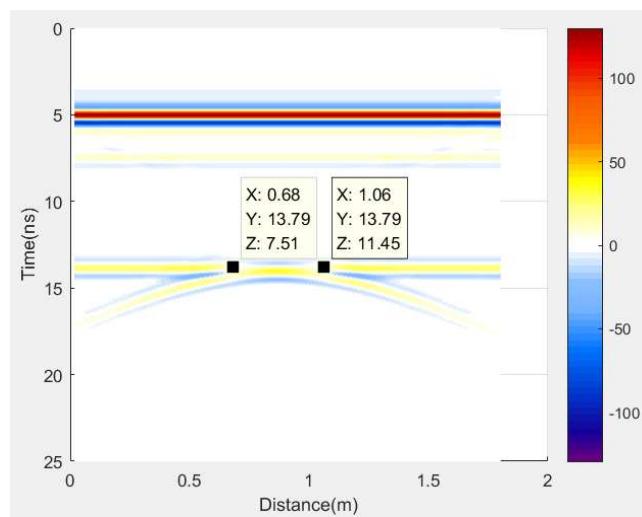

图 4.66 B 扫描脱空水平尺寸（0.12m 长度 0.1m 高度充气脱空，800MHz 天线）

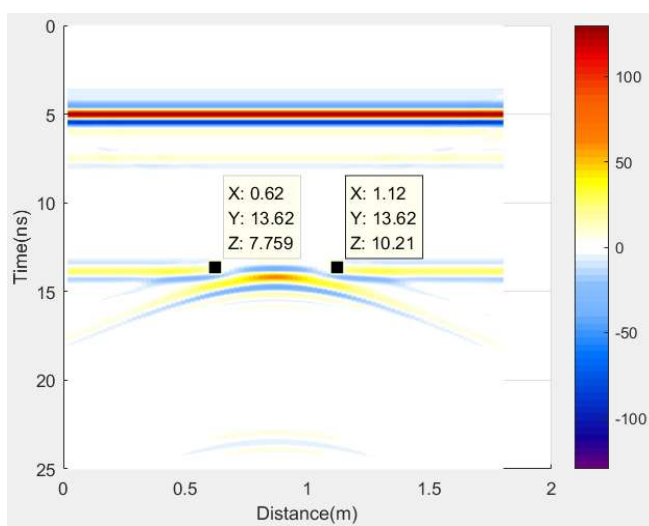

图 4.67 B 扫描脱空水平尺寸（0.28m 长度 0.1m 高度充气脱空，800MHz 天线）

根据表 4.6 中数据，对于 0.08m 到 0.52m 长度范围内的脱空，进行一元线性回归分析。得到线性回归方程

$$p(d) = 0.8077d + 0.2877 \tag{4.8}$$

式中 $p(d)$ 为脱空水平尺寸的计算值，$d$ 为特征长度，即脱空位置两侧的基层-土基界面反射的边缘间距。

图 4.68 是脱空水平尺寸计算回归方程和真实值的图像。计算得到相关系数为 0.99，说明脱空水平尺寸与特征长度具有很好的线性相关性，该回归方程可以准确计算脱空的水平尺寸。根据 4.3.1 中的结论，该公式应当也适用于其他结构层厚度和材料电磁特性参数的情况。

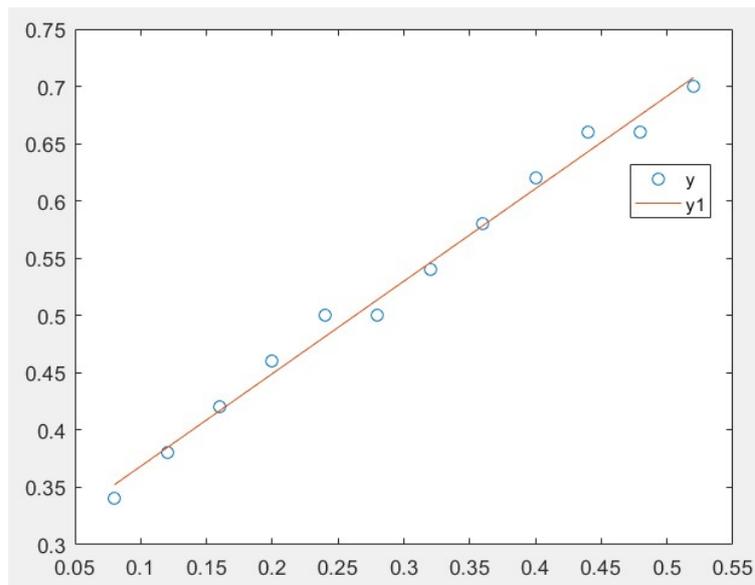

图 4.68 脱空水平尺寸计算回归公式

## 4.4 检测及评价整体流程

以上讨论了不同填充情况的半刚性基层层底脱空的识别和高度、水平尺寸的估算方法。在实际工程检测项目中，操作流程按先后顺序可分为脱空的识别、脱空尺寸的估算、注浆修复后的评价。为便于将本章的成果用于实际工程，以下讨论脱空检测及评价的流程，以及各个步骤采用的方法。

（1）脱空的识别

在脱空位置未知的情况下，需要在道路上大范围测量，寻找脱空位置。此阶段可以采用介电常数法或解卷积法。

介电常数法：寻找雷达图像上的异常反射，根据其反射波幅计算异常位置的介电常数，据此进一步判断是否存在脱空以及脱空的填充情况（充气或充水）。

解卷积法：适用于充气脱空，因此只适用于持续一段时间干燥天气下得到的检测数据。通过解卷积法分离重叠的反射波幅，进而判断是否存在充气脱空。

（2）脱空高度的估算

由于充水脱空的下界面反射非常微弱无法识别，因此想要估算脱空高度就必须在持续干燥天气检测。此阶段可以采用解卷积法或最小二乘系统辨识法。

解卷积法：处理速度快，适用于大范围处理，但只能估算高度不小于 0.1m 的脱空的高度。

最小二乘辨识法：处理速度慢，适用于局部精细计算，估算精度能够达到 1cm。

（3）脱空水平尺寸的估算

使用 4.3 中的回归公式，处理持续干燥天气下得到的检测结果，得到脱空的

水平尺寸。

（4）脱空注浆修复后的评价

注浆修复后，需要检测注浆材料是否有效填充了原有脱空。此阶段可以采用介电常数法。

介电常数法：注浆后，在采用了注浆修复措施的原有脱空位置上方，再次使用探地雷达检测，根据反射波幅计算脱空位置的介电常数，据此进一步判断是否脱空被注浆材料填充或未被注浆材料填充。

## 4.5 本章小结

（1）通过数值仿真，研究了脱空在 A 扫描上的特征。充气脱空的特征为前后两个正负相反的反射子波，文中将图像上两子波叠加情况随脱空高度减小的变化分成三个阶段，并得到使用 800MHz 或 1200MHz 天线时三个阶段相对应的脱空高度范围。由于充水脱空内的电磁耗损明显，脱空底部的反射非常微弱；注浆修复脱空由于介电常数与下方土基的介电常数相近，其底部的反射也不明显。

（2）针对充气脱空上下反射波叠加的特点，提出最小二乘系统辨识的方法。生成脱空高度为 0.01m-0.3m 共 30 个脱空道路模型后，任取一个脱空高度的模型，添加高斯白噪声，然后使用最小二乘系统辨识的方法识别其脱空高度。结果表明，信噪比低至 2 时这种方法仍可准确识别脱空高度。但这种方法要求待测重叠波与已知重叠波在时间轴上的位置对齐，且处理时间长，因此适合小范围的充气脱空高度的精确分析。

（3）采用解卷积的方法消除充气脱空反射子波重叠的影响，并采用吉洪诺夫正则化方法解决解卷积的不适定问题。结果表明，解卷积的方法能够把 0.1m-0.2m 脱空高度原本重叠的信号分离，但对于小于 0.1m 的脱空高度，这种方法无法分离重叠的反射波。这种方法对原始图像的要求低于最小二乘系统辨识法，可用于大范围的充气脱空识别及尺寸估测。

（4）针对充水脱空及注浆修复脱空的脱空底部反射微弱的特点，使用介电常数法进行脱空识别。结果表明充水和注浆脱空的反射子波重叠不明显：充水脱空高度不小于 0.02m 时脱空顶部反射波幅与脱空高度无关，注浆脱空高度不小于 0.03m 时脱空顶部反射波幅与脱空高度无关。实际的电磁耗损对介电常数估算值影响很大，使用介电常数法之前需要先根据现场数据对理论公式进行修正。

（5）选取 B 扫描上脱空位置两侧的基层-土基界面反射的边缘间距作为脱空水平尺寸估算的指标，回归得到脱空水平尺寸与估算指标之间的相关关系，二者呈良好的线性相关性。

# 第五章 探地雷达信号处理技术

在对探地雷达的数据及图像的分析中，常常会根据需要编辑整理数据文件，以及压抑某些信号、加强另外一些信号，从而达到突出某些特征的目的。这些分析手段统称为信号处理技术。信号处理技术的需求来源于两个方面，一是探地雷达信号中干扰的存在，二是雷达信号传递过程中自身的衰减效应等。

与前两章使用的建模仿真信号不同，在实际的工程检测中，探地雷达采集到的数据含有各种各样的干扰。其中既有外部来源的电磁波的干扰，也有设备自身产生的杂波。这些干扰有可能会被误认为是地下目标物，如水平噪声易被当做道路内部分界面。较强的噪声也可能将较弱的真实反射波掩盖。此外，由于实际操作条件限制，现场采集的数据文件常常需要整理，比如现场操作差错导致的一些废道需要被剔除。因此，上文第三、四章得到的脱空识别方法不能直接用于处理现场采集到的数据，雷达数据的处理首先需要去除现场数据中的干扰。

另一方面，由于探地雷达天线发射的信号呈扩散发射状，距离发射天线越远，信号强度越小；此外，各介质界面的反射和介质内部的损耗也导致信号在传播过程中不断减弱。在第三章的数值仿真中可以看出，土基顶面的反射子波强度远远小于道路表面反射子波强度。因此，需要采用一定方法加强真实界面上的弱反射的显示效果。

本章首先分析探地雷达数据中干扰信号的来源，之后按照处理流程的先后，研究各个步骤的探地雷达信号处理技术。本章以某段沥青路面上采集到的实测数据为例，验证各信号处理技术的效果。

## 5.1 探地雷达数据干扰来源

探地雷达的干扰主要有两种来源。一种被称为噪声，包括探地雷达接收天线接收到的各种外部电磁波，比如操作者的手机信号、对讲机发出的无线电信号以及无线电台的广播信号等。由于探地雷达使用的是超宽带信号（Ultra Wide Band，UWB），覆盖的电磁波频率范围很广，因此很容易受到外来电磁波的影响。正因如此，在探地雷达检测操作中对环境有较高要求，如操作员应当将手机关闭。

另一种来源被称为杂波，指由探地雷达的发射信号引起、但不利于分析解读雷达数据的无用信号。有的杂波在探地雷达设备内部产生，比如对于发射天线和接收天线分离的双站式雷达，发射天线发射的电磁波经空气直接传播至接收天线产生高强度的"直达波"，又比如质量不佳的雷达设备，发射波会在天线内壁发生反射被接收天线接收。杂波也可能因探测目标物产生。在探地雷达道路扫描图

像中常常可以看到路表反射的相同形状线条多次重复出现,这种情况一般是由于雷达信号在天线和路表之间多次反射引起的。如果道路内部存在上下两个较强反射面,也会导致雷达波被"困"在这两个反射面中间多次反射,反映在雷达图像上为道路内部同一反射多次重复出现。

这些信号干扰有时会很显著,导致强度较弱的真实反射信号无法被识别。空耦式天线相对于地耦式天线更容易受外界信号的干扰。除了这些干扰信号以外,数据序列与现场位置的不对应、信号的衰减及地表的不平整等,都会造成雷达图像分析的困难。因此,需要采用一定的信号处理技术。

## 5.2 数据的编辑整理

从现场采集回数据后,第一步工作是数据的编辑和整理。道路检测的一个基本要求为检测数据和现场位置能够对应。不同检测手段之间的综合分析、道路状况的准确评价和修复方案的提出,都离不开雷达数据的准确定位。在探地雷达检测中,一般采用等间距扫描模式,即通过安装在检测车轮上的距离传感器记录前进的距离,并控制探地雷达按照预设的采样频率等间隔发射、接收信号。探地雷达检测的这种线性定位方式与道路的里程定位一致,因此较为方便,但由于探地雷达设备的前进路线不是绝对的直线,因此需要每完成一定距离的测量,记录下对应的路边的里程桩号。此外,探地雷达设备本身无法记录设备在路面上的横向位置,需要人工记录,并保证测量过程中不发生明显的横向距离的移动。一些较先进的探地雷达设备配置了 GPS 定位系统,但由于民用 GPS 系统的精度较低,在不使用 GPS 基站的情况下还是需要人工记录 GPS 的横向距离以及用于校正的路侧桩号。

工程实际中,路面上的单一测线往往不能满足要求。因此,常常需要在横向上设置多条测线,若要生成三维 GPR 图像则需要设置较密的测线。出于测量的方便考虑,检测设备的移动一般按照往返路线。这种情况下,需要在调头前后分别采集雷达信号,并在内业工作中把与目标移动放向相反的雷达数据作反转处理。

当测线太长,电池电量不足或因其他原因,而将一条测线的数据分成几个文件储存时,需要在内业中合并文件。

在信号采集过程中,如果设备移动速度过快,会出现数据道的空缺。如果存在外部暂时性的噪声或设备的暂时故障,采到的数据会存在废道,需要剔除。这两种情况都需要根据相近的数据道进行填补,以保证数据的完整以及数据道的等间距。

## 5.3 信道调整

由于探地雷达设备的自身限制，需要对信道内部或信道之间进行一定调整。

### 5.3.1 振幅恢复

探地雷达接收天线所能接收的信号强度有一定上限，当地表反射的强度超过该限值时，设备接收到的信号发生"截断"，如图 5.1 所示，第一个反射子波的波峰和波谷都超过了接收天线的限值，均发生部分截断。这种现象会导致回波信号中的最大波幅无法正常显示。可以使用样条差值的方法，根据已知的发射波的波形，将被截断的强反射子波恢复。

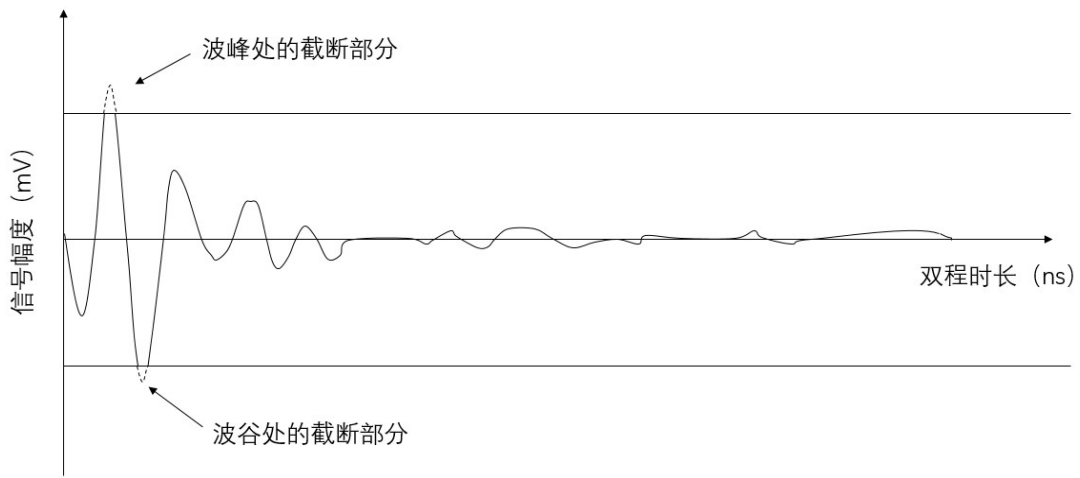

图 5.1 信号的"截断"效应

### 5.3.2 时间零点校正

理想的 B 扫描探地雷达图像中，路表等连续的界面在图像上应当也是一条连续的线条。但未经处理的图像上常常出现某段图像的纵向"跳跃"，影响数据的处理和图像的解读。这种现象是由天线与路表距离变化（常由颠簸引起）、电子元器件不稳定等因素造成的。

在后续的信号处理中，一个前提条件为不同信道的时间起点一致，因此需要对原始数据处理。可以选择地表反射波作为校正的参考，比如各信道找出地表反射的波峰，并将整个信道沿时间轴平移，使各信道的地表反射波峰处于同一条水平线上。如果噪声与地表反射叠加会导致地表反射波形变化，时间零点校正失效；为保证地表反射的波形不受直达波的干扰，应当保证探地雷达天线与路表保持足够距离。

### 5.3.3 道间均衡

由于探地雷达设备的不稳定，发射天线发射的信号强度可能随时间有所变化。探地雷达信号分析的一个前提条件是发射信号强度稳定，因此需要道间均衡。具体操作过程为：计算各信道的平均振幅 $\overline{A_i}$，进而求得所有信道的平均振幅 $\overline{A}$，各信道的数据分别乘以 $\overline{A} / \overline{A_i}$，各信道能量便可得到均衡。

## 5.4 去直流漂移和去直达波

### 5.4.1 信号中的直流漂移及其去除

去直流漂移是探地雷达信号处理不可缺少的一个步骤。直流漂移是指由信号中的直流成分导致的信号的正负分布不均衡（均值不为零），A 扫描的图像整体上偏向于坐标轴的一侧的现象，如图 5.2 所示。这种现象会导致 B 扫描的部分深度处整体呈现某种颜色，影响图像的可视效果；低频干扰的存在会导致整个信道的频谱发生变化，影响之后的频谱分析技术。

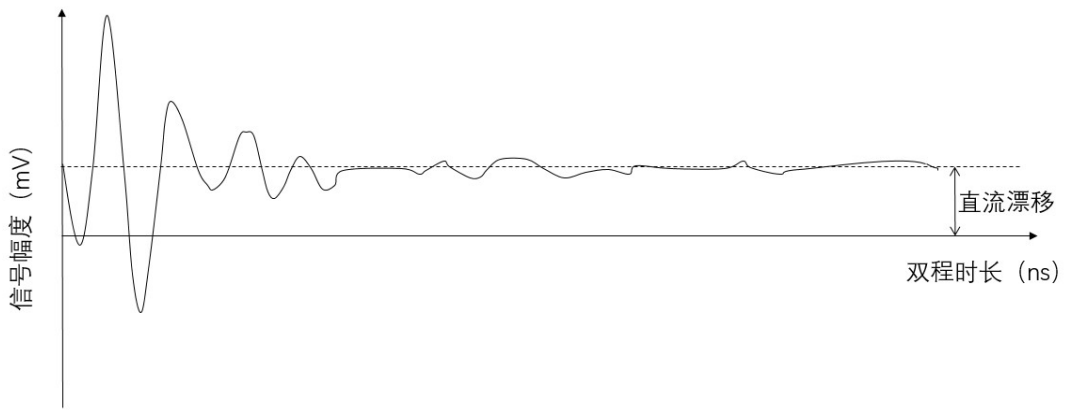

图 5.2 A 扫描中的直流漂移

去除直流漂移最简单的方法是计算某信道的时间域信号平均值，然后所有数据减去该均值。这种方法假设直流漂移是恒定的，与大多数实际情况不符。一种改进的方法是将一条信道划分为若干个区段，各段分别求平均值、减平均值。这种改进提高了精度，但不同区段的交界处可能出现不连续。去直流漂移还可以使用中值滤波、低频截断等滤波技术。去除直流漂移后的 B 扫描颜色更为均衡，图像特征更容易被识别。

### 5.4.2 信号中的耦合波及其去除

根据常用的双站式探地雷达的构造，发射天线发出的电磁波有一部分直接通过空气传播至接收天线，在雷达图像上显示为一个与时窗起点很接近的雷达子波，被称为直达波或耦合波。

耦合波如果与路表反射部分重叠，会增加数据分析的困难，也会导致无法进行零点校正。为避免这一问题，可以在采集雷达信号时保证天线与路表之间有足够的距离，避免两个子波的叠加。

为便于后续信号处理以及实际雷达图像与数值仿真图像的对比分析，需要去除耦合波。当雷达的发射天线朝向天空扫描时，接收天线接收到的只有耦合波而没有反射波。因此，将发射天线朝向天空扫描一段时间，对采集到的数据取平均，并从道路扫描数据中减去该平均值，即可得到去除耦合波的雷达信号。需要注意的是，去除耦合波后仍会有少量残留。

另一种避免直达波影响的方法，是在正式的雷达扫描操作前，尝试在有突起或坑槽的地面扫描并观察信号窗口变化，分辨出图像上的直达波和地表反射。调整时窗的起点，使时窗范围内不包含直达波。

在图 5.3 的采集过程中，通过移动时窗，排除了大部分直达波，仅有少部分残余位于时窗起点附件。之后，时窗位置参数不变，将天线朝向天空发射，得到图 5.4 所示的直达波图像；由于时窗的移动，图内采集的只是直达波波峰之后的残余部分。在图 5.3 的基础上减去图 5.4，残余的直达波也被去除，得到图 5.5，可以看到时窗起点附近的直达波被除去。

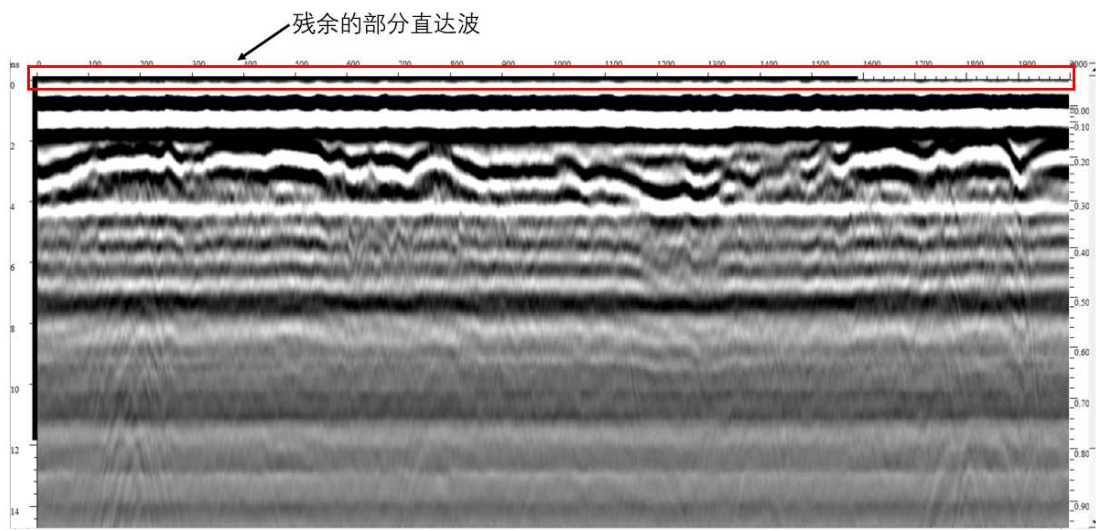

图 5.3 文件相减之前雷达图像上的残留直达波

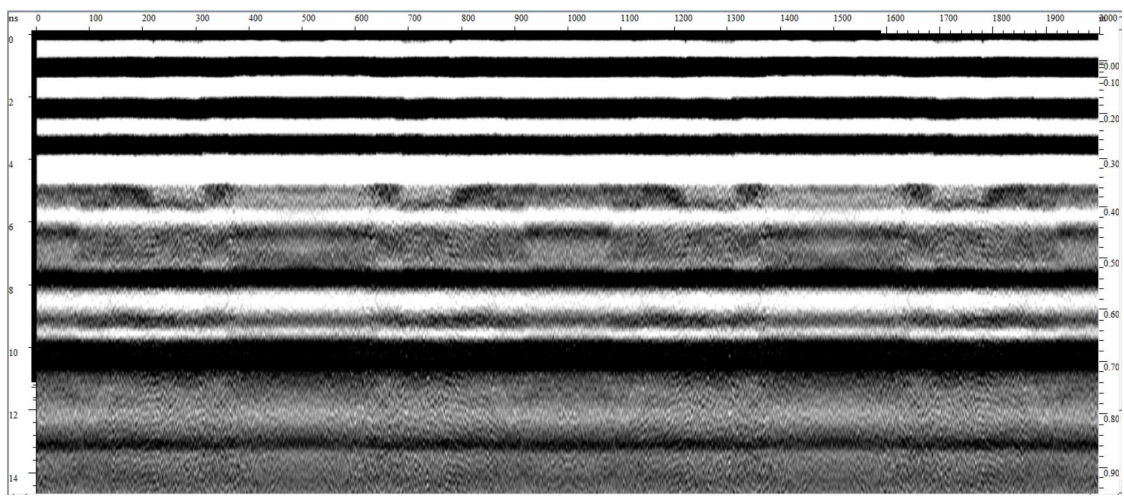

图 5.4 天线朝向天空采集到的图像

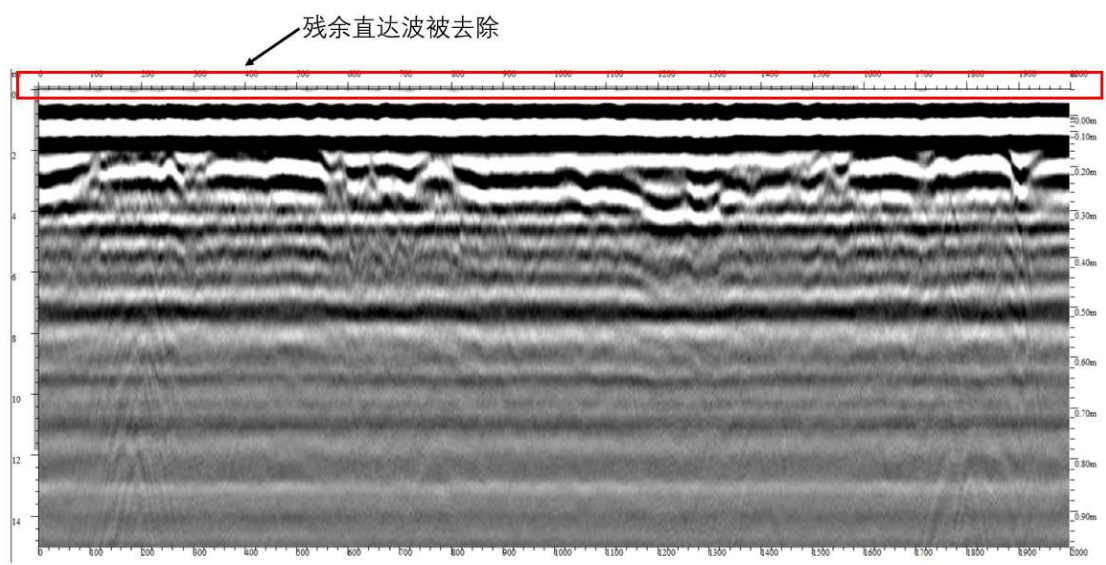

图 5.5 文件相减之后残留直达波被去除

## 5.5 滤波

滤波是探地雷达信号处理中常用的一大类手段。滤波主要有两方面作用。一方面，滤波可以去除外来的或设备内生的信号干扰，5.1 节中已经分析了信号干扰的来源。另一方面，滤波也被用于抑制某特定类型的信号特征或突出特定类型的信号特征。以下按照滤波的类型加以讨论。

### 5.5.1 时域和频域滤波

时域滤波指在时间域上对单个信道的滤波，频域滤波指在频率域上对信道的

滤波。主要包括以下类型：

（1）低通滤波、高通滤波

低通滤波和高通滤波分别指只允许低于（或高于）某一频率阈值的信号通过的滤波。低通滤波常用于去除高频噪声，如大多数 4G 手机信号频段在 2GHz 以上，大于一般的探地雷达设备使用的频段。高通滤波有助于去除直流漂移等低频信号成分。

（2）带通滤波

带通滤波是低通滤波和高通滤波的组合，只允许高、低频率界限中间的信号成分通过，兼有低通和高通滤波的功能。如图 5.6 所示。带通滤波在探地雷达信号处理中非常常用。

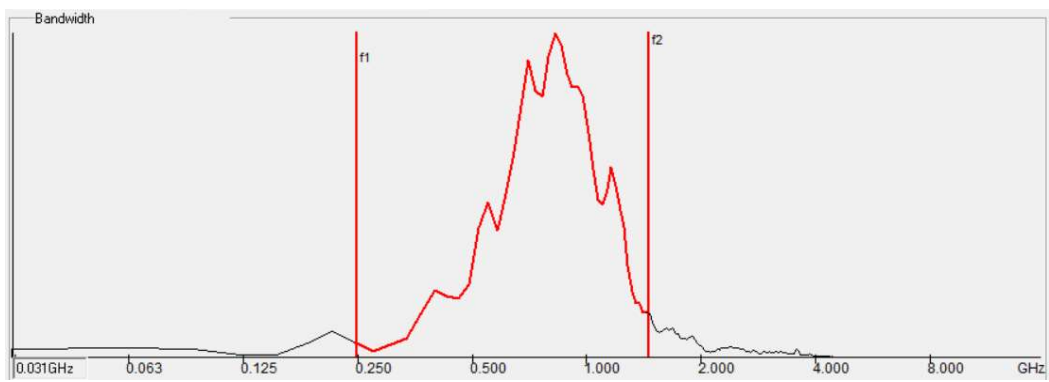

图 5.6 带通滤波器频谱示意图

带通滤波的一个关键指标是频率区间界限的选择。若频率界限过窄，虽然噪声去除可能更彻底，但真实信号也被部分去除。若频率界限过宽，则可能导致噪声去除效果不明显。根据工程经验，一般选用宽度为发射波中心频率 1.5 倍的频率界限。例如此次测量使用中心频率为 1000MHz 的天线，可以选用 1500MHz 宽度的频率区间，即下限（1000-750=）250MHz，上限（1000+750=）1750MHz。

图 5.7 为经过带通滤波后的探地雷达图像。经过带通滤波处理，雷达图像上较深位置的高频噪声被去除，图 5.8 为 8-14 纳秒时间范围内、100-400 道扫描图像的带通滤波处理前后对比，可以明显看出原图像上的"颗粒"（高频噪声）被去除。

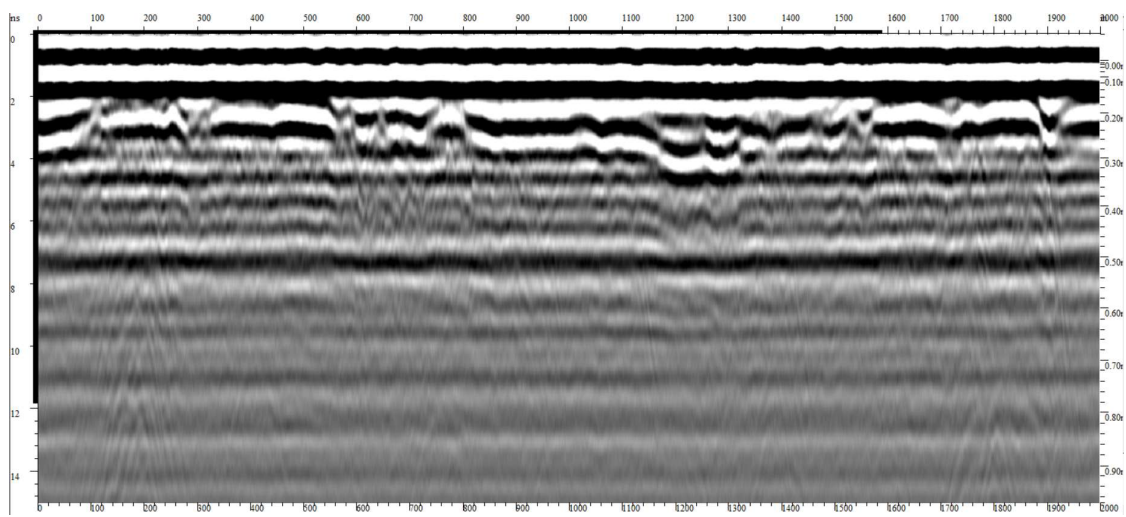

图 5.7 带通滤波处理后的探地雷达图像

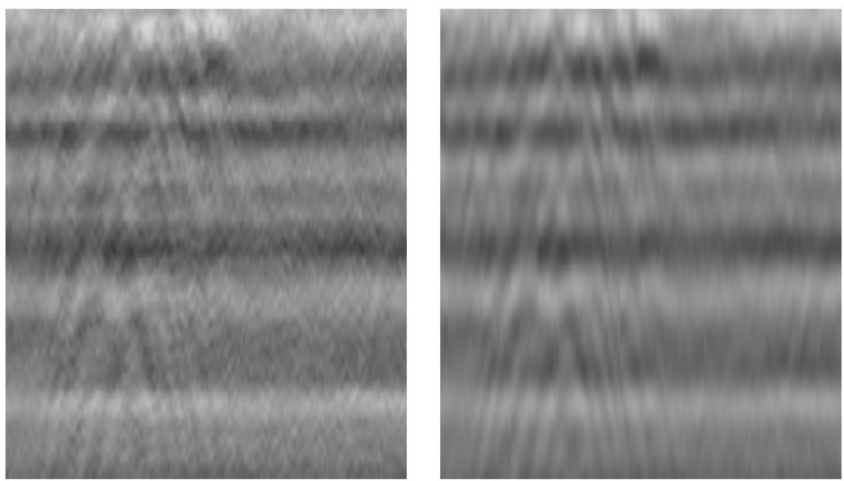

图 5.8 带通滤波前、后局部图像对比

（3）时变滤波

上述的低通、高通及带通滤波，均是对整条信道采用同样的频率界限。而探地雷达发射波在实际的道路材料中，随着传播深度的增加，信号的高频成分迅速衰减而低频成分衰减速度相对较慢（这也是中心频率越高的天线探测深度越浅的原因）。因此，在较浅的探测范围内（即时窗的前面部分），带通滤波的高频界限不能设置过低，以免大量过滤掉真实反射信号的高频成分；而在较深的探测范围内（即时窗的后面部分），真实信号的高频成分已经几乎损耗殆尽，接收天线采集的信号高频成分基本为噪声，因此可以降低带通滤波的高频界限。这种在时窗不同位置采用不同频率界限的带通滤波被称为时变滤波。为避免滤波处理后出现信号的不连续，时变滤波的频率界限的变化应当采用渐变的方式。

（4）均值滤波

也被称为均值平滑，指设定一个滤波时窗长度，时窗内所有采样点的信号幅值取其均值，该滤波时窗沿雷达信号时窗移动，重复取平均值的操作。该滤波被用于去除信号中尤外来干扰等引起的异常波动值。

（5）中值滤波

也被称为中值平滑，类似于均值滤波，不同在于滤波时窗内所有采样点取中指。类似地，中值滤波也可以去除噪声的"毛刺"，但也会导致真实信号的分辨率降低。因此，均值滤波和中值滤波应当只在探地雷达图像中有毛刺现象时使用，而且滤波时窗不应过长。

图 5.9 为该路段扫描中某位置上的 A 扫描，可以看出并不存在"毛刺"现象，因此无需采用均值滤波或中值滤波。

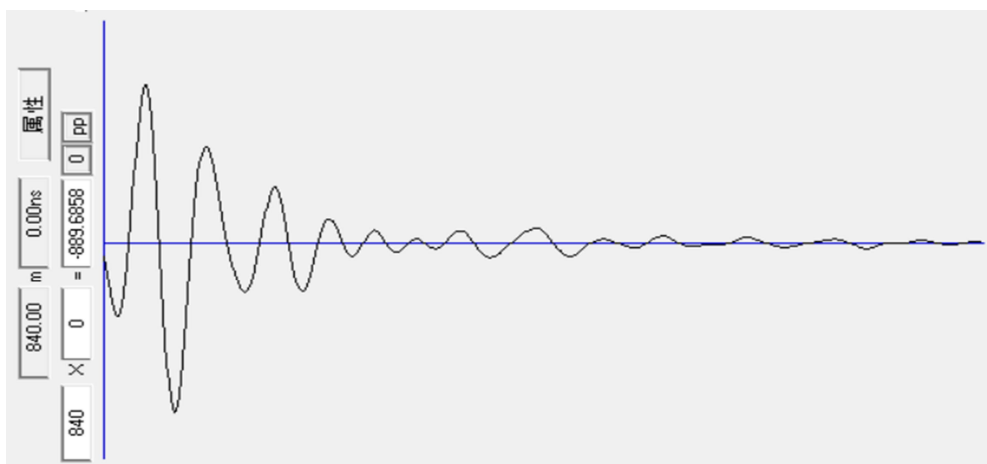

图 5.9 该路段第 840 道数据的 A 扫描图像

## 5.5.2 空域滤波

空域滤波指在空间域上对一系列信道的滤波。主要包括以下类型：

（1）滑动平均

空间域上的滑动平均是将一定距离范围内的若干道数据的同一时刻的数值取平均，然后该距离范围内的各道数据均取这个均值。滑动平均的作用在于减小雷达图像在同一深度处水平方向上的变异性，从而突出图像中的水平反射面的特征。

（2）背景去除

背景去除是指将所有或一部分信道的雷达数据取平均，再从各条信道中把这个平均值去除。背景去除实际上是去除了同一深度处各信道之间的共同部分，保留了信道之间的差异。因此，背景去除能够去除或抑制雷达图像上的水平特征，而相对突出竖向特征。背景去除对于消除振铃等水平干扰有较好的效果。

　　背景去除会使雷达图像中的所有水平特性都被消除或抑制，因此有一定副作用。道路扫描图像中的路表反射一般为水平线（天线高度固定或经过时间零点校正），而且道路内部界面也有可能近似为水平面，这些特征都会被背景去除抑制。

　　图 5.10 为采用背景去除前的该路段探地雷达图像，图 5.11 为背景去除后的探地雷达图像。可以明显看出，图像的中间和下部的很多特征原本被"振铃"现象掩盖，经过背景去除，这些特征变得容易识别。但应当注意，处理后图像的下部仍有部分"振铃"现象残余，不能将其判断为真实的反射物。

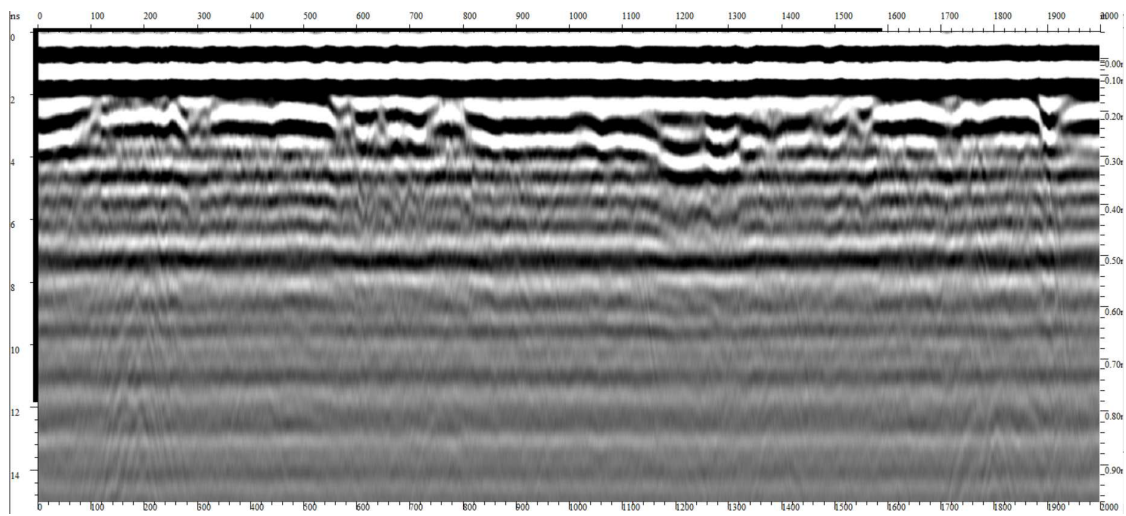

图 5.10 背景去除前的探地雷达图像

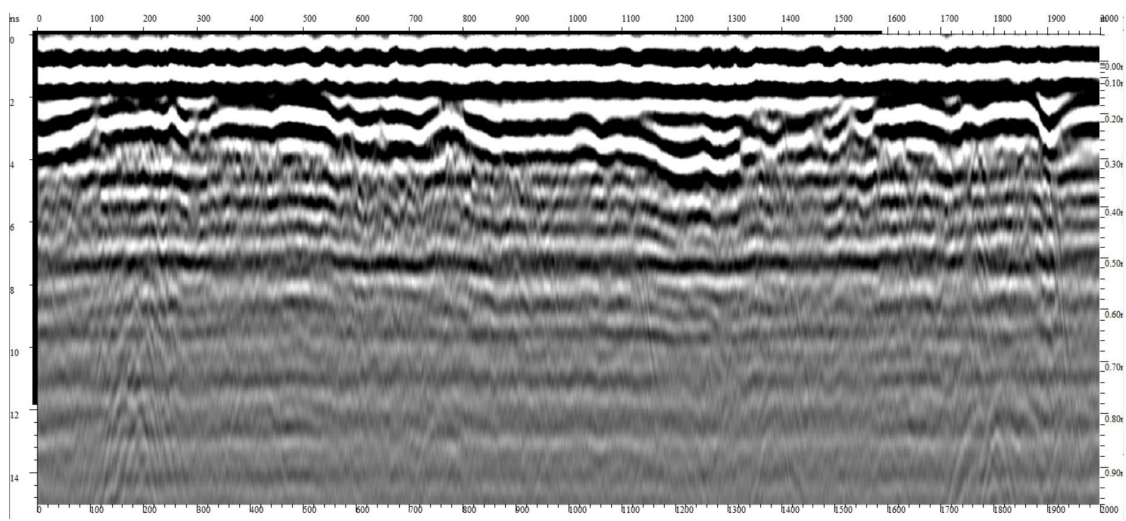

图 5.11 背景去除后的探地雷达图像

### 5.5.3 二维滤波

　　二维滤波是同时在时域和空域上对雷达数据进行处理的滤波。主要包括以下类型：

（1）t-x 平均

t-x 平均可看做时域的均值滤波和空域的滑动平均的结合。t-x 平均可以用于去除整体噪声。

（2）t-x 中值

类似于 t-x 平均，区别在于 t-x 中值的二维窗口范围内各点全部取中值。

（3）F-K 滤波

F-K 滤波指频率波数滤波，是将信号数据转换到频率波数域（frequency-wavenumber domain）进行处理。

## 5.6 信号增益

探地雷达图像处理中一个常用的可视效果优化工具是信号增益。由于信号扩散、界面反射、介质损耗等原因，探地雷达信号在传播过程中不断减弱，在时窗的靠后部分变得很不明显。信号增益是指按照一定的规则，提高时窗靠后部分的信号幅值，从而减小时窗中信号幅值的差异，增强路面下较深处反射的显示效果。增益函数可以分为以下几类：

（1）振幅恢复增益

振幅恢复增益的设计目的在于消除电磁波的传播扩散造成的振幅变小。因此，该种增益的关键在于估计电磁波的扩散导致的耗散程度。波的扩散后强度应当与 $1/r^2$ 成正比，r 为传播距离，增益函数的参数需要在工程中实际标定。增益曲线如图 5.12 所示，处理效果如图 5.13 所示。

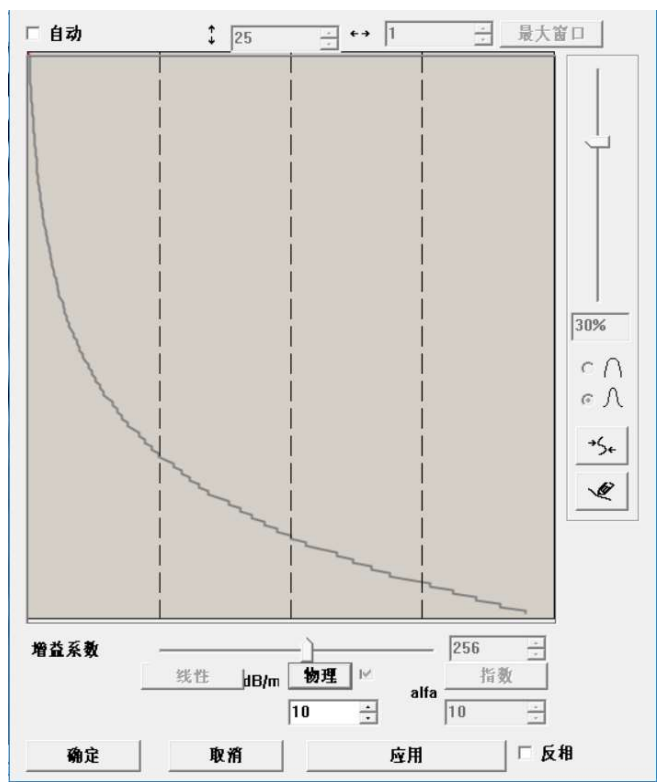

图 5.12 振幅恢复增益曲线

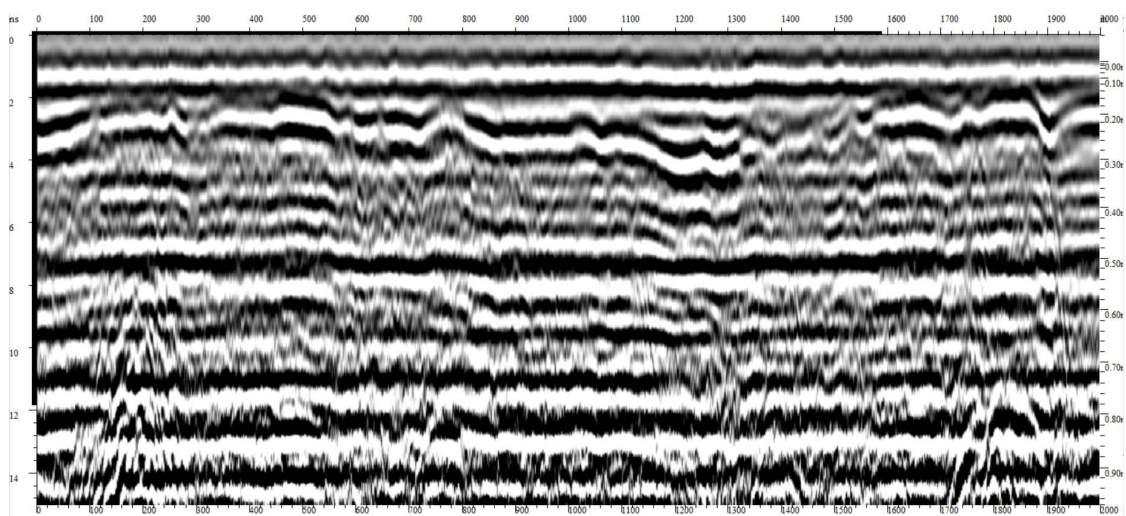

图 5.13 振幅恢复增益模式效果图

（2）自动增益

自动增益（automatic gain function，AGC）是一种常用的增益函数，其设计目的在于使时窗内的信号振幅保持在同一个水平。其处理过程为，将信号时窗划分为若干个处理时窗，设定一个增益比值，将每个处理时窗内的波幅均值与整个信号时窗内的最大信号幅值加以比较，按照设定的增益比值，将各处理时窗内的信号平均幅值扩大到相同的水平。处理效果如图 5.14 所示。

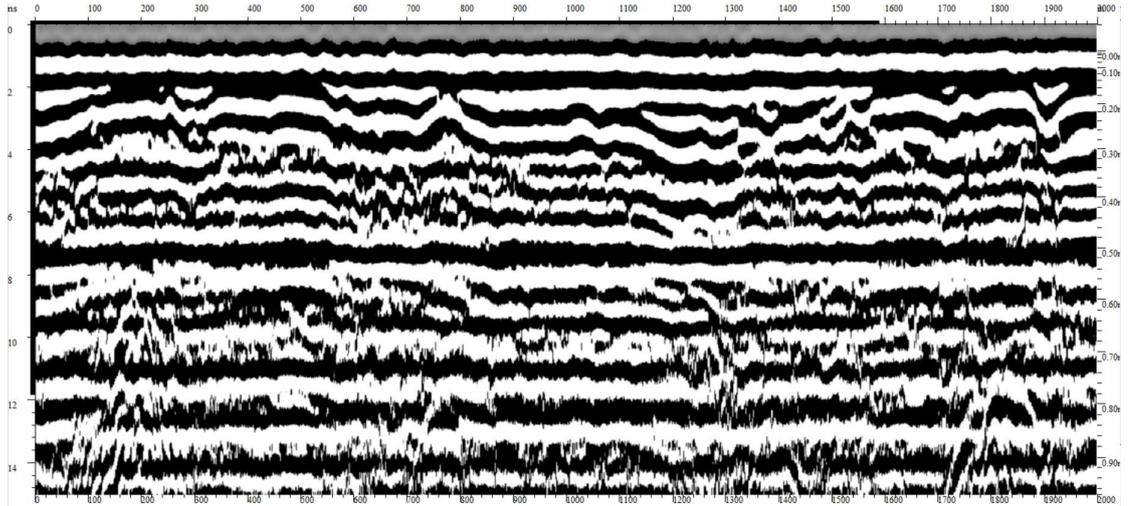

图 5.14 自动增益模式效果图

（3）自定义增益

还有一种增益的类型为自定义增益，数据处理人员按照经验及需要，自行设定增益函数。这种增益模式允许各种线性及非线性增益函数，甚至可以在确定道路各结构层分界面位置后，专门在分界面附件采用较大增益比例，突出界面反射的显示效果。这里以最基本的线性增益为例，增益曲线如图 5.15 所示，处理效果如图 5.16 所示。

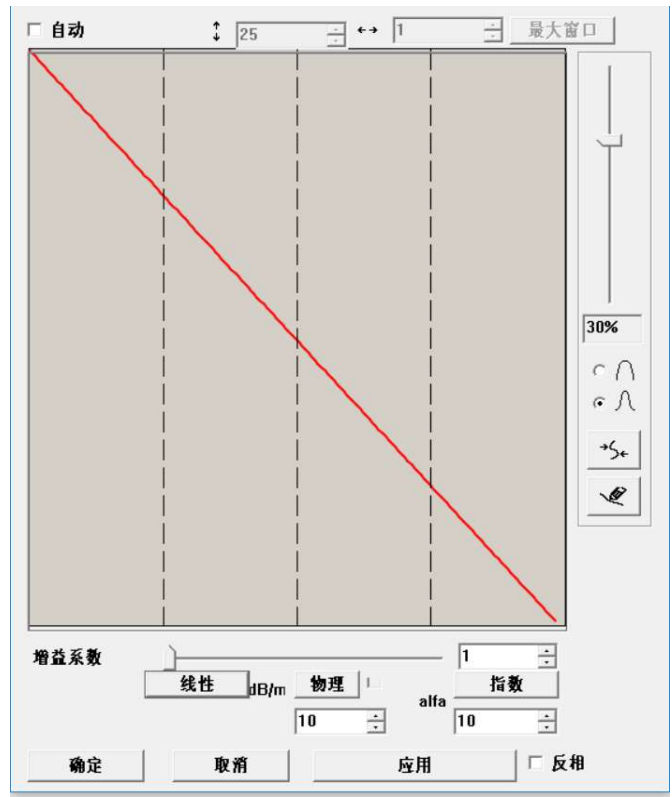

图 5.15 线性增益曲线

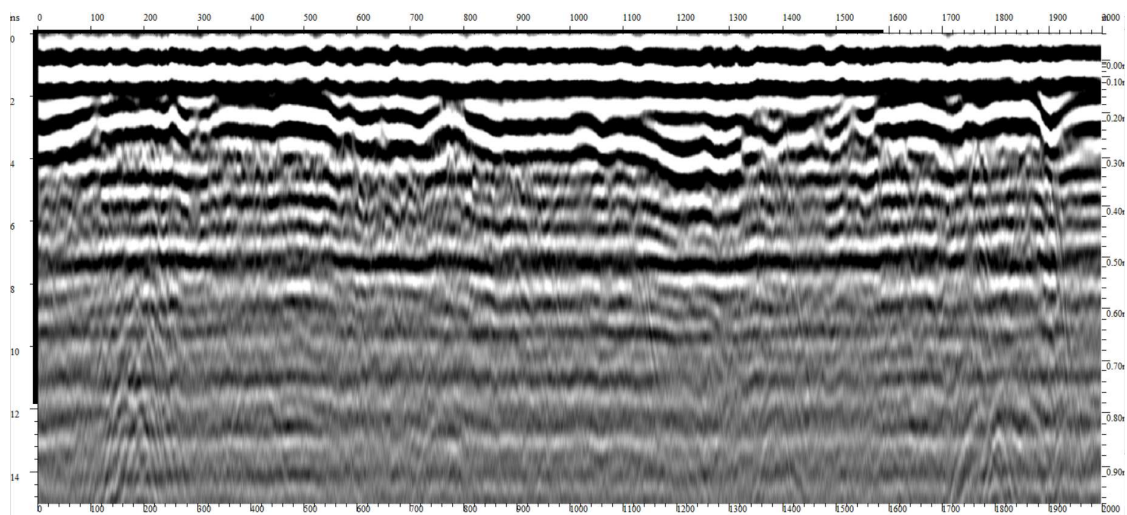

图 5.16 线性增益模式效果图

## 5.7 本章小结

（1）在采用第四章中的脱空识别及尺寸估测方法前，需要先对采集到的信号进行处理。这一方面是因为探地雷达信号中存在干扰，另一方面是因为信号传递过程中自身的衰减效应。信号干扰有两种来源，包括外界环境中的噪声，以及探地雷达设备自身或探测目标引起的杂波。

（2）首先对数据编辑整理，使检测数据与实际道路位置相对应，并根据需要作反转、合并、剔除处理。

（3）由于探地雷达设备的自身限制以及设备在探测过程中的颠簸，需要对数据进行信道调整，包括振幅恢复、时间零点校正以及道间均衡。接着，需要去除信号中的直流漂移和耦合波。

（4）研究了时域滤波、频域滤波、空域滤波及二维滤波的处理效果。具体来说，低通、高通及带通滤波器能够有效去除外界环境的高频或低频噪声，均值滤波或中值滤波能够去除时窗内信号上的"毛刺"；空域的滑动平均能够突出图像中水平反射面的特征，背景去除对于消除"振铃"现象有较好效果。

（5）由于信号扩散及介质损耗等，探地雷达信号在传播过程中不断衰弱。可以使用信号增益增强较深位置处的信号反射。增益方式分为振幅恢复增益、自动增益、自定义增益。

# 第六章 结论与展望

本文首先研究了探地雷达设备的类型、组成及性能评价指标，分析了电磁波的传播、反射、折射理论。然后，使用基于 FDTD 方法的 gprMax 软件数值模拟发射波在道路中的传播，通过理论分析和数值模拟研究了基底脱空在探地雷达图像上的特征。接着，研究了三种基于 A 扫描的基底脱空的判别及高度估算方法，以及基于 B 扫描的基底脱空水平尺寸的估算方法。最后，针对实际扫描信号中的干扰及衰减，研究了探地雷达信号处理流程。本文提出的道路基底脱空判识及尺寸估算方法对于道路脱空病害检测及养护具有重要意义。

## 6.1 主要研究结论

1. 讨论了探地雷达设备的类型、组成及性能评价指标，分析了电磁波的传播、反射、折射理论。

（1）按照调制技术，探地雷达可分为调频式、步进式及脉冲式；按照天线与地面的相对位置，可分为空耦式和地耦式。探地雷达设备由雷达主机、收发天线、电源、存储和显示设备等部件组成。

（2）衡量探地雷达性能水平的指标包括信噪比、信号稳定性、长时间信号稳定性、穿透性能等。

（3）麦克斯韦方程组和本构方程组描述了电磁场在同种介质中传播任意位置的电磁场量。电磁波的反射和折射理论描述了电磁波在不同介质接触面上的反射、折射规律。

（4）道路是一个具有多个水平反射面的层状体系，因此接收天线采集到的信号在时域上可以看作是一系列经过振幅缩小、添加时延后的发射信号的叠加，如式（6.1）所示。

$$y_r(t) = \sum_{i=0}^{N-1} A_i x(t - \tau_i) + n(t) \tag{6.1}$$

2. 使用基于 FDTD 方法的 gprMax 软件数值模拟发射波在道路中的传播，通过理论分析和数值模拟研究了基底脱空在探地雷达图像上的特征。

（1）从理论上分析了基底脱空的图像特征。A 扫描上，空气脱空的特征为相邻的一负、一正两波峰，充水脱空和注浆修复脱空的特征为相邻的一正、一负两波峰，脱空高度较小时两波峰重叠。B 扫描上，较小范围的脱空的图像特征为等轴双曲线的一支，实际的脱空位置在双曲线的焦点上；较大范围脱空的图像特

征类似于道路结构层分界面反射。

（2）gprMax 软件数值仿真的模型建立过程包括：规定空间范围、空间精度、边界条件，设定道路的几何及材料参数，选定发射波类型、收发天线的移动方式及采集时窗。

（3）仿真得到不同脱空参数、不同天线频率的一系列道路模型的 B 扫描和中心位置 A 扫描，用于研究脱空在探地雷达图像上的特征及脱空高度和水平尺寸的计算。

（4）仿真图像验证了对于基底脱空的图像特征的理论分析；此外，水平坐标和竖向坐标的比例关系会影响双曲线的渐近线斜率，且基层-土基界面的反射与脱空反射重叠。

3. 研究了三种基于 A 扫描的基底脱空的判别及高度估算方法，以及基于 B 扫描的基底脱空水平尺寸的估算方法。

（1）A 扫描上，充气脱空的特征为相邻的一负、一正两波峰，两波幅叠加情况随脱空高度减小的变化可分成三个阶段。A 扫描上充水、注浆脱空底部的反射不明显，前者是由于充水脱空内的电磁耗损显著，后者是由于介电常数与下方土基的介电常数相近。

（2）最小二乘系统辨识是估算充气脱空高度的一种有效方法。针对 0.01-0.3m 脱空高度范围内的测试结果表明，信噪比低至 2 时这种方法仍可准确识别脱空高度。但这种方法要求待测重叠波与已知重叠波在时间轴上的位置对齐，且处理时间长，因此适合小范围的充气脱空高度的精确分析。

（3）解卷积是分离重叠的充气脱空上下界面反射子波的一种方法，吉洪诺夫正则化可以解决解卷积的不适定问题。结果表明，解卷积的方法能够把脱空高度大于 0.1m 的原本重叠的信号分离，但对于小于 0.1m 的脱空高度，这种方法无法分离重叠的反射波。这种方法对原始图像的要求低于最小二乘系统辨识法，可用于大范围的充气脱空识别及高度估测。

（4）使用介电常数法对充水脱空及注浆修复脱空进行脱空判别。结果表明，充水脱空高度不小于 0.02m 时脱空顶部反射波幅与脱空高度无关，注浆脱空高度不小于 0.03m 时脱空顶部反射波幅与脱空高度无关，因此这种方法可以有效判别这两种脱空类型。实际的电磁耗损对介电常数估算值影响很大，需要先根据现场数据对理论公式进行修正。

（5）选取 B 扫描上脱空位置两侧的基层-土基界面反射的边缘间距作为脱空水平尺寸估算的指标，回归得到脱空水平尺寸与估算指标之间的相关关系，二者呈良好的线性相关性，如式（6.2）所示，相关系数为 0.99。

$$p(d) = 0.8077d + 0.2877 \qquad (6.2)$$

4. 针对实际扫描信号中的干扰及衰减，研究了探地雷达信号处理流程。

（1）信号处理有两方面作用，一是消除或减少探地雷达信号中的干扰，二是补偿信号传递过程中的衰减。信号干扰的来源包括外界环境中的噪声以及探地雷达设备自身或探测目标引起的杂波。

（2）数据编辑整理是为了使检测数据与实际道路位置相对应，并根据需要作反转、合并、剔除处理。

（3）由于探地雷达设备的自身限制以及设备在探测过程中的颠簸，需要对数据进行信道调整，包括振幅恢复、时间零点校正以及道间均衡。接着，需要去除信号中的直流漂移和耦合波。

（4）研究了时域滤波、频域滤波、空域滤波及二维滤波的处理效果，以及 IIR 和 FIR 两种滤波器设计类型的特点。具体来说，低通、高通及带通滤波器能够有效去除外界环境的高频或低频噪声，均值滤波或中值滤波能够去除时窗内信号上的"毛刺"；空域的滑动平均能够突出图像中水平反射面的特征，背景去除对于消除"振铃"现象有较好效果。FIR 滤波器虽然对计算机处理能力要求更高，但具有稳定性良好、线性相位等优点。

（5）由于信号扩散及介质损耗等，探地雷达信号在传播过程中逐渐衰弱。振幅恢复增益的增益函数依据电磁波的传播扩散及介质中的损耗确定，自动增益是使时窗内的信号振幅保持在同一个水平，自定义增益的增益函数由数据处理人员按照经验及处理需要自行设定。

## 6.2 本文创新点

1. 通过理论分析和数值模拟，考虑不同的脱空高度、水平尺寸和天线频率，以及充气、充水以及注浆修复三种不同的填充情况，研究了基底脱空在探地雷达 A 扫描图像和 B 扫描图像上的特征。

2. 提出基底脱空的判别及高度估算方法。对于充气脱空，将最小二乘系统辨识的方法和吉洪诺夫正则化解卷积的方法成功应用于脱空判别及高度估算。对于充水脱空和注浆修复脱空，研究了基于反射波幅的介电常数法在脱空判别中的效果。

3. 提出基底脱空的水平尺寸的估算方法。依据 800MHz 天线扫描充气脱空道路结构的仿真结果，提出水平尺寸估算的指标，并回归得到计算公式，脱空水平尺寸与估算指标呈良好的线性相关性。

4. 讨论了信号处理的流程，并依据实测数据研究了各种信号处理技术在去除干扰及补偿衰减等方面的实际效果。

## 6.3 有待进一步研究的问题

本研究在以下几方面有待进一步探讨和改进：

1. 本文关于脱空判别及尺寸估算的研究主要基于数值模拟的结果，限于实验设备条件未能在实际工程中验证。尽管本研究通过人为添加噪声模拟了实际采集的数据，但后续研究仍应结合实际工程数据验证本文结论。

2. 本文研究了信号处理流程在去噪等方面的效果，并初步讨论了信号处理对于基底脱空在图像上的特征的影响。由于脱空的判别及尺寸估算建立在脱空的图像特征的基础上，因此后续研究需要进一步研究信号处理的各种手段对于脱空判别及尺寸估算的影响。

3. 本文提出了脱空判别及尺寸估算方法并研究了其有效性，但处理过程较为复杂，耗时较长。为了将其应用于工程实际，还需要进一步研究这些方法的程序自动实现。

# 致谢

本文是在本人导师钱劲松副教授的悉心指导下完成的。从论文选题、制定研究思路到研究识别算法、现场试验等，本研究的各个部分均离不开钱老师的宝贵建议和大力支持。同时，在钱老师的带领下，本人在硕士期间参与了多个科研项目，培养了一定的学术能力，为本文的顺利完成打下了必要的基础。此外，课题组的凌建明老师、赵鸿铎老师、袁捷老师、黄琴龙老师、杨戈老师也对本研究提供过宝贵的指导和帮助，在此一并致以最衷心的感谢和敬意。

本文的完成也得益于同门师兄弟姐妹的帮助。感谢肖文兴师兄在探地雷达技术方面的指导；感谢马鲁宽师兄在本文选题和拟定研究思路方面提供的宝贵建议；感谢李雪垠、蔡爵威、李鹏辉、古小明、金辰、解一鸣、赵策同学在现场试验方面的支持。本文也得到了课题组其他多位同学的建议和帮助，难以一一列出，在此一并致以最衷心的感谢。

在同济度过的三年青春，写在纸上的是本文，写在自己身上心上的是成长。钱劲松副教授等课题组的各位恩师治学严谨、师德高尚，教会了本人许多治学做事和为人处世的道理；同级的陈俊君、陈欣然、陈峙昂、纪更占、唐睿、王宇翔、韦福禄、吴荻非、吴世涛、杨壮、张磊、张瑞菊、郑亚奇诸友，三年来相互鼓励，相互支持，朝夕相处，情同手足；杜增明、韩秉烨、林盛、刘媚、罗鑫、马鲁宽、陶泽峰、肖文兴、袁远、郑毅、朱立国等师兄师姐博学多才，远见卓识，倾力相助，本人受益匪浅；组里的各位同窗如同一家人，切磋学艺，畅谈人生，互为解难，并肩奋战；父母虽在千里之外，但对本人的关心和督促如在耳畔，使我不敢松懈，奋力向前。在此，对关心和支持本人的所有人表达最真诚的感谢。

最后，向硕士期间对本人予以无私帮助的所有人致以最真挚的祝福。祝愿钱老师等课题组诸位恩师桃李满天下，功业遍九州；祝愿组里各位兄弟姐妹会当凌绝顶，一览众山小；祝愿父母身体健康，事事顺心，希望儿子能够成为你们的骄傲。

# 参考文献

[1] Ni J C, Cheng W C. Trial grouting under rigid pavement: a case history in Magong Airport, Penghu[J]. Journal of Testing and Evaluation, 2011, 40(1): 1-12.

[2] 陈南, 曹长伟, 凌建明, 等. 基于弯沉盆的半刚性基层脱空判识方法[J]. 同济大学学报（自然科学版）, 2014, 42(5): 695-700.

[3] 陈南. 沥青路面半刚性基层底脱空响应及判识方法[D]. 上海:同济大学, 2014.

[4] 赵军, 唐伯明, 谈至明, 等. 基于弯沉指数的水泥混凝土路面板角脱空识别[J]. 同济大学学报 (自然科学版), 2006, 34(3): 335-339.

[5] 唐伯明. 刚性路面板脱空状况的评定与分析——落锤式弯沉仪 (FWD) 应用研究[J]. 中国公路学报, 1992, 1: 006.

[6] 张蓓, 赵欢, 钟燕辉,等. 基于弯沉率的水泥混凝土路面板角脱空面积分析方法[J]. 中外公路, 2010, 30(5):90-93.

[7] 张宁, 钱振东, 黄卫. 水泥混凝土路面板下地基脱空状况的评定与分析[J]. 公路交通科技, 2004, 21(1): 4-7.

[8] 韩西, 陈上均, 钟厉, 等. 砼路面板脱空检测方法综述[J]. 重庆交通大学学报 (自然科学版), 2006, 25(4): 73-76.

[9] Lahouar S, Al-Qadi I L. Automatic detection of multiple pavement layers from GPR data[J]. Ndt & E International, 2008, 41(2):69-81.

[10] Al-Qadi I L, Lahouar S. Measuring layer thicknesses with GPR – Theory to practice[J]. Construction & Building Materials, 2005, 19(10):763-772.

[11] Bastard C L, Baltazart V, Wang Y, et al. Thin-Pavement Thickness Estimation Using GPR With High-Resolution and Superresolution Methods[J]. IEEE Transactions on Geoscience & Remote Sensing, 2007, 45(8):2511-2519.

[12] Liu H, Sato M. In situ measurement of pavement thickness and dielectric permittivity by GPR using an antenna array[J]. NDT & E International, 2014, 64: 65-71.

[13] Loizos A, Plati C. Accuracy of pavement thicknesses estimation using different ground penetrating radar analysis approaches[J]. NDT & e International, 2007, 40(2): 147-157.

[14] Rmeili E, Scullion T. Detecting Stripping in Asphalt Concrete Layers Using Ground Penetrating Radar[J]. Transportation Research Record Journal of the Transportation Research Board, 1997, 1568(1):165-174.

[15] Hammons M, Quintus H V, Geary G, et al. Detection of Stripping in Hot-Mix Asphalt[J]. Transportation Research Record Journal of the Transportation Research Board, 2006, 1949(1):20-31.

[16] 张龙. 基于探地雷达技术的沥青路面早期水损害评价[D]. 广州:华南理工大学, 2013.

[17] 张蓓, 王复明, 刘俊. 沥青路面剥落的路面雷达电磁波模拟[J]. 公路交通科技, 2008, 25(4): 7-10.

[18] Alqadi I L, Leng Z, Larkin A. In-Place Hot Mix Asphalt Density Estimation Using Ground Penetrating Radar[J]. Transportation Research Record Journal of the Transportation Research Board, 2011, 2152(-1):19-27.

[19] Saarenketo T. Using Ground-Penetrating Radar and Dielectric Probe Measurements in

Pavement Density Quality Control[J]. Transportation Research Record Journal of the Transportation Research Board, 1997, 1575(1):34-41.

[20] 李想堂, 王端宜, 张肖宁, 等. 探地雷达在高速公路沥青路面施工质量检测中的应用[J]. 中外公路, 2007, 27(1): 70-73.

[21] Sebesta S, Wang F, Scullion T, et al. New infrared and radar systems for detecting segregation in hot-mix asphalt construction[R]. 2006.

[22] Sebesta S, Scullion T. Application of infrared imaging and ground-penetrating radar to detect segregation in hot-mix asphalt overlays[J]. Transportation Research Record: Journal of the Transportation Research Board, 2003 (1861): 37-43.

[23] Diamanti N, Redman D. Field observations and numerical models of GPR response from vertical pavement cracks[J]. Journal of Applied Geophysics, 2012, 81: 106-116.

[24] Benedetto A. A three dimensional approach for tracking cracks in bridges using GPR[J]. Journal of Applied Geophysics, 2013, 97: 37-44.

[25] Diamanti N, Redman D, Giannopoulos A. A study of GPR vertical crack responses in pavement using field data and numerical modelling[C]//Ground Penetrating Radar (GPR), 2010 13th International Conference on. IEEE, 2010: 1-6.

[26] Solla M, Lagüela S, González-Jorge H, et al. Approach to identify cracking in asphalt pavement using GPR and infrared thermographic methods: Preliminary findings[J]. NDT & E International, 2014, 62: 55-65.

[27] Xia T, Xu X, Vekatachalam A, et al. Development of a high speed UWB GPR for rebar detection[C]//Ground Penetrating Radar (GPR), 2012 14th International Conference on. IEEE, 2012: 66-70.

[28] Li J, Xing H, Chen X, et al. Extracting rebar's reflection from measured GPR data[C]//Ground Penetrating Radar, 2004. GPR 2004. Proceedings of the Tenth International Conference on. IEEE, 2004, 1: 367-370.

[29] Chang C W, Lin C H, Lien H S. Measurement radius of reinforcing steel bar in concrete using digital image GPR[J]. Construction and Building Materials, 2009, 23(2): 1057-1063.

[30] Pérez-Gracia V, García F G, Abad I R. GPR evaluation of the damage found in the reinforced concrete base of a block of flats: A case study[J]. NDT & e International, 2008, 41(5): 341-353.

[31] Scullion T, Saarenketo T. Ground penetrating radar technique in monitoring defects in roads and highways[C]//Symposium on the Application of Geophysics to Engineering and Environmental Problems 1995. Society of Exploration Geophysicists, 1995: 63-72.

[32] Sheftick D E, Bartoski T A. Study of void detection methods and slab stabilization procedures[R]. 1997.

[33] Yanqing X, Xiaoming H, Wei Z, et al. Improvement on Recognition Method of Void beneath Slab Based on Nondestructive Testing Technologies[M]//Sustainable Construction Materials 2012. 2013: 77-91.

[34] 魏国杰. 混凝土路面脱空雷达检测及处治关键技术研究[D]. 重庆:重庆交通大学, 2014.

[35]JTJ 073.1-2001, 公路水泥混凝土路面养护技术规范[S]. 北京:人民交通出版社出版,2001.

[36] JTG E60-2008, 公路路基路面现场测试规程[S]. 北京: 人民交通出版社, 2008.

[37] 张宁,钱振东,黄卫等.水泥混凝土路面板下地基脱空状况的评定与分析[J].公路交通科技,2004,21(1):4-7,21.

[38] 王陶.基于遗传算法的刚性路面脱空判定[J].中国公路学报,2003,16(3):23-26.

[39] Kofman L, Ronen A, Frydman S. Detection of model voids by identifying reverberation phenomena in GPR records[J]. Journal of Applied geophysics, 2006, 59(4): 284-299.

[40] 万捷. 水泥混凝土路面板底脱空检测及防治技术研究[D]. 西安:长安大学, 2007.

[41] 柴福斌. 基于探地雷达的水泥混凝土路面板底脱空检测技术[D]. 西安:长安大学, 2009.

[42] 冯晋利. 路面雷达在刚性路面脱空识别中的应用研究[D]. 郑州:郑州大学, 2007.

[43] Kalinski M E, Karem W A. Use of Ground Penetrating Radar to Locate Voids and Estimate Lateral Extent of Grouting Within a Pavement System[C]//Transportation Research Board 89th Annual Meeting. 2010 (10-1813).

[44] Chen D H, Tang C, Xiao H B, et al. Utilizing electromagnetic spectrum for subsurface void detection—case studies[J]. Arabian Journal of Geosciences, 2015, 8(9): 7705-7717.

[45] 阳恩慧, 向可明, 廖志勇. 用弯沉对路基压力注浆效果结构性评价[J]. 路基工程, 2009, (1): 42-43.

[46] 柴震林, 袁捷, 罗勇. 机场复合道面注浆工程效果评价方法研究[J]. 中国民航大学学报, 2013, 31(5): 38-42.

[47] 曾胜, 赵健, 邹金锋, 等. 水泥混凝土路面板底脱空注浆的有效性检验指标[J]. 中国公路学报, 2010, 23(6): 7-15.

[48] 李燕清, 向阳开, 熊潮波. 水泥混凝土路面板底脱空灌浆材料研究[J]. 重庆交通大学学报（自然科学版）, 2012, 31(5): 957-961.

[49] Ni J C, Cheng W C. Quality Control for Grouting Under Rigid Pavement[M]//Pavements and Materials: Recent Advances in Design, Testing and Construction. 2011: 183-191.

[50] Ni J C, Cheng W C. Trial grouting under rigid pavement: a case history in Magong Airport, Penghu[J]. Journal of Testing and Evaluation, 2011, 40(1): 1-12.

[51] 黄宏伟, 杜军, 谢雄耀. 盾构隧道壁后注浆的探地雷达探测模拟试验[J]. 岩土工程学报, 2007, 29(2): 243-248.

[52] Xie X, Zeng C. Non-destructive evaluation of shield tunnel condition using GPR and 3D laser scanning[C]//Ground Penetrating Radar (GPR), 2012 14th International Conference on. IEEE, 2012: 479-484.

[53] Liu H, Xie X, Sato M. Accurate thickness estimation of a backfill grouting layer behind shield tunnel lining by CMP measurement using GPR[C]//Ground Penetrating Radar (GPR), 2012 14th International Conference on. IEEE, 2012: 137-142.

[54] Scullion T, Lau C L, Saarenketo T. Performance specifications of ground penetrating radar[C]//Sixth International Conference on Ground Penetrating Radar. 1996: 341-346.

[55] Texas Transportation Institute, Specification NO. TxDOT-845-49-62, October 1997.

[56] Li Q, Xia Z, Shao J, et al. Non-destructive survey of pavement layer thicknesses with ground penetrating radar[C]//TENCON 2013-2013 IEEE Region 10 Conference (31194). IEEE, 2013: 1-4.

[57] Annan A P. Ground Penetrating Radar Workshop Notes[J]. Tech.rep.sensors & Software Inc, 1999, 8029(1):453-457.

[58] Friedman S P. A saturation degree-dependent composite spheres model for describing the effective dielectric constant of unsaturated porous media[J]. Water Resources Research, 1998, 34(11): 2949-2961.

[59] Tsui F, Matthews S L. Analytical modelling of the dielectric properties of concrete for subsurface radar applications[J]. Construction and Building Materials, 1997, 11(3): 149-161.

[60] Al-Qadi I L, Lahouar S, Loulizi A. Ground-Penetrating Radar Calibration at the Virginia Smart

Road and Signal Analysis to Improve Prediction of Flexible Pavement Layer Thicknesses[R]. Virginia Center for Transportation Innovation and Research, 2005.

[61] Warren C, Giannopoulos A, Giannakis I. gprMax: Open source software to simulate electromagnetic wave propagation for Ground Penetrating Radar[J]. Computer Physics Communications, 2016, 209: 163-170.

[62] 王淼. 高等级公路路面路基 GPR 量化检测技术研究[D]. 青岛:中国海洋大学, 2011.

[63] Warren C, Giannopoulos A. (April 13, 2017) "gprMax User Guide - Release 3.0.19". [Online] Available: https://media.readthedocs.org/pdf/gprmax/latest/gprmax.pdf (April 26, 2017)

[64] 尹光辉. 基于 GprMax 的道路空洞探地雷达图像正演模拟[D]. 西安:长安大学, 2015.

[65] Xie X, Zeng C, Wang Z. GPR signal enhancement using band-pass and K–L filtering: a case study for the evaluation of grout in a shielded tunnel[J]. Journal of Geophysics and Engineering, 2013, 10(3): 034003.

[66] Zhao S, Shangguan P, Al-Qadi I L. Application of regularized deconvolution technique for predicting pavement thin layer thicknesses from ground penetrating radar data[J]. NDT & E International, 2015, 73: 1-7.

[67] 刘金琨, 沈晓蓉, 赵龙. 系统辨识理论及 MATLAB 仿真[M]. 电子工业出版社, 2013.

[68] 罗鹏飞,杨世海.数字信号处理实践方法[M].2 版.北京:电子工业出版社,2004

[69] 王宏钺, 林治铖. 信号处理中的不适定问题[J]. 信号处理, 1985(3):39-47.

[70] Tikhonov A N, Arsenin V I A. Solutions of ill-posed problems[M]. Washington, DC: Winston, 1977.

# 个人简历、在读期间发表的学术论文与研究成果

**个人简历：**

郑家麒，男，1993 年 1 月 30 日出生。

2014 年 7 月毕业于长安大学道路、桥梁与渡河工程专业，获得学士学位。

2014 年 9 月进入同济大学道路与铁道工程专业，攻读硕士研究生。

**已发表论文：**

[1]郑家麒.GPR 信号处理技术研究及在道路沥青注浆评价中的应用[J]. 交通科技,2017,(02):143-146.